\documentclass{emulateapj}

\newcommand{\iso}{{\em ISO}}
\newcommand{\iras}{{\em IRAS}}
\newcommand{\mum}{\ifmmode{\rm \mu m}\else{$\mu$m}\fi}
\newcommand{\figpath}{./}

\submitted{submitted to The Astronomical Journal}
\journalinfo{ }

\begin{document}

\title{Spectral Calibration in the Mid-Infrared:  Challenges
and Solutions}

\author{
G.~C.~Sloan\altaffilmark{1},
T.~L.~Herter\altaffilmark{2},
V.~Charmandaris\altaffilmark{3,4,5},
K.~Sheth\altaffilmark{6},
M.~Burgdorf\altaffilmark{7}, \&
J.~R.~Houck\altaffilmark{2}
}
\altaffiltext{1}{Cornell University, Center for Radiophysics and Space
  Research, Ithaca, NY 14853-6801, sloan@isc.astro.cornell.edu}
\altaffiltext{2}{Cornell University, Astronomy Department,
  Ithaca, NY 14853-6801}
\altaffiltext{3}{Department of Physics and ITCP, University of Crete,
  GR-71003, Heraklion, Greece}
\altaffiltext{4}{Institute for Astronomy, Astrophysics, Space Applications \& 
  Remote Sensing, National Observatory of Athens, GR-15236, Penteli, Greece}
\altaffiltext{5}{Chercheur Associ\'e, Observatoire de Paris, F-75014, Paris, 
  France}
\altaffiltext{6}{National Radio Astronomy Observatory, 520 Edgemont Road, 
  Charlottesville, VA 22903, USA}
\altaffiltext{7}{HE Space Operations, Flughafenallee 24, D-28199 Bremen,
  Germany}

\begin{abstract}

We present spectra obtained with the Infrared Spectrograph 
(IRS) on the {\it Spitzer Space Telescope} of 33 K giants and 
20 A dwarfs to assess their suitability as spectrophotometric
standard stars.  The K giants confirm previous findings that 
the strength of the SiO absorption band at 8~\mum\ increases 
for both later optical spectral classes and redder 
($B-V$)$_0$ colors, but with considerable scatter.  For K 
giants, the synthetic spectra underpredict the strengths of 
the molecular bands from SiO and OH.  For these reasons, the 
assumed true spectra for K giants should be based on neither 
the assumption that molecular band strengths in the infrared 
can be predicted accurately from optical spectral class or 
color nor synthetric spectra.  The OH bands in K giants grow 
stronger with cooler stellar temperatures, and they are
stronger than predicted by synthetic spectra.  As a group, 
A dwarfs are better behaved and more predictable than the K 
giants, but they are more likely to show red excesses from 
debris disks.  No suitable A dwarfs were located in parts of 
the sky continuously observable from {\it Spitzer}, and with 
previous means of estimating the true spectra of K giants 
ruled out, it was necessary to use models of A dwarfs to 
calibrate spectra of K giants from observed spectral ratios 
of the two groups and then use the calibrated K giants as 
standards for the full database of infrared spectra from {\it 
Spitzer}.  We also describe a lingering artifact that affects
the spectra of faint blue sources at 24~\mum.
\end{abstract}


\section{Introduction} \label{s.intro} % Sec. 1.0

The Infrared Spectrograph \citep[IRS,][]{hou04} on the
{\it Spitzer Space Telescope} \citep{wer04} has produced a 
rich legacy of scientific results, due in large part to its 
unprecedented sensitivity.  One of the few drawbacks of such
a powerful telescope is that bright objects saturate the 
detectors, eliminating most well understood infrared standard 
stars from consideration as calibrators.

Spectral calibration generally proceeds with the following 
equation:
\begin{equation} % Eq. 1
  S_T = \frac{S_o}{C_o} C_T,
\end{equation}
where all of the quantities are functions of wavelength.  
$S_o$ and $S_T$ are the observed and true spectra of the 
science target, respectively, and $C_o$ and 
$C_T$ are defined similarly for the calibrator (or standard 
star).  The quantity $C_T$ is the assumed spectrum of the
calibrator, referred to hereafter as its {\it truth} 
spectrum.  For ground-based spectrometers, the calibration 
varies with atmospheric conditions, requiring spectra of the 
standard and science target to either be corrected for 
differences in atmospheric transmission or be obtained at 
similar airmasses and at roughly the same time.

For space-based missions there is no atmosphere, and the
calibration is more stable.  The problem then becomes one of 
determining a general spectral correction applicable as long as 
the responsivity of the instrument does not change.  Previous 
satellite missions have defined a relative spectral response 
function $RSRF=C_o/C_T$ which can be used to calibrate the 
spectra of science targets: $S_T = S_o/RSRF$.

Any errors in the assumed truth spectra of the standards will
propagate as artifacts into the entire database of calibrated
spectra.  For calibration of an instrument on a space-based 
mission, the challenge is to choose proper standards and 
define good truth spectra.  The improved sensitivity of 
each new mission has revealed that we have more to learn 
about the stars chosen as standards.

The calibration of the {\it Infrared Astronomical Satellite}
(\iras) is a good example of the challenges.  The \iras\ 
team quickly discovered that $\alpha$~Lyr (A0 V), the primary
photometric standard at all wavelengths, had an infrared
excess beyond $\sim$20~\mum\ due to a circumstellar debris 
disk \citep{aum84}.  Another surprise followed.  
\cite{coh92b} showed that the original calibration of the 
Low-Resolution Spectrometer (LRS) on \iras\ produced an 
emission artifact in all of the spectra, due to the standard 
assumption at the time that K giants had featureless continua 
in the infrared.

The problem of assuming correct truth spectra for standard 
stars remains fundamental to infrared spectroscopy, despite 
the two decades that have passed between the \iras\ and
{\it Spitzer} missions.  For the IRS, this issue was 
particularly acute.  To avoid the danger of saturation, a 
standard star for the low-resolution IRS modules needed to be 
fainter than $\sim$2 Jy at 12~\mum, which ruled out all of 
the previously calibrated standard stars.  Thus, we had to
start over with new standards.

The spectra calibrated by the {\it Spitzer} Science Center 
(SSC) met all pre-launch calibration requirements.  This
paper describes the effort by IRS team members at Cornell and
the SSC to improve on that effort.  It concentrates on the 
observational findings made as part of that effort, both to 
justify some of the decisions made and to guide the 
calibration of future infrared space telescopes.  The 
sections below describe how we chose the standards for the 
IRS, determined their truth spectra, and analyzed the larger 
sample of candidate standards.  The approach was iterative, 
with multiple definitions of truth spectra, tests, and 
improvements, making a linear presentation challenging.

In Section 2, we review methods used to estimate truth 
spectra of infrared standards before {\it Spitzer} launched.
In Section 3, we describe how we chose candidate standards
and how the calibration plan evolved during the {\it Spitzer} 
mission.  Section 4 describes the observations, the Cornell 
data pipeline, and the construction of truth spectra and 
spectral corrections to calibrate all low-resolution IRS 
spectra.  Section 5 examines actual coadded spectra for the 
standard stars observed most frequently and compares them to 
various truth spectra.  A key finding is that our method does 
not propagate any identifiable artifacts into the IRS 
database.  We also point out some of the shortcomings with 
traditional methods of estimating truth spectra.  Section 6 
focuses on the K giants, especially the SiO and OH absorption 
bands which must be properly accounted for if they are to be 
used as standards.  Section 7 concentrates on the photometric 
properties of the standards, both K giants and A dwarfs.  
Section 8 focuses spectra of the A dwarfs, while Section 9 
investigates the properties of the debris disks uncovered in 
our sample.  Finally, Section 10 discusses our findings and 
draws some conclusions.

\section{Truth spectra} % Sec. 2.0

\subsection{Composite spectra} \label{s.composites} % Sec. 2.1

Prior to the work by \cite{coh92b}, K giants were commonly
used as standards, with the assumption that their spectra 
could be modeled with a 10$^4$ K Planck function.  The high 
temperature of the assumed Planck function mimics the effect 
of the H$^-$ ion, which has an opacity that increases 
smoothly with wavelength, pushing the photosphere to cooler 
layers of the star at longer wavelengths.  \cite{eng92} 
showed that a better fit to the continuum could be obtained 
with the Engelke function, which makes the brightness 
temperature of the star a smooth function of wavelength.

\cite{coh92b} uncovered the presence of molecular absorption
bands in the infrared spectra of late-type giants and
documented their impact on the calibration of the \iras\
spectra.  To recalibrate the LRS database, \cite{coh92a} 
started with A dwarfs, which can be modeled in a relatively 
straightforward manner.  They based their calibration on 
$\alpha$~Lyr (A0~V) and $\alpha$~CMa (A1~V), relying on
$\alpha$~CMa beyond 20~\mum\ due to the red dust excess in
the spectrum of $\alpha$~Lyr.  Kurucz models 
\citep{kur79}\footnote{\cite{kur79} describes the technique,
but the individual models have been distributed outside the 
refereed literature.  These models assume plane-parallel 
geometry and local thermodynamic equilibrium, as explained in
more detail in his Sec.\ IV (b).} of these two stars served 
as the truth spectra for the calibration of other bright 
infrared standards.

\cite{coh92b} produced a continuous and fully calibrated
spectrum of $\alpha$~Tau (K5 III), primarily using 
$\alpha$~CMa as the calibrator and its Kurucz model as 
the truth spectrum.  They built a composite spectrum of 
$\alpha$~Tau from spectral segments observed with a variety 
of ground- and space-based telescopes.  They ensured 
photometric accuracy and continuity in the composite by 
pinning each spectral segment to infrared photometry.  The 
resulting spectrum of $\alpha$~Tau showed multiple molecular 
absorption bands from the overtone and fundamental modes in 
CO and SiO.  \cite{coh92b} also published 
corrections for the LRS database by comparing their new 
spectrum of $\alpha$~Tau to the original LRS calibration.  

\cite{coh95} followed up with additional composite spectra
of $\alpha$~Boo (K1.5~III) and four other giants, calibrated
from Kurucz models of $\alpha$~Lyr and $\alpha$~CMa and the
existing composite spectrum of $\alpha$~Tau.  These composite
spectra then served as the calibrators for further composite
spectra \citep{coh96a,coh96b,coh03}.  Table~\ref{t.cohen} 
lists the 16 composite spectra produced using this method.  
These spectra have the advantage that the spectral features 
are as actually observed, based on spectral ratios with a 
fidelity that can be traced back to the original Kurucz 
models of $\alpha$~Lyr and $\alpha$~CMa.  

\begin{deluxetable}{llrlcc} % Table 1
\tablenum{1}
\tablecolumns{6}
\tablewidth{0pt}
\tablecaption{Primary infrared standards with composite spectra}
\label{t.cohen}
\tablehead{
  \colhead{Standard} & \colhead{Spectral} & \colhead{F$_{\nu}$ at 12} &
  \colhead{Cohen et} & \colhead{Used for} & \colhead{SWS} \\
  \colhead{Star}     & \colhead{Type}     & 
  \colhead{\mum\ (Jy)\tablenotemark{a}} &
  \colhead{al. Ref.} & \colhead{Template} & \colhead{Update\tablenotemark{b}}
} 
\startdata
$\alpha$ Lyr   & A0 V        &  28.7 & 1992a        & Y       & \nodata \\
$\alpha$ CMa   & A1 V        &  98.7 & 1992a        & Y       & \nodata \\
$\alpha^1$ Cen & G2 V        & 153.2 & 1996a        & \nodata & Y       \\
$\beta$ Gem    & K0 III      &  85.9 & 1995         & Y       & Y\tablenotemark{c}\\
$\alpha$ Boo   & K1.5 III    & 547.0 & 1995, 1996b  & Y       & Y       \\
$\alpha$ TrA   & K2 III      &  99.3 & 1996a        & \nodata & \nodata \\
$\alpha$ Hya   & K3 II-III   & 108.7 & 1995         & Y       & \nodata \\
$\epsilon$ Car & K3 III      & 169.7 & 1996a        & \nodata & \nodata \\
$\beta$ UMi    & K4 III      & 110.6 & 2003         & \nodata & Y       \\
$\alpha$ Tau   & K5 III      & 482.6 & 1992b        & Y       & Y       \\
$\gamma$ Dra   & K5 III      & 107.0 & 1996b        & \nodata & Y       \\
$\beta$ And    & M0 III      & 197.7 & 1995         & \nodata & Y       \\
$\mu$ UMa      & M0 III      &  69.6 & 1996b        & \nodata & Y       \\
$\alpha$ Cet   & M1.5 III    & 161.9 & 1996b        & \nodata & Y       \\
$\beta$ Peg    & M2.5 II-III & 267.1 & 1995         & \nodata & Y       \\
$\gamma$ Cru   & M3.5 III    & 596.8 & 1996b        & \nodata & Y       \\
\enddata
\tablenotetext{a}{Photometry from the IRAS Point-Source Catalog 
  \citep[PSC;][]{psc} and color corrected by dividing by 1.45}
\tablenotetext{b}{Short-Wavelength Spectrometer; \cite{eng06}.}
\tablenotetext{c}{$\beta$ Gem was not observed by the SWS; this spectrum
  was built from SWS observations of similar stars.}
\end{deluxetable}

\subsection{Spectral templates} \label{s.templates} % Sec. 2.2

All of the sources in Table~\ref{t.cohen} are too bright 
for the mid-infrared calibration of the instruments on {\it 
Spitzer.}  However, they can still serve as the basis for 
calibrating {\it Spitzer} through the process of spectral 
templating.  Using the composite spectra of sources in 
Table~\ref{t.cohen} as prototypes, \citep{coh99} created 422 
spectral templates of fainter sources.  Each templated 
spectrum is based on the composite spectrum of the bright 
standard with the same optical spectral class, after 
adjusting for differences in photometry and interstellar 
reddening.  Sources with discrepant colors were excluded.
The accuracy of the spectral template depends on the 
assumption that the infrared spectrum of a source can be 
accurately predicted from its optical spectral class.  In 
preparation for the {\it Spitzer} mission, \cite{coh03} 
expanded the wavelength coverage of their templates using 
optical spectroscopy for shorter wavelengths and Engelke 
functions for longer wavelengths.  The resulting products, 
called supertemplates, were planned as the basis of the 
calibration of the Infrared Array Camera (IRAC) on {\it 
Spitzer}.  They are generated using the same methods and 
assumptions as the earlier templates.

The spectral templates suffer from two limitations.  First, 
because they are based on the actual observations which make 
up the composite spectra, they contain some noise.  This
noise is a minor issue for photometric calibration, but can 
be more significant for spectroscopic calibration.  Second, 
some stars of a given spectral type may not have identical 
spectra to their prototype. \cite{her02} found considerable 
variation in the strengths of the CO and SiO bands in the 
sample of late-type stars observed by the Short Wavelength 
Spectrometer (SWS) on the {\it Infrared Space Observatory} 
(\iso).  Again, these variations affect spectroscopic 
calibration more than photometric calibration.  The SWS 
sample contained a range of luminosity classes at most of the 
spectral classes examined, leaving open the possibility that 
variations in luminosity were responsible for the range of 
band strengths within each spectral class, but the size of 
the sample was too small to search for systematic trends.

\subsection{Synthetic spectra} \label{s.models} % Sec. 2.3

The calibration of the SWS relied on many of the standard 
stars in Table~\ref{t.cohen} \citep{shi03,sch96}, but
with synthetic spectra used for the truth spectra at shorter
wavelengths.  The synthetic spectra are free from noise,
which is strong enough in the composite spectra to propagate 
through the calibration and into the SWS database.  

However, \cite{pri02} revealed that the synthetic spectra
underestimated the strength of the SiO molecular band at 
8~\mum, and that the difference between observed and truth
spectrum could propagate as an emission artifact into the
entire SWS database.  The difficulty with synthetic spectra 
of K giants lies in the atmospheric models on which they are 
based.  The molecular absorption bands arise in rarified and 
extended layers of the atmosphere that have resisted accurate 
modeling (D.\ Carbon, private communication, 2002).  The 
errors in the modeled temperatures and densities as a 
function of radius lead to inaccurate profiles for the 
absorption bands (we return to this point in 
Section~\ref{s.oh}).

Synthetic spectra of earlier-type stars, such as A dwarfs,
are more reliable.  The observational problem is that A 
dwarfs are relatively rare compared to K giants among bright 
infrared targets, making it less likely they are located in 
readily accessible parts of the sky.  For {\it Spitzer}, no 
A dwarfs suitable as standards could be found in the 
continuous viewing zones.

\section{The calibration plan} % Sec. 3.0

\subsection{Choice of stellar classes for standards} \label{s.choice} % Sec. 3.1

When planning the spectral calibration of the IRS on {\it
Spitzer}, we had to balance multiple challenges.  K giants
have two strong molecular absorption bands in the IRS
wavelength range, the SiO fundamental at 8~\mum\ and 
the CO fundamental at 5~\mum.  Using A dwarfs as calibrators 
would avoid these bands, but the hydrogen recombination lines 
in their spectra present their own problems.  Any difference 
in the strength, position, or profile of these lines between 
the synthetic and observed spectra introduces narrow 
artifacts in the calibration.  To match the line profiles in 
A dwarfs, one must convolve the synthetic spectrum with the 
correct unresolved line profile, which is both a function of 
wavelength and position in the slit.  In addition, an offset 
of the position of the star in the dispersion direction 
(perpendicular to the long axis of the spectroscopic slit) 
will displace the lines in wavelength space.  

Prior to the start of the {\it Spitzer} mission, we decided
to use both the A dwarfs and K giants, using one group to 
mitigate for specific weaknesses in the other, since the K 
giants do not show atomic absorption lines in the infrared, 
and the A dwarfs do not show molecular absorption bands.  We 
excluded M giants because of their deeper and more complex 
absorption bands and increased likelihood of variability.  
We did not consider solar analogues as potential standards
because they are relatively uncommon and they are likely to
show both molecular and atomic absorption features.

\subsection{The sample of standard stars} \label{s.sample} % Sec. 3.2

\begin{deluxetable*}{llrrccllclrr} % Table 2
\tablenum{2}
\tablecolumns{12}
\tablewidth{0pt}
\tablecaption{The sample of K giants}
\label{t.kstd}
\tablehead{
  \colhead{Adopted} & \colhead{HR}   & \colhead{HD}         & \colhead{HIP} &
  \colhead{RA}      & \colhead{Dec.} & \multicolumn{3}{c}{Spectral Type}    &
  \colhead{F$_{\nu}$ at 12}  & \multicolumn{2}{c}{Observations} \\
  \colhead{Name}    & \colhead{ }    & \colhead{ }          & \colhead{ }   &
  \multicolumn{2}{c}{(J2000)}        & \colhead{Literature} & 
  \colhead{Reference\tablenotemark{a}} & \colhead{Modified\tablenotemark{b}} & 
  \colhead{\mum\ (Jy)\tablenotemark{c}} & \colhead{Total} & \colhead{Used}
}
\startdata
HD 41371   & \nodata &  41371  &  28420 & 06 00 07.71 & $-$64 18 36.0 & K0 III & MC1      &      K0    & 0.372                  &   1 &  1 \\
HR 7042    & 7042    & 173398  &  91606 & 18 40 56.41 & $+$62 44 58.1 & K0 III & H55, E62 &      K0    & 0.800                  &   2 &  2 \\
HD 51211   & \nodata &  51211  &  32813 & 06 50 25.27 & $-$69 59 10.6 & K0 III & MC1      &      K0    & 0.476                  &   2 &  2 \\
HR 6606    & 6606    & 161178  &  86219 & 17 37 08.88 & $+$72 27 20.9 & G9 III & H55      & {\bf K0}   & 1.143                  &  64 & 36 \\
HR 6348    & 6348    & 154391  &  83289 & 17 01 16.93 & $+$60 38 55.5 & K1 III & H55      & {\bf K0}   & 0.755                  &  84 & 57 \\
HR 2712    & 2712    &  55151  &  34270 & 07 06 14.31 & $-$68 50 15.3 & K0 III & MC1      &      K0    & 0.670                  &   4 &  4 \\
\\
HR 1815    & 1815    &  35798  &  24256 & 05 12 25.76 & $-$81 32 30.2 & K1 III & MC1      &      K1    & 0.743                  &   3 &  3 \\
HD 59239   & \nodata &  59239  &  35809 & 07 23 06.83 & $-$70 38 11.8 & K1 III & MC1      &      K1    & 0.574                  &   5 &  5 \\
HD 156061  & \nodata & 156061  &  84494 & 17 16 27.69 & $-$25 18 19.6 & K1 III & MC4      &      K1    & 0.534\tablenotemark{d} &   2 &  2 \\
HR 6790    & 6790    & 166207  &  88732 & 18 06 53.48 & $+$50 49 22.2 & K0 III & H55, E62 & {\bf K1}   & 0.797                  &   4 &  3 \\
HD 39567   & \nodata &  39567  &  27436 & 05 48 35.14 & $-$65 10 06.6 & K2 III & MC1      & {\bf K1}   & 0.315                  &   2 &  1 \\
HR 7341    & 7341    & 181597  &  94890 & 19 18 37.87 & $+$49 34 10.0 & K1 III & H55      &      K1    & 0.996                  &  74 & 29 \\
\\
HD 52418   & \nodata &  52418  &  33306 & 06 55 40.95 & $-$68 30 20.6 & K2 III & MC1      &      K2    & 0.589                  &   1 &  1 \\
HD 130499  & \nodata & 130499  &  72238 & 14 46 23.43 & $+$56 36 59.0 & K2 III & Y61      &      K2    & 0.672                  &   2 &  2 \\
HD 39577   & \nodata &  39577  &  27564 & 05 50 16.54 & $-$55 03 12.0 & K2 III & MC1      &      K2    & 0.383                  &   2 &  2 \\
HD 115136  & \nodata & 115136  &  64522 & 13 13 28.03 & $+$67 17 16.7 & K2 III & S60      &      K2    & 0.815                  &   3 &  3 \\
HD 50160   & \nodata &  50160  &  32396 & 06 45 49.77 & $-$70 31 21.6 & K2 III & MC1      &      K2    & 0.361                  &   4 &  4 \\
%$\xi$ Dra & 6688    & 163588  &  87585 & 17 53 31.73 & $+$56 52 21.5 & K2 III & R52, M53 &      K2    & 11.53                  &   0 &  0 \\
\\
HD 56241   & \nodata &  56241  &  34381 & 07 07 42.68 & $-$76 02 57.2 & K3 III & MC1      &      K3    & 0.551                  &   5 &  5 \\
HD 42701   & \nodata &  42701  &  28970 & 06 06 50.55 & $-$67 17 00.0 & K3 III & MC1      &      K3    & 1.121                  &   2 &  2 \\
HD 44104   & \nodata &  44104  &  29820 & 06 16 48.05 & $-$54 37 01.2 & K3 III & MC1      &      K3    & 0.493                  &   2 &  1 \\
HD 23593   & \nodata &  23593  &  17329 & 03 42 34.46 & $-$64 11 43.1 & K3 III & MC1      &      K3    & 0.738                  &   2 &  2 \\
HD 214873  & \nodata & 214873  & 112021 & 22 41 25.65 & $-$12 13 47.8 & K2 III & MC4      & {\bf K3}   & 1.132                  &   1 &  1 \\
\\
HD 19241   & \nodata &  19241  &  14188 & 03 02 55.91 & $-$60 47 51.9 & K5 III & MC1      & {\bf K4}   & 0.859                  &   3 &  3 \\
HD 99754   & \nodata &  99754  &  55981 & 11 28 23.34 & $-$23 49 36.1 & K4 III & MC4      &      K4    & 0.753                  &   2 &  2 \\
HD 166780  & \nodata & 166780  &  88877 & 18 08 38.85 & $+$57 58 46.9 & K5 III & M50      & {\bf K4}   & 0.777                  &  41 & 26 \\
HD 214046  & \nodata & 214046  & 111551 & 22 35 53.95 & $-$20 56 05.2 & K4 III & MC4      &      K4    & 0.689                  &   2 &  2 \\
HD 38214   & \nodata &  38214  &  26751 & 05 41 01.53 & $-$54 44 22.1 & K5 III & MC1      & {\bf K4}   & 0.484                  &   3 &  3 \\
BD+62 1644 & \nodata & \nodata &  91691 & 18 41 52.39 & $+$62 57 41.2 & K3 III & C02      & {\bf K4}   & 0.272                  &   3 &  3 \\
\\
HD 53561   & \nodata &  53561  &  34258 & 07 06 05.03 & $+$13 59 08.8 & K5 III & M50      &      K5    & 0.855\tablenotemark{d} &   2 &  2 \\
HD 15508   & \nodata &  15508  &  11364 & 02 26 23.35 & $-$68 38 07.2 & K4 III & MC1      & {\bf K5}   & 0.576                  &   2 &  2 \\
HD 173511  & \nodata & 173511  &  91673 & 18 41 40.59 & $+$61 32 47.1 & K5 III & M50      &      K5    & 0.814                  & 132 & 84 \\
HD 34517   & \nodata &  34517  &  24559 & 05 16 10.06 & $-$41 14 57.2 & K5 III & MC2      &      K5    & 1.247                  &   3 &  3 \\
HD 39608   & \nodata &  39608  &  27516 & 05 49 36.46 & $-$60 40 34.7 & K5 III & MC1      &      K5    & 1.001                  &   1 &  1 \\
\enddata
\tablenotetext{a}{The listed references are not meant to be complete.  C02 = 
  Cohen, M., private comm.\ (2002), E62 = \cite{egg62}, H55 = \cite{hal55}, 
  MC1, 2, 4 are the Michigan catalogue, vol.\ 1, 2, and 4 
  \citep{mss75,mss78,mss88}, M50 = \cite{mp50}, 
  % M53 = \cite{mor53}, R52 = \cite{rom52}, 
  S60 = \cite{ste60}, and Y61 = \cite{yos61}.}
\tablenotetext{b}{Changes are in bold; see Sec.~\ref{s.ksp}.}
\tablenotetext{c}{Photometry from the IRAS Faint-Source Catalog 
  \citep[FSC;][]{fsc} and color corrected by dividing by 1.45, unless noted 
  otherwise.}
\tablenotetext{d}{From the IRAS PSC and color corrected as the FSC data.}
\end{deluxetable*}

\begin{deluxetable*}{llrrcclllrr} % Table 3
\tablenum{3}
\tablecolumns{11}
\tablewidth{0pt}
\tablecaption{The sample of A dwarfs}
\label{t.astd}
\tablehead{
  \colhead{Adopted} & \colhead{HR}   & \colhead{HD}         & \colhead{HIP} &
  \colhead{RA}      & \colhead{Dec.} & \multicolumn{2}{c}{Spectral Type}    &
  \colhead{F$_{\nu}$ at 12}          & \multicolumn{2}{c}{Observations} \\
  \colhead{Name}    & \colhead{ }    & \colhead{ }          & \colhead{ }   &
  \multicolumn{2}{c}{(J2000)}        & \colhead{Literature} & 
  \colhead{Reference\tablenotemark{a}} & 
  \colhead{\mum\ (Jy)\tablenotemark{b}} & \colhead{Total}   & \colhead{Used} 
}
\startdata
HR 1014        & 1014    &  20888 &  15353 & 03 17 59.07 & $-$66 55 36.7 & A3 V     & E64, MC1      & 0.148             &  5 &  4 \\
$\nu$ Tau      & 1251    &  25490 &  18907 & 04 03 09.38 & $+$05 59 21.5 & A1 V     & S54, C69      & 0.783             &  2 &  2 \\
$\eta^1$ Dor   & 2194    &  42525 &  28909 & 06 06 09.38 & $-$66 02 22.6 & A0 V     & MC1           & 0.119             & 66 & 44 \\
HD 46190       & \nodata &  46190 &  30760 & 06 27 48.62 & $-$62 08 59.7 & A0 V     & MC1           & 0.077             &  1 &  1 \\
21 Lyn         & 2818    &  58142 &  36145 & 07 26 42.85 & $+$49 12 41.5 & A1 IV    & S54           & 0.386             &  3 &  3 \\
\\
26 UMa         & 3799    &  82621 &  47006 & 09 34 49.43 & $+$52 03 05.3 & A2 V     & M53, S54, E62 & 0.536             &  2 &  2 \\
HR 4138        & 4138    &  91375 &  51438 & 10 30 20.13 & $-$71 59 34.1 & A1 V     & MC1           & 0.418             &  9 &  7 \\
$\tau$ Cen     & 4802    & 109787 &  61622 & 12 37 42.16 & $-$48 32 28.7 & A2 V     & L72, MC2      & 0.937             &  7 &  5 \\
$\xi^1$ Cen    & 4933    & 113314 &  63724 & 13 03 33.31 & $-$49 31 38.2 & A0 V     & B60, MC2      & 0.335             &  3 &  3 \\
HR 5467        & 5467    & 128998 &  71573 & 14 38 15.22 & $+$54 01 24.0 & A1 V     & C69           & 0.115             & 49 & 38 \\
\\
HR 5949        & 5949    & 143187 &  78017 & 15 55 49.62 & $+$58 54 42.4 & A0 V     & C69           & 0.080             &  4 &  4 \\
$\delta$ UMi   & 6789    & 166205 &  85822 & 17 32 13.00 & $+$86 35 11.3 & A1 V     & S54, E60      & 0.539             & 43 & 33 \\
HD 163466      & \nodata & 163466 &  87478 & 17 52 25.37 & $+$60 23 47.0 & A2       & SIMBAD        & 0.090             & 48 & 34 \\
HR 7018        & 7018    & 172728 &  91315 & 18 37 33.50 & $+$62 31 35.7 & A0 V     & C69           & 0.134             & 27 & 18 \\
$\lambda$ Tel  & 7134    &  93148 & 175510 & 18 58 27.77 & $-$52 56 19.1 & B9 III   & V57, B58      & 0.294             &  3 &  3 \\
\\
HD 165459      & \nodata & 165459 &  88349 & 18 02 30.74 & $+$58 37 38.2 & A2       & SIMBAD        & 0.076             &  2 &  2 \\
29 Vul         & 7891    & 196724 & 101865 & 20 38 31.34 & $+$21 12 04.3 & A0 V     & S54, E62, C69 & 0.311             & 17 &  9 \\
$\epsilon$ Aqr & 7950    & 198001 & 102618 & 20 47 40.55 & $-$09 29 44.8 & A1 V     & E50, J53, M53 & 0.906             &  3 &  3 \\
$\mu$ PsA      & 8431    & 210049 & 109285 & 22 08 23.01 & $-$32 59 18.5 & A1.5 IVn & G06           & 0.460             &  3 &  3 \\
$\alpha$ Lac   & 8585    & 213558 & 111169 & 22 31 17.50 & $+$50 16 57.0 & A2 V     & S54, E57 & 0.959\tablenotemark{c} & 28 & 19 \\
\enddata
\tablenotetext{a}{The listed references are not meant to be complete.  
  B58 = \cite{bm58}, B60 = \cite{bm60}, C69 = \cite{cow69}, E50 = \cite{egg50},
  E57 = \cite{egg57}, E60 = \cite{egg60}, E62 = \cite{egg62}, 
  E64 = \cite{eva64}, G06 = \cite{gra06}, J53 = \cite{jm53}, 
  L72 = \cite{lev72}, M53 = \cite{mor53}, MC1 and MC2 are the Michigan 
  catalogue, vol.\ 1 and 2, \citep{mss75,mss78}, S54 = \cite{sle54}, and 
  V57 = \cite{dev57}.  For the two entries labelled ``SIMBAD'', no further
  reference information about the spectral type could be found.}
\tablenotetext{b}{Photometry from the IRAS FSC and color 
  corrected by dividing by 1.45, unless noted otherwise.}
\tablenotetext{c}{From the IRAS PSC; and color-corrected like 
  the FSC data.}
\end{deluxetable*}

The sample of IRS standards was developed with multiple purposes 
in mind.  Our primary goal was to observe a sufficient number 
to guarantee that after rejecting sources found to be 
inadequate for any reason, we still had enough sources 
to test the calibration for self-consistency.  Second, we 
wanted to observe a statistically significant sample of stars 
of each spectral class, including A0 and A1 dwarfs and each 
subclass of K giant, in order to study how their spectral 
properties varied within each subclass and from one subclass 
to the next.  

We selected several K giants in the continuous viewing zones
(CVZs), so that they could be observed in repeated IRS 
campaigns and help diagnose any variations in instrumental
responsivity.  With no suitable A dwarfs in the CVZs, we
chose A dwarfs positioned around the ecliptic so that one or 
more would always be available.  To provide sufficient 
temporal coverage and to sample spectral subclasses, we 
selected 20 A dwarfs.  Because of uncertainties in the 
spectral classes, our A dwarfs included sources from B9 to 
A3.  We did not observe later subclasses because they were 
not included in the original set of composite spectra.  To 
properly sample the K giants, we observed $\sim$5--6 of each 
subclass from K0 to K5, for a total of 33 K giants and 53 
standards.\footnote{This count does not include some fainter 
sources added later in the mission to test linearity or 
brighter sources used for the high-resolution modules.}

Candidate standards had to meet a number of criteria.  First,
they had to have a reliable spectral type, with both a 
spectral class and a luminosity class and with little or no
disagreement among published sources.  For K giants, we
required a luminosity class of ``III'', with no subtypes
indicated, to see if we could reduce the scatter in band
strengths observed by \cite{her02}.  Photometry at B, V, and 
in the 12~\mum\ \iras\ filter had to agree with the colors 
expected for the star's spectral class.  The [12]$-$[25] 
\iras\ color had to be consistent with a naked star and not 
indicate a possible debris disk.  For most of the standards 
observed during the scientific verification phase of the 
mission,\footnote{In 2003 November and December, between 
in-orbit checkout and the beginning of normal operations.} 
M.\ Cohen prepared spectral templates, using the methods 
described above.  He fitted the templates to the available 
photometry, and some candidates were rejected at this stage.  

Tables~\ref{t.kstd} and \ref{t.astd} present the K giants and 
A dwarfs ultimately observed by the IRS as potential 
standards and considered in this paper.  The K giants are 
ordered by their dereddened ($B$$-$$V$)$_o$ values as 
explained in Section~\ref{s.ksp}.  The A dwarfs are listed in 
order of right ascension.  

\subsection{Evolution of the plan} \label{s.evo} % Sec. 3.3

When {\it Spitzer} was launched, two plans were in place for 
a parallel calibration of the spectra, one using synthetic 
spectra and one using spectral templates.  As described 
above, each has strengths and weaknesses.  Cornell focused on 
the spectral templates, while the SSC focused on the 
synthetic spectra.  The Cornell team settled on three stars 
which behaved as expected, were relatively bright, and yet 
still faint enough to observe with the IRS Red Peak-up (PU) 
sub-array, which gave us photometry simultaneously with our 
spectra:  HR~6348, HD~166780, and HD~173511.  All were in or 
close enough to the northern continuous viewing zone to be 
observable from {\it Spitzer} most of the time.  

HR~6348 became our primary standard because it had the 
weakest fundamental SiO band of the three.  The SiO bandhead 
lies at 7.5~\mum, right at the boundary of the two spectral 
orders in the Short-Low (SL) module, and choosing a standard 
with a weak band reduced any problems which might arise from 
inaccuracies in the band profile in our truth spectrum.
\footnote{The 1-$\sigma$ pointing uncertainty of
0\farcs4 could produce a shift in the dispersion direction
as large as 0.013~\mum\ in SL1, which for HR~6348 would
result in an apparent 0.1\% artifact, due to an apparent
shift between the actual bandhead and where it appeared in
an observed spectrum.  For HD~173511, which has a stronger
SiO band, the effect would be 0.5\% of the continuum, which
is enough to be noticeable.}

It soon became apparent that the spectral template for 
HR~6348 (and the other standards) was insufficient as a truth
spectrum.  The apparent structure between 6 and 8~\mum\ in 
the template for HR~6348 has a peak-to-peak amplitude of 
$\sim$3\%, and this structure propagated from the template, 
which served as our initial truth spectrum to calibrate SL, 
to everything we calibrated with it.  It was readily apparent 
in all IRS spectra with high signal/noise ratios (SNRs).  
This structure originated in noise in the spectra used to 
construct the composite spectrum of $\beta$~Gem, the 
prototype for the K0 spectral template.  

We did not have synthetic spectra for HR~6348, HD~166780, or 
HD~173511, but the IRS spectra of standards which had been 
modeled, HR~6606 and HR~7341, confirmed our previous
concerns about the ability of the synthetic spectra to
accurately model the strength of the molecular absorption
bands (see Section~\ref{s.stdk2}).  Thus neither the 
templates nor the synthetic spectra would serve as truth 
spectra.

Fortunately, the large sample of potential IRS standards 
included two reasonably well-behaved and repeatedly observed 
A dwarfs, $\alpha$~Lac and $\delta$~UMi, putting us in a 
position to repeat the methodology of \cite{coh92b}.  Where 
they calibrated $\alpha$~Tau from $\alpha$~Lyr and 
$\alpha$~CMa, we calibrated HR~6348 (K0~III) from 
$\alpha$~Lac and $\delta$~UMi.  We then used the spectrum of 
HR~6348 to calibrate HD~173511, and with these two, all of 
the low-resolution spectra from the IRS on {\it Spitzer}.
The following sections show that this approach produces 
self-consistent results for all of the standard stars 
observed.

We started with Kurucz models of a generic A2 dwarf, scaled 
to the photometry for $\alpha$~Lac and $\delta$~UMi.  We
estimate errors in the overall slope of the spectrum from
differences between the assumed model and the actual stars
by inferring blackbody temperatures from their $B-V$ colors 
and the temperature calibration of \cite{kh95}.  Across SL
(i.e., 5--14~\mum), these errors would be less than 0.5\%,
and across the full spectral range (5--37~\mum), they would
be 0.4\% or less.  As explained in Section~\ref{s.obs},
other systematic issues will dominate this error.

In order to calibrate HR~6348 with A dwarfs, we must mitigate
for differences in the assumed and actual strengths and
profiles  of the hydrogen recombination lines in the A 
dwarfs, which will propagate to the spectrum of HR~6348.  In 
our first attempt, we used polynomials to smooth the spectrum 
of HR~6348 in SL.  In LL, we found that the shape of the 
spectrum was consistent with an Engelke function, but with 
structure due in part to noise and fringing.  To remove these 
apparent artifacts, we replaced the LL spectrum with an 
Engelke function \citep{eng92}.  Unfortunately, this step also
removed real structure in the spectrum of HR~6348, primarily 
from OH bands in the 12--20~\mum\ range (see 
Section~\ref{s.oh}).  As a consequence, we finally decided to 
use the actual observed spectral ratios of HR~6348 to the A 
dwarfs at all wavelengths, mitigating for artifacts in the 
vicinity of recombination lines or from fringing only where 
these were clearly identifiable, as described in 
Section~\ref{s.truth}.

\begin{figure} % Fig. 1
\includegraphics[width=3.4in]{\figpath 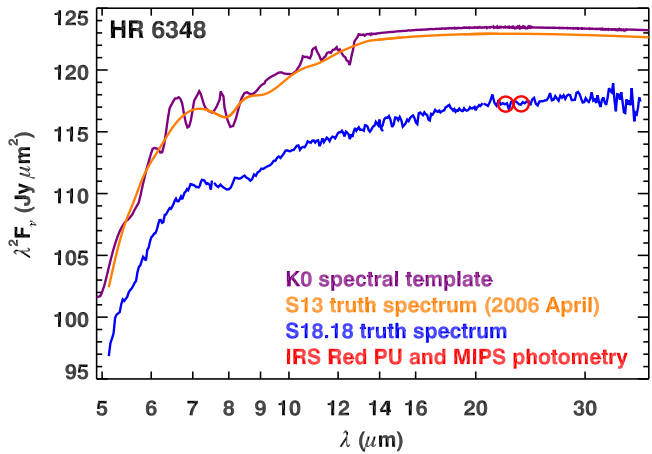}
\caption{The evolution of the truth spectrum of HR~6348,
our primary standard for the low-resolution modules.  This 
and most other figures are plotted in Rayleigh-Jeans flux 
units, so that the Rayleigh-Jeans tail of a Planck function 
will be horizontal.  The spectral template of a K0 giant 
(purple or darker upper trace) was the starting point.  We 
shifted to a truth spectrum based on a heavily smoothed ratio 
between an A dwarf and HR~6348 (orange or lighter upper 
trace).  The final truth spectrum for HR 6348 is shown on the 
lower trace, along with the photometry to which we forced 
it.\label{f.hr6348}}
\end{figure}

Concurrently with improvements in the quality of the truth
spectra, we have also improved the photometric calibration of
the spectra.  Throughout, we have forced the spectra to be
consistent with their Red PU photometry, but the PU 
calibration has changed.  As described by \cite{sl11b}, the 
Red PU calibration is now tied to the calibration of the 
24~\mum\ photometry from the Multi-band Imaging Photometer 
for {\it Spitzer} (MIPS) for HR~6348, HD~166780, and 
HD~173511 \citep{eng07}, accounting for the difference in
central wavelength (22.35 versus 23.675~\mum). We originally 
forced the three spectra to the mean of the delivered 
spectral templates, shifting each to account for the relative 
PU photometry, then later to the mean \iras\ FSC photometry 
\citep{fsc} for these three sources.  However, this mean was 
5\% too high compared to the 24~\mum\ photometry, as shown in 
Figure~\ref{f.hr6348}.  The 5\% shift ties the current IRS 
calibration directly to the 24~\mum\ MIPS calibration 
\citep{rie08}.

\section{Data} % Sec. 4.0

\subsection{Observations} \label{s.obs} % Sec. 4.1

\begin{deluxetable}{lc} % Table 4
\tablenum{4}
\tablecolumns{2}
\tablewidth{0pt}
\tablecaption{Low-Resolution IRS Spectral Orders}
\label{t.orders}
\tablehead{
  \colhead{Module}    & \colhead{Usable Wavelength} \\
  \colhead{\& Order } & \colhead{Range (\mum)}
}
\startdata
SL2      &  5.10--7.53  \\
SL-bonus &  7.73--8.39  \\
SL1      &  7.53--14.20 \\
LL2      & 13.95--20.54 \\
LL-bonus & 19.28--21.23 \\
LL1      & 20.56--37.00
\end{deluxetable}

The first set of usable observations of standard stars came
in IRS Campaign P, which was executed 2003 November 14--16.  
Normal operations began on 2003 December 14, and every IRS 
campaign from then until cryogens were exhausted on 2009 May 
15 included several observations of standards as part of what 
became known as the ``calsfx'' program.

This paper concentrates on the observations obtained with 
the two low-resolution IRS modules, Short-Low (SL) and
Long-Low (LL), which obtained spectra with resolutions
($\lambda$/$\Delta \lambda$) of roughly 80--120.  Each
module included separate apertures for first-order and
second-order spectra.  These are referred to, in order
of increasing wavelength, as SL2, SL1, LL2, and LL1.  When
the star was placed in the SL2 or LL2 aperture, a small piece 
of first-order spectrum was also obtained.  These pieces are 
referred to as the bonus orders (SL-b or LL-b).  
Table~\ref{t.orders} gives the usable wavelength ranges for 
each of the orders.

All spectra were observed in the standard staring mode of
the IRS, which placed the target in two nod positions, one
third and two thirds along each of the four low-resolution
spectrographic slits. A typical low-resolution observation 
consisted of at least eight pointings, with two nods per 
aperture in both of the low-resolution modules.  

Wherever possible, we peaked up on the standard itself, using 
the IRS Red PU sub-array (centered at 22.35~\mum)\footnote{The
peak-up operation uses the IRS imaging mode to determine the
offsets necessary to properly center a target in the
spectroscopic slits at the nod positions.}.  For faint 
targets, we used the Blue PU sub-array (16~\mum).  Targets 
brighter than $\sim$1.0~Jy at 12~\mum\ would saturate the PU 
sub-arrays, and for these, we generally peaked up with an 
offset star in the IRS PU sub-arrays.  In all cases, we used 
the highest precision PU mode available, which generally 
placed the target within $\sim$0\farcs4 of the nominal nod 
position.  Self PU was the preferred mode because it required 
less accurate coordinates.  As long as the target was the 
brightest source within $\sim$45\arcsec of the requested 
position,\footnote{This distance was reduced during the 
mission.} the IRS would center it in the PU field and then 
shift it to the desired spectroscopic aperture.  

Any pointing error in the dispersion direction (i.e.\ 
perpendicular to the long axis of the slit) would lead to 
truncation of part of the point-spread function (PSF) by the 
edges of the slit.  This problem is known as spectral 
pointing-induced throughput error (SPITE), and it has been 
the subject of multiple IRS technical reports 
\cite[e.g.][]{slo03, sl12a}.  The increasing size of the PSF
with wavelength means that the slit throughput is a function
of wavelength as well as position of the source within the 
slit.  The general tendency is for more red flux than blue to 
be lost in the spectrum of a mispointed source, but the 
interaction of the slit edges and the Airy rings complicates
this behavior.

SPITE is a significant issue for SL, because the typical
pointing error ($\sim$0\farcs4) is a significant fraction of 
the distance from the center of the slit to its
edge (1\farcs8).  The problem was negligible in LL due to the 
larger slit width (10\arcsec).

To help reduce the impact of SPITE, we observed most of the 
standards described in this program at least two times.  The 
only exceptions were the handful observed in Campaign P and 
not revisited during normal science operations.  Some of the 
standards were observed much more frequently.  HD~173511, for 
example, was observed in Campaign P and all 61 IRS campaigns 
during normal operations.  Several other standards were 
observed frequently throughout the cryogenic mission (as can 
be seen in the last two columns of Tables~\ref{t.kstd} and 
\ref{t.astd}.)

\subsection{Data reduction} \label{s.reduce} % Sec. 4.2

We started with the S18.18 version of the SSC pipeline, with 
the basic calibrated data (BCD) files generated from the data 
cubes transferred from the spacecraft.  These flatfielded 
images are the result of several steps in the SSC pipeline, 
most notably the removal of dark currents, mitigation for 
cosmic ray hits, and fitting of slopes to the signal ramps 
for each pixel.  To generate a continuous spectrum from 5 to 
37~\mum, we follow a standard reduction algorithm that has 
now been applied to thousands of spectra published by the 
IRS Team at Cornell and the IRS Disks Team at Rochester.  

First, suitable background images were subtracted from the
on-source images.  In SL, the default was to use the image
with the source in the other aperture, referred to as an
aperture difference.  Thus, SL1 images served as the 
background for SL2 and vice versa.  HD~46190 was the only 
exception, because the integration times in SL1 and SL2 
differed.  In LL, nod differences were the default, because 
the distance between the long slit length (150\arcsec) and 
the higher background levels relative to the emission from 
the stars left us more susceptible to residuals from 
background gradients.

Background-subtracted images were cleaned, using the {\sc 
imclean} software developed at Cornell\footnote{This 
algorithm was included in the {\sc irsclean} package 
distributed by the SSC.} and the masks of rogue pixels 
generated for each campaign by the SSC.  Rogue pixels are 
those with unstable responsivities or dark currents, 
preventing a proper calibration when subtracting the dark 
current from the data cubes or flatfielding the images.  The 
SSC delivered one rogue mask per campaign, identifying all 
pixels identified as rogues during that campaign.  We used 
those rogue pixel masks to generate cumulative rogue masks, 
which we updated every few campaigns.  Any pixel identified 
as a rogue in at least two rogue masks from previous 
campaigns was marked as a rogue by us and cleaned.  The 
number of rogue pixels generally increased during the mission 
due to the effects of space weathering.  In addition to rogue 
pixels, we used the bit masks associated with each image to 
identify and clean pixels with invalid data, whether in the 
image with the source or the background.

To extract spectra from the images, we used the standard
tapered-column extraction available with CUPID, the 
Customizable User Pipeline for IRS Data available from the
SSC.  This software is a copy of the offline pipeline.  We
used the {\sc profile}, {\sc ridge}, and {\sc extract}
modules to identify the position of the source in the slit,
map it onto the image, and extract it.  For each wavelength
element, a quadrilateral of roughly one pixel height was
defined, tilted with respect to the row axis of the array
to follow a line of constant wavelength.  The width of
the quadrilateral increases linearly with wavelength to
follow the expanding size of the PSF.  The downside of
this algorithm is that it integrates noise from off-source
pixels along with the source itself, but this is only a
problem for faint targets, and our sample consisted primarily
of bright stars.

We considered using the optimal extraction method developed
at Cornell \citep{leb10}, which fits a PSF to the data at
each wavelength to extract a spectrum.  For sources as bright 
as ours, the gain in SNR is small, while the undersampled 
character of the PSF in SL2 and LL2 can produce small 
artifacts clearly visible at high SNRs, degrading the quality 
of the spectra more than improving it.  For fainter sources,
the optimal extraction did improve the SNR of the resulting
spectra, but for consistency, we used tapered-column
extraction for all sources.

We extracted a spectrum from each image in a given nod
position and then coadded them.  We then calibrated the
coadded spectra using the RSRF defined in the introduction.
Spectra from the two nod positions were combined using a 
spike-rejection algorithm that identified and ignored any
structure in one of the two spectra not present in the other.
These steps gave us spectra in the four apertures:  SL2 
(plus SL-b), SL1, LL2 (plus LL-b), and LL1.  

To determine uncertainties in the flux densities at each
wavelength element, we measured the standard deviation 
$\sigma$ when coadding all spectra in a given nod.  The 
assigned uncertainty is the uncertainty in the mean 
($\sigma/\sqrt{N}$).  We re-evaluated the uncertainty when 
combining the nods, taking the larger of the propagated 
uncertainty and the uncertainty in the mean calculated from 
the data in the two nod positions.

To combine these spectral segments, we followed a
``stitch-and-trim'' algorithm.  We applied scalar 
multiplicative corrections to each spectral segment to remove 
discontinuities between them, always normalizing upward to 
the (presumably) best-centered spectral segment.  This 
usually meant that we scaled the two SL segments up to 
match LL.  The bonus orders provided the overlap needed to 
scale SL2 to SL1 and LL2 to LL1.  To scale SL1 to LL2, we 
used the small overlap of usable data (Table~\ref{t.orders})
between them.   We then collapsed the bonus order data into 
the other orders by averaging where they overlapped and were 
within the range of valid wavelengths.  All data outside the 
ranges given in Table~\ref{t.orders} were trimmed from the
spectra to produce the final spectra.

The last two columns in Tables~\ref{t.kstd} and \ref{t.astd} 
list the total number of spectra obtained in standard staring 
mode for each standard star as part of the spectral 
calibration program and the number we used when coadding 
spectra.  The total number does not include spectra of some 
sources taken for other purposes, because these observations 
usually differed from the standard staring mode and were not 
suitable for spectroscopic calibration.  The number used 
reflects spectra rejected for several possible 
reasons.  The primary problem was reduced fluxes because of 
pointing errors, which could distort the shape of the 
spectra \citep{sl12a}.  Other spectra were rejected because 
they showed evidence of residual images from previously 
observed bright sources, data drop-outs, or other artifacts 
which indicated some problem with the observation.

\subsection{Constructing truth spectra} \label{s.truth} % Sec. 4.3

To generate the truth spectra, we started with HR~6348, 
$\alpha$~Lac, and $\delta$~UMi, generating ratios of
observed spectra and calibrating them with scaled Kurucz
models of the A dwarfs.  \cite{sl12b} describe our method
for SL, and \cite{sl12c} treat LL and the combination
of SL and LL into a single low-resolution truth spectrum.
We refer the reader to these reports for any questions not
answered by the briefer description provided here.

\subsubsection{Short-Low} \label{s.truthsl} % Sec. 4.3.1

The SL wavelength grid is finer than in LL, requiring more
attention to possible artifacts arising from differences in 
assumed and observed profiles of the hydrogen recombination 
lines in the A dwarfs.  For the spectra from each nod 
position, we calibrated HR~6348 separately with each of the 
two A dwarfs, then combined them using a spline to force them 
to their mean shape and strength while preserving detailed 
differences between them.  When one spectrum showed more 
structure than the other, we dropped the affected wavelengths 
from that spectrum from consideration.  We then combined the 
spectra from the two nods similarly, repeating the same 
sequence for SL1, SL2, and the SL-bonus order.  

The resulting SL spectrum of HR~6348 shows some structure in 
the vicinity of the recombination lines, most notably in the 
immediate vicinity of 7.46~\mum, where Pfund-$\alpha$ blends
with three other recombination lines.  We generated a 
correction to the spectrum of HR~6348 at Pfund-$\alpha$ based
on how the artifact propagated to spectra of the asteroids 
Amalia and Isara and the ultraluminous infrared galaxy IRAS 
07598+6508.  These three sources should show a smooth red
spectrum in this position, and with them, we were able to
reduce the deviations at Pfund-$\alpha$ to $\sim$0.8\% of the
continuum (from $\sim$1.5\%).  What remains could be an 
artifact, or it could be intrinsic to the spectrum of 
HR~6348.  We were unable to reduce the apparent structure in 
the vicinity of other recombination lines, which can reach 
$\sim$0.5\% of the continuum.  Thus, our truth spectra
should propagate no apparent spectral structure into other
IRS spectra up to a SNR of $\sim$120 at 7.5~\mum\ and 200
elsewhere.

\subsubsection{Long-Low} \label{s.truthll} % Sec. 4.3.2

In LL, the spectra from the two nod positions showed no 
significant differences, so we first combined them with a
simple average and then concentrated on the differences in 
ratios of HR~6348 with $\alpha$~Lac and $\delta$~UMi which
arise from an apparent red excess in $\alpha$~Lac (Section 
\ref{s.stda} below).  When calibrated with $\alpha$~Lac, the 
LL1 spectrum of HR~6348 is bluer than a Rayleigh-Jeans tail, 
which is unphysical, while the spectrum calibrated with 
$\delta$~UMi behaves more reasonably.  The spectra in LL2 
and the bonus order also showed similar differences depending 
on which A dwarf was used.  These results confirm the red
excess in $\alpha$~Lac.  To mitigate, we used a spline 
function to force the shape of the two calibrations of 
HR~6348 to match that of the calibration with $\delta$~UMi
while preserving the finer differences between the two.  We 
applied the same procedure to each of the three spectral 
orders.

Many of the LL1 spectra in the IRS database are affected by
fringing, and despite the many spectra coadded to produce
the truth spectrum of HR~6348, this fringing did not cancel
out.  We corrected for the fringing by fitting a polynomial 
to the spectra of two red sources expected to show a smooth 
continuum at these wavelengths, IRAS~07598+6508 and Mrk~231,
and assuming that the differences between the polynomial and 
the spectra arise from fringing.  Because the fringing in the 
red sources is roughly twice as strong as in HR~6348, we 
reduced the correction by a factor of two before applying it.  
We also limited the correction to wavelengths below 32~\mum\ 
because the data and the correction grow more unreliable to 
the red.

% Vassilis would like a fringing example

The dominant remaining structure in the LL spectrum of
HR~6348 is real and arises from OH absorption, a point 
discussed more fully in Section~\ref{s.oh} below.  The OH 
band structure is strongest in LL2, where it produces an 
apparent corrugation similar to the fringing in LL1.  For 
that reason, we avoiding using Fourier analysis to defringe, 
since that would remove any periodic structure from the 
spectrum.

Early in the mission, we assumed that an Engelke function
was a suitable model for the long-wavelength spectrum of 
HR~6348 (and other K giants).  The omission of the OH
contribution introduced an artifact in LL2 in early Cornell
calibrations, which while small was noticeable in spectra 
with high SNRs.

\subsubsection{The combined low-resolution truth spectrum} % Sec. 4.3.3

Normally, the stitch-and-trim step is the final step, leaving
us with a spectrum normalized to what is presumably the 
best-centered spectral segment.  However, when generating
truth spectra, we can force them to match the Red PU 
photometry by shifting them multiplicatively.  As explained
in Section 3.3, \cite{sl11b} have calibrated the Red PU
photometry to align with the calibration of MIPS-24 
\citep{rie08}, so this last step cross-calibrates the IRS
directly with MIPS.

The scalar multiplicative corrections made in the 
stitch-and-trim step and when shifting the spectrum to
match the PU photometry do not correct for any distortions 
in the shape of the spectra from pointing-related errors
(SPITE; Section~\ref{s.obs}).  These distortions will 
ultimately limit the absolute calibration of the spectra.  
\cite{sl12c} estimate that the overall fidelity of the shape 
of the SL portion of the truth spectrum of HR~6348 is good to 
within 2\%.  This fidelity should not be confused with the
point-to-point, or spectroscopic uncertainties, which are
smaller, as noted in Section~\ref{s.truthsl}.  \cite{sl12c}
estimate that the fidelity in LL is better than 0.5\% to
the blue of $\sim$29~\mum\ and $\sim$1\% to the red, 
primarily due to limitations in the SNR.

\cite{sl12c} generated a truth spectrum for HD~173511, using
HR~6348 as the calibrator, based on data up to IRS Campaign 
44 (2007 October 7).  The detector settings of LL changed 
with the beginning of IRS Campaign 45 (2007 October 29), 
requiring an independent calibration for LL spectra from that 
time to the end of the cryogenic mission.  The fidelity of
this spectrum is similar to that of HR~6348.

\subsection{Spectral corrections} \label{s.rsrf} % Sec. 4.4

To use a truth spectrum to calibrate data from the IRS, we 
created a spectral correction, or RSRF, as defined in
Section~\ref{s.intro}.  This correction is a full vector 
correction, meaning that each wavelength is treated 
independently, and it differs from the corrections used in 
the SSC pipeline.  The SSC corrections are polynomials 
fitted to the full vector, are smoother, and do not account
as well for wavelength-to-wavelength variations.

At the beginning of the {\it Spitzer} mission, we adopted 
separate spectral corrections for spectra obtained in the two 
nod positions in all low-resolution orders.  We anticipated 
that as the quality of the flatfield corrections improved, 
these differences would work their way into the flatfield and 
we could eventually adopt a single spectral correction for 
all pointings in a given order.  Nonetheless, at the end of 
the mission, the differences in the raw nod spectra are still 
significant enough to warrant independent corrections. 

To generate spectral corrections, we used HR~6348 for all SL
data and for all LL data up to Campaign 44.  HR~6348 was not
observed enough times after the changes to LL to generate a
spectral correction with a sufficiently high SNR.  
Consequently, we use both HR~6348 and HD~173511 to calibrate
LL data from Campaigns 45--61.

The spectral corrections currently available in 
SMART\footnote{The Spectroscopy Modeling Analysis and 
Reduction Tool; see \cite{hig04}.} are generated as described
above.  The spectral corrections are available for the 
tapered-column extraction in SMART, which is functionally 
equivalent to the method used by CUPID.  They are also 
available for optimal extraction \citep{leb10}.  Spectral 
corrections can be generated for any extraction method, as
long as the standards are extracted identically to the
science targets.

\subsection{Spectral atlas} % Sec. 4.5

The spectra of all of the stars listed in Tables~\ref{t.kstd}
and \ref{t.astd} are also publicly available.  They can be 
obtained from the Infrared Science Archive (IRSA), VizieR, or 
directly from the first author's website.  Each spectral file
has the following columns:  wavelength (\mum), flux density
and uncertainty (Jy), and spectral order number (2=SL2, 1=SL1,
5=LL2, 4=LL1).

\section{Checking the calibration with other standards} \label{s.std} % Sec. 5

\subsection{$\alpha$ Lac and $\delta$ UMi} \label{s.stda} % Sec. 5.1

\begin{figure} % Fig. 2
\includegraphics[width=3.4in]{\figpath 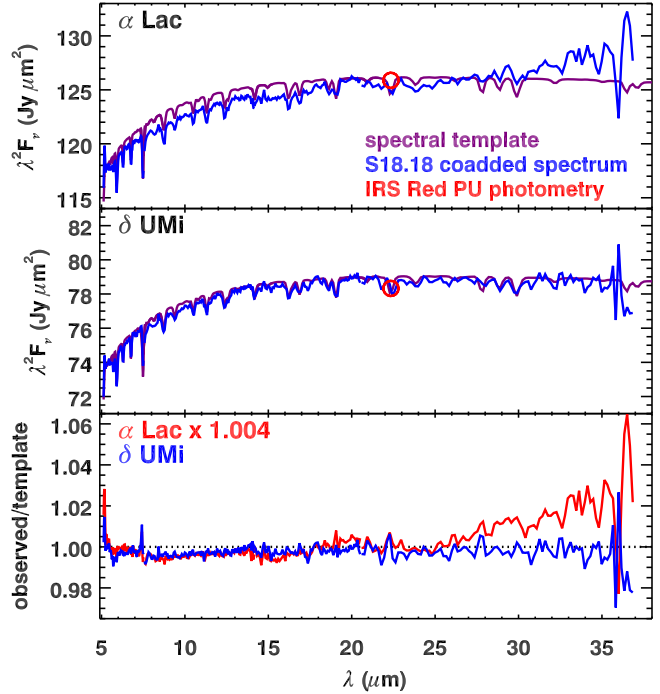}
\caption{The templates and coadded low-resolution IRS
spectra of the A dwarfs $\alpha$~Lac (top) and $\delta$~UMi
(middle).  The template is a scaled Kurucz model, and the
differences between it and the coadded spectra reflect the 
fidelity of the shape of the IRS spectra for well-pointed 
sources.  The red excess in $\alpha$~Lac beyond 
25--30~\mum\ is a different matter.  The bottom panel
compares the ratios of the two spectra to their templates.
We have shifted the spectrum of $\alpha$~Lac up by 0.4\% to 
demonstrate that the SL spectrum (below 14~\mum) is 
unaffected by this excess, which we suspect arises from a 
low-contrast debris disk.\label{f.adwarf}}
\end{figure}

Figure~\ref{f.adwarf} compares the coaddition of all of the 
unrejected spectra of the A dwarfs $\alpha$~Lac and 
$\delta$~UMi to the Kurucz models with which the calibration 
started.  At a minimum, we should get back what we started
with, and for $\delta$~UMi, we do, to within better than 
$\sim$0.5\% at all wavelengths, except at Pfund~$\alpha$, 
which differs by $\sim$1\% in total depth.  

The observed coadded spectrum of $\alpha$~Lac differs from
the Kurucz model in two ways.  The difference in overall
photometric level is attributable to the accumulated impact
of small pointing errors, but, as described by \cite{sl12c}, 
the observed spectrum shows a small red excess.  If measured 
starting at 22~\mum, it has a strength of $\sim$2.5\% by 
35~\mum.  The cause is most likely a low-contrast debris disk 
around $\alpha$~Lac.  Section~\ref{s.truthll} explains how 
we addressed this problem when calibrating LL.

\subsection{HD 173511 and HD 166780} \label{s.stdk1} % Sec. 5.2

\begin{figure} % Fig. 3
\includegraphics[width=3.4in]{\figpath 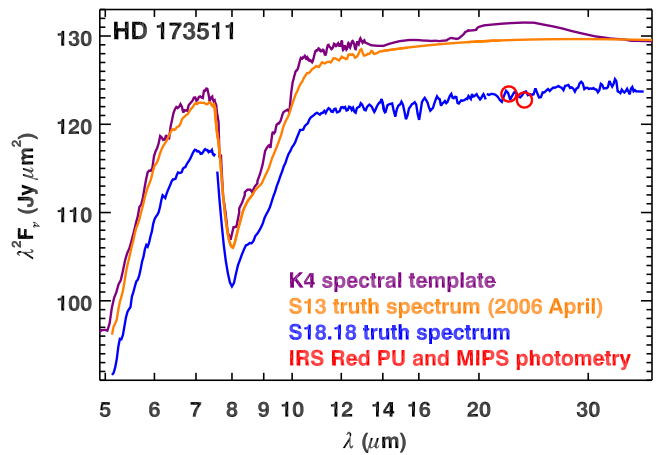}
\caption{A comparison of the various truth spectra of
HD~173511.  The shift between early versions and the final 
truth spectrum reflects the shift in photometric calibration.  
The structure in the S18.18 truth spectrum between 10 and 
20~\mum\ is real and arises from OH.  The spectral template 
was based on a K4 giant, but we now classify the star as 
K5~III.\label{f.hd173511}}
\end{figure}

\begin{figure} % Fig. 4
\includegraphics[width=3.4in]{\figpath 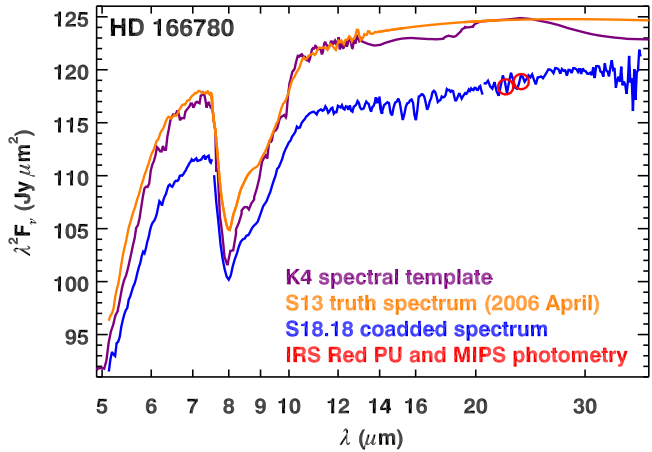}
\caption{The evolution of the spectrum of HD~166780.  For
the S18.18 spectrum, we use a coadded spectrum from all of
the unrejected observations because we did not generate
a final truth spectrum.  The agreement with the photometry
is fortuitous.\label{f.hd166780}}
\end{figure}

Early in the mission, HD~166780 and HD~173511, both late
K giants, were used to supplement HR~6348 when calibrating
LL, due to the lower SNR of blue sources in LL compared to
SL and the limited number of HR~6348 observations available
at the time.  As described by \cite{sl11a}, HD~166780
exhibited more variability in the Red PU data than the 
other standards considered, and as a consequence we dropped
it as an IRS standard.  The MIPS 24~\mum\ photometry also 
hints at possible variability.

Figure~\ref{f.hd173511} illustrates the evolution of the
assumed truth spectra for HD~173511.  Prior to launch we
reclassified HD~173511 as a K4~III.\footnote{Based on its 
$(B-V)_0$ color, we reclassified HD~173511 back to K5, as 
explained in Section~\ref{s.ksp}.}  The K4 spectral template 
is based on an average of the prototypes for K3 and K5, 
$\alpha$~Hya and $\alpha$~Tau, and both stars are bright 
enough to produce a composite spectrum with an excellent 
SNR.  The template matches the depth and shape of the 
observed SiO profile well.  The primary deficiency in the 
template occurs at wavelengths past 13~\mum, where the 
template shows a broad plateau at $\sim$20~\mum\ not apparent 
in the S18.18 truth spectrum.  The older S13 truth spectrum 
in the figure used an Engelke function past 14~\mum\ and is 
thus not a good test.  It is possible that the apparent dip 
in the template from 13 to 20~\mum\ could reflect in part the 
effect of OH band absorption in $\alpha$~Tau, but the 
downturn past 23~\mum\ is not real.

Figure~\ref{f.hd166780} compares the spectra of HD~166780.
We reclassified it as K4 prior to launch, like HD~173511, 
which means that their spectral templates only differ
photometrically.  Because HD~166780 is no longer an IRS 
standard, we do not plot an S18.18 truth spectrum; rather, we 
have presented the coaddition of unrejected spectra with no 
other adjustments.  The issues past 13~\mum\ are similar to 
HD~173511.  More importantly, the SiO absorption band at
8~\mum\ in HD~166780 is weaker than in the template, showing
that the templates cannot be relied upon to match the profile 
of the absorption bands.

\subsection{HR 6606 and HR 7341} \label{s.stdk2} % Sec. 5.3

\begin{figure} % Fig. 5
\includegraphics[width=3.4in]{\figpath 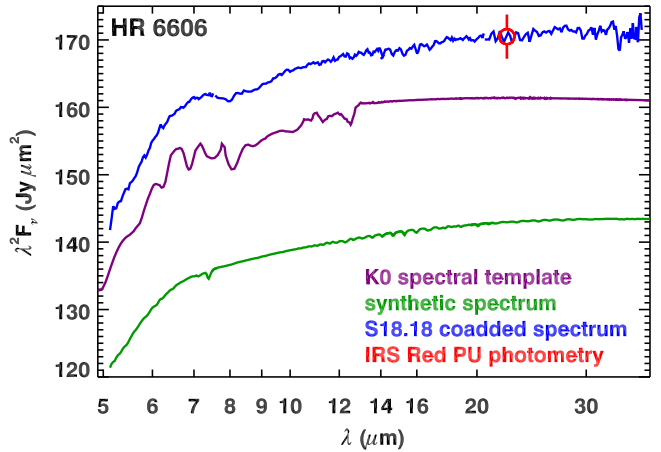}
\caption{Candidate truth spectra for HR~6606.  The synthetic 
spectrum is from \cite{dec04}.\label{f.hr6606}}
\end{figure}

\begin{figure} % Fig. 6
\includegraphics[width=3.4in]{\figpath 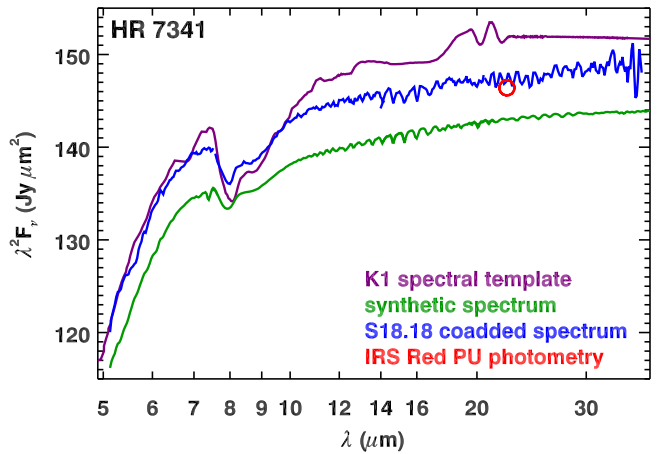}
\caption{Candidate truth spectra for HR~7341.  The synthetic 
spectrum is from \cite{dec04}.  The spectral template
overpredicts the strength of the SiO band at 8~\mum, while
the synthetic spectrum underpredicts it; neither is 
suitable as a truth spectrum.\label{f.hr7341}}
\end{figure}

Figures~\ref{f.hr6606} and \ref{f.hr7341} illustrate the
challenges of predicting the infrared spectral properties of
K giants from their photometry.  The spectral template
of HR~6606 is based on $\beta$~Gem (like HR~6348), while the
template for HR~7341 is based on a weighted average of
$\beta$~Gem (K0) and $\alpha$~Boo (K1.5).  The synthetic
spectra are from \cite{dec04}.\footnote{All of the synthetic
spectra utilized in this paper are available at
http://irsa.ipac.caltech.edu/data/SPITZER/docs/irs/calibrationfiles/decinmodels/.  
These spectra were synthesized using plane-parallel models 
for the G9 giants and spherically symmetric models for the 
later spectral classes \citep{dec03b}.} The S18.18 spectra 
are a coaddition of all good spectra observed by the IRS.  
The alignment with the PU photometry indicates the 
photometric quality of the coadded spectra.

We rejected HR~6606 early in the mission because it was
unusually bright compared to expectations, and it was 
brighter than recommended for peaking up with the IRS Red PU 
array.  We did manage to obtain one PU measurement, which is 
shown, in Figure~\ref{f.hr6606}.  The spectral template 
contains the same pattern noise seen in HR~6348.

The synthetic spectrum of HR~6606 predicts no measurable
SiO absorption, but the observed IRS data clearly show this
band.  Similarly, in HR~7341, the synthetic spectrum
underpredicts the strength of the band.  Section~\ref{s.sio}
shows more quantitatively that the observed SiO bands are
always deeper than expected from the synthetic spectra.  We
conclude that the synthetic spectra should not be used as
truth spectra for spectroscopic standards.

The SSC adopted HR~7341 as their primary low-resolution IRS 
standard.  Their use of its synthetic spectrum for a truth
spectrum propagated an SiO artifact into the SSC calibration 
of the full database of SL spectra.  While the size of the
artifact is small, only $\sim$1.5\% at 8~\mum, it was large 
enough to be noticed by \cite{ard10}, who published an atlas 
of stellar spectra observed by SL and LL.  They found the
artifact in all of their spectra and corrected it with a
polynomial fitted to the ratio of the observed spectrum of
$\alpha$~Lac to its spectral template (i.e.\ a Kurucz model).  
The spectral template for HR~7341 would not solve this 
problem, because it overpredicts the strength of the SiO 
band, making it equally unsuitable.  The SiO artifact remains 
in all spectra in the S18.18 pipeline from the SSC.

The synthetic spectra of both HR~6606 and HR~7341 show a 
narrow absorption feature centered at 7.39~\mum\ not 
identified by their creators.  This is close to the 
Pfund~$\alpha$ complex at 7.46~\mum, but shifted enough to 
the blue that it would have to be blended with another 
feature or be a different feature entirely.  While our truth 
spectrum of HR~6348 does have weak Pfund~$\alpha$ absorption, 
it is likely to be an artifact from the use of A dwarfs to 
calibrate it.  Our calibration of the K giants, which is 
based on the observed ratios of A dwarfs and HR~6348, does 
not reproduce this 7.39~\mum\ feature.

We have described the shortcomings of the synthetic spectra 
and spectral templates, but the coadded spectra also have
problems.  Past 20~\mum, the spectra of both HR~6606 and 
HR~7341 show a periodic structure with a peak-to-peak 
amplitude of $\sim$1\%.  The synthetic spectra of both stars 
show that the structure due to OH should be much weaker at 
these wavelengths compared to the OH band structure at 
$\sim$14--16~\mum.  While it is possible that OH could be 
contributing to the observed structure past 20~\mum, it is 
far more likely that it is due to fringing in LL1.

Thus, we have no perfect solutions to the problem of 
generating truth spectra for K giants.  While truth spectra
based on observed spectral ratios with A dwarfs (and 
propagated through HR~6348 as an intermediate calibrator) 
offer the best solution at wavelengths below 20~\mum, to the
red, fringing limits the fidelity of the IRS data to
$\sim$1\%.  

The 24~\mum\ dip remains an unsolved calibration problem with
IRS data.  While this problem does not appear to affect the 
four K giants with $F_{22}$ $\la$ 120~mJy in our sample, the
small sample size requires some caution in drawing general
conclusions.  \cite{ard10} applied a correction to all 
sources in their sample fainter than 100~mJy (at 24~\mum).  
Better understanding of this artifact awaits a more rigorous 
analysis of the fainter stellar spectra observed by the IRS.

\section{The K Giants} % Sec. 6.0

\subsection{Spectra} \label{s.k} % Sec. 6.1

\begin{figure} % Fig. 7
\includegraphics[width=3.4in]{\figpath 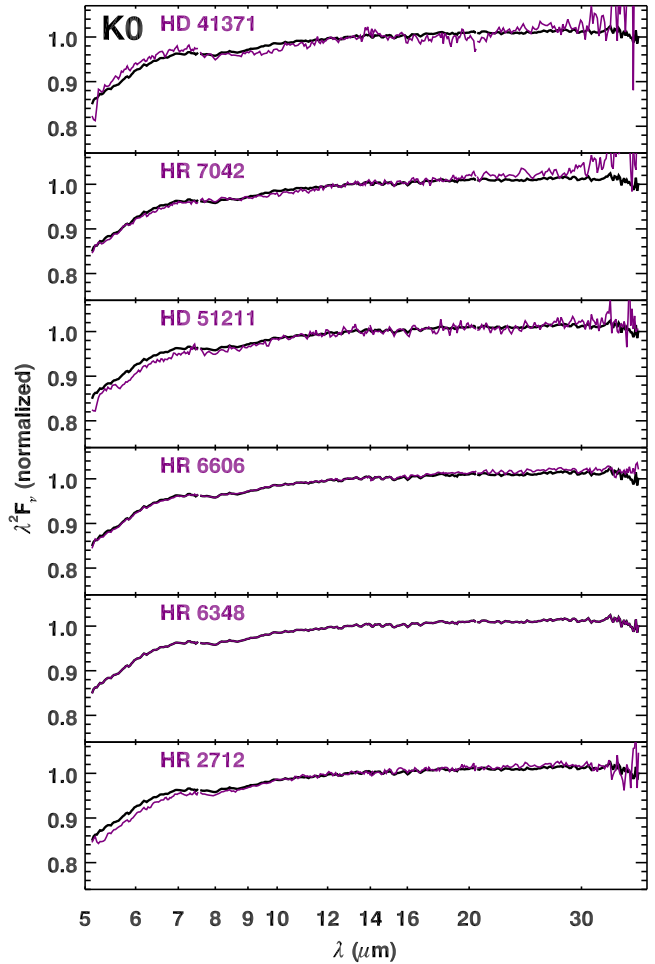}
\caption{IRS spectra of the K0 giants in Rayleigh-Jeans
units, and with the spectrum of HR~6348 overplotted as the
prototype in each panel in black.\label{f.k0}}
\end{figure}

\begin{figure} % Fig. 8
\includegraphics[width=3.4in]{\figpath 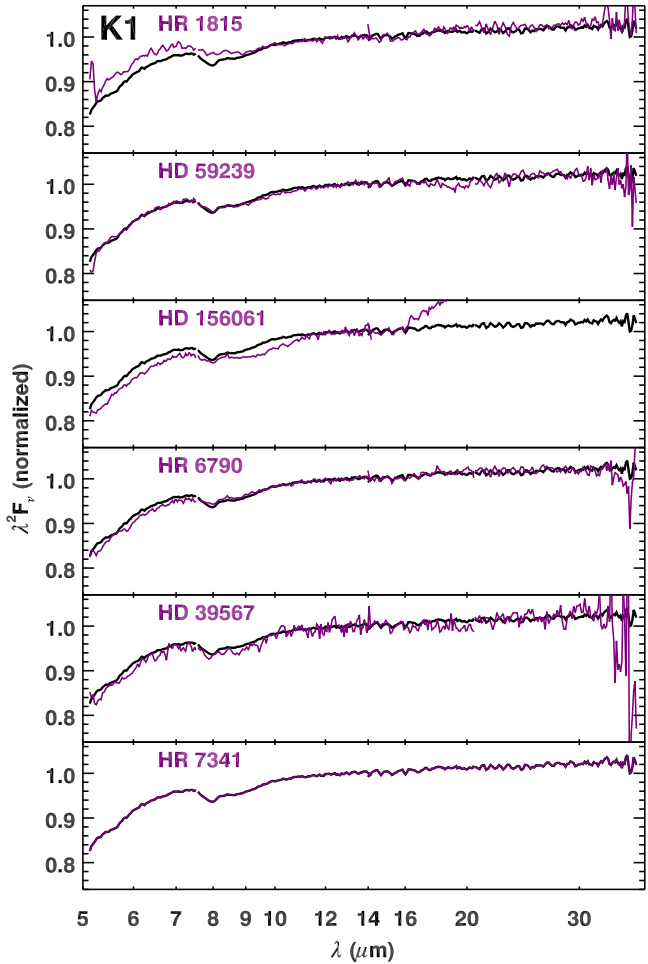}
\caption{IRS spectra of the K1 giants, plotted as in 
Fig.~\ref{f.k0}.  HR~7341 is the prototype and is 
plotted in black in each panel.  The spectrum
of HD~156061 shows a strong red excess which may arise from
a debris disk (see Section~\ref{s.debris}).\label{f.k1}}
\end{figure}

\begin{figure} % Fig. 9
\includegraphics[width=3.4in]{\figpath 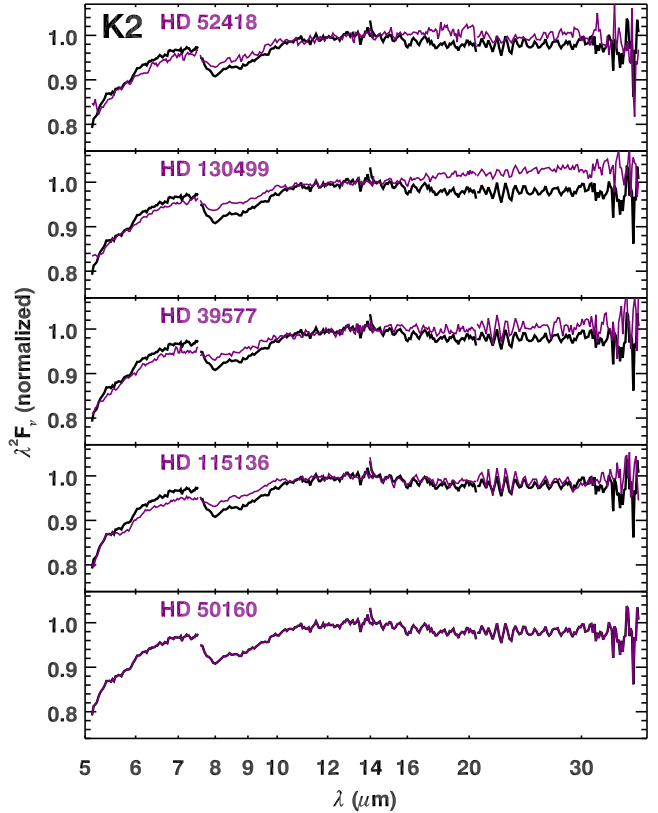}
\caption{IRS spectra of the K2 giants, plotted as in
Fig.~\ref{f.k0}.  HD~50160 appears in each panel in
black; is not an ideal prototype,
due to the fringing in LL1, but none of the other
spectra are as suitable.\label{f.k2}}
\end{figure}

\begin{figure} % Fig. 10
\includegraphics[width=3.4in]{\figpath 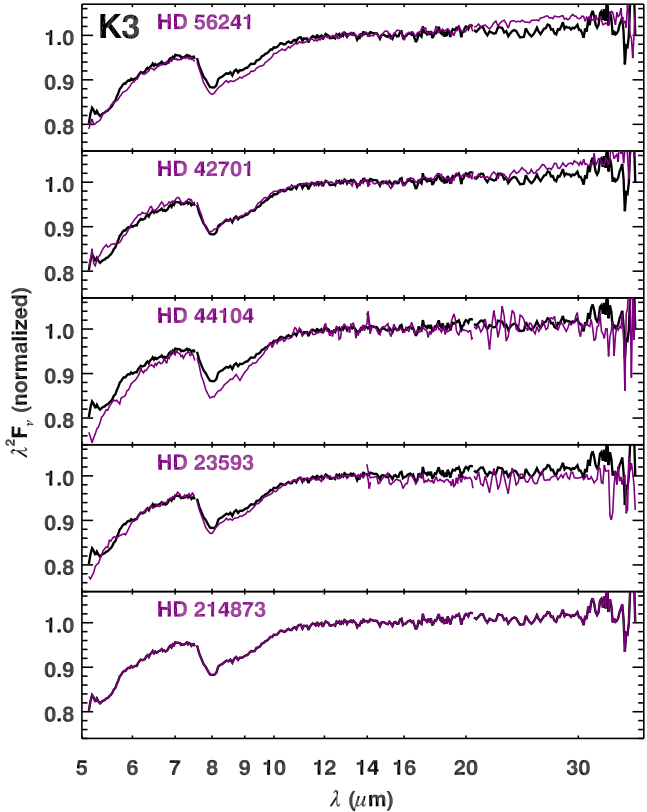}
\caption{IRS spectra of the K3 giants, plotted as in
Fig.~\ref{f.k0}.  The prototype here is HD~214873,
although it was observed only once; it appears in
black in each panel.  The structure
in its spectrum at 5.2~\mum\ is probably not 
real.\label{f.k3}}
\end{figure}

\begin{figure} % Fig. 11
\includegraphics[width=3.4in]{\figpath 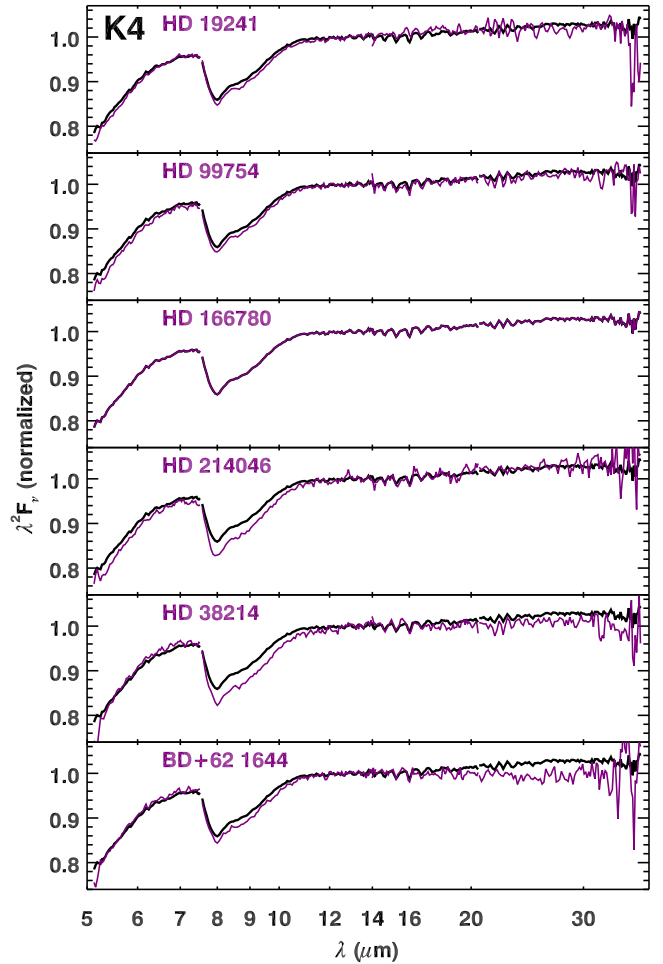}
\caption{IRS spectra of the K4 giants, plotted as in
Fig.~\ref{f.k0}.  HD~166780 is the prototype, overplotted
in black in each panel.\label{f.k4}}.
\end{figure}

\begin{figure} % Fig. 12
\includegraphics[width=3.4in]{\figpath 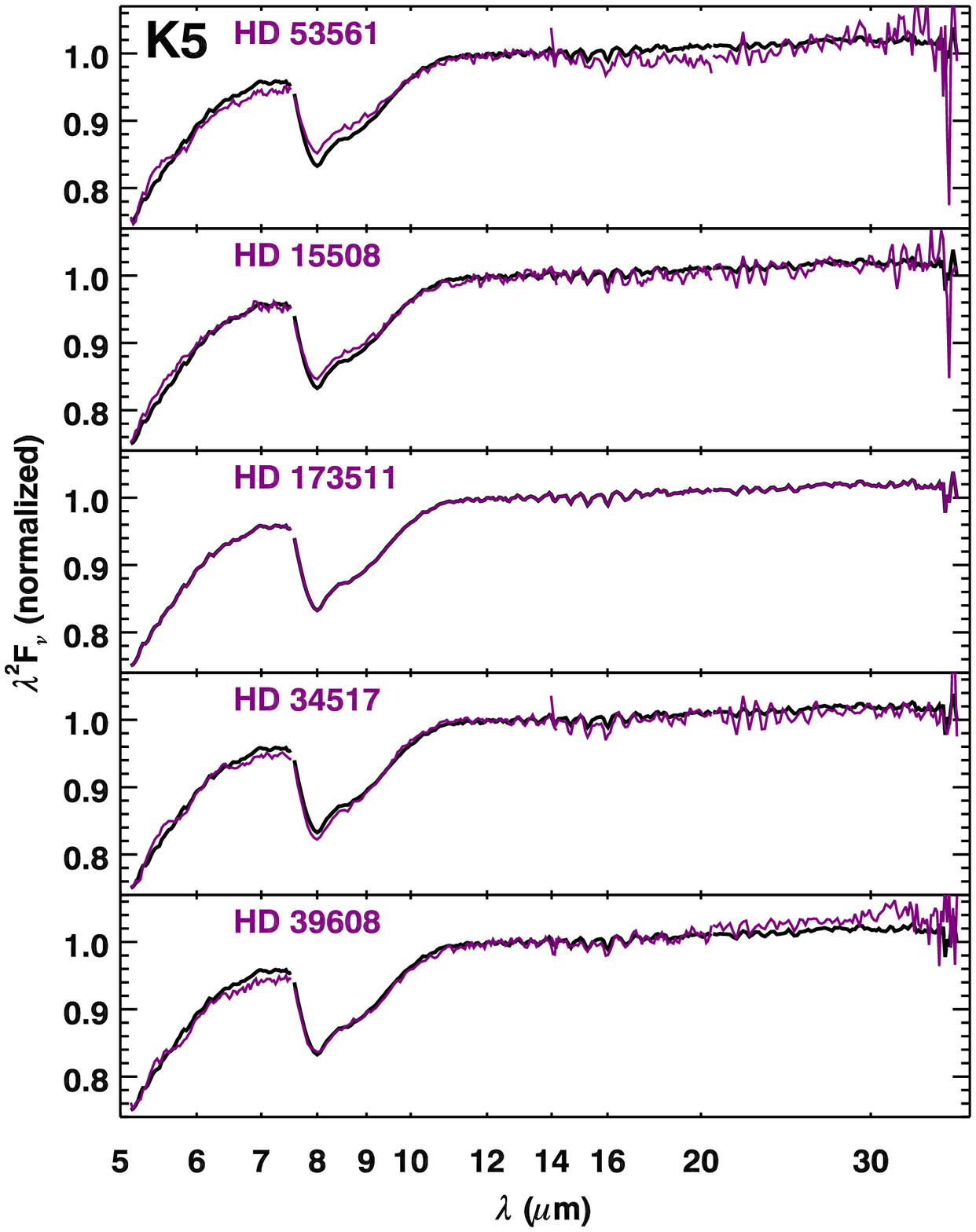}
\caption{IRS spectra of the K5 giants, plotted as in
Fig.~\ref{f.k0}.  HD~173511 is the prototype, overplotted
in each panel in black.\label{f.k5}}.
\end{figure}

Figures~\ref{f.k0} through \ref{f.k5} plot the spectra from
the IRS of all 33 of our K giants in Rayleigh-Jeans units, 
organized by modified spectral class (as defined in 
Section~\ref{s.ksp}).  For each spectral subclass, a 
prototype is overplotted in all panels for comparison.  This 
prototype is usually one of the spectra we observed 
repeatedly as a standard, but we have no such standard for K2 
and K3.  

HR~6348 is the prototype for K0.  It and HR~6606, the other
well-observed source in this group, are virtually identical
out to 20~\mum.  The deviations in HD~41371 may be 
attributable to pointing errors and the resulting throughput
variations in the slit (SPITE, see Section~\ref{s.obs}),
as this source was only observed once.  More generally, the
fewer times a spectrum was observed, the more likely these
throughput errors will affect the shape of the spectrum
especially in SL.  This effect can be seen in several spectra 
in these and other figures.  For sources observed many times
(see the last column of Table~\ref{t.kstd}), individual
pointing deviations average out and the shape of the spectrum
becomes more reliable.

HR~7341 is the prototype for K1.  The variations among the
spectra are more noticeable than for K0.  HD~156061 stands 
out with its strong red excess, possibly from a debris disk,
which would be unusual for a K giant.  We discuss this excess
further in Section~\ref{s.debris} below.

None of the K2 giants were observed more than four times, and 
the quality of the spectra reflect that, with strong fringing 
in LL1 and more obvious noise.  HD~50160 is the prototype 
because it was observed more than the others and has a better
behaved spectrum.  It also has the strongest SiO absorption, 
which is consistent with its redder $B-V$ color 
(Section~\ref{s.ksp} below).

The K3 giants also show noisy spectra with LL1 fringing, due
to the relatively low number of observations.  HD~214873 is 
the prototype for the group because it has the best behaved 
spectrum, even though it was observed only once.  Because 
this observation came in Campaign P, before normal science
operations had even begun, our confidence in the calibration
is limited, making it likely that the structure at 5.2~\mum\
is not real.  The other spectra show considerable variation 
in the strength of the SiO band, with HD~42701 and HD~44104 
representing the extremes.  Two spectra, HD~56241 and 
HD~42701, show a measurable red excess compared to HD~214873.  
%Pointing errors have little effect on LL1, and in any event, 
%HD~56241 was observed five times, so these excesses are 
%probably real.

HD~166780 is the prototype for the K4 giants.  This group
exhibits a range of SiO absorption strengths, with HD~166780
showing the weakest band.  The fringing in LL1 (past 20~\mum)
is quite noticeable in some of the spectra and can be
distinguished from the OH band structure from 14 to 18~\mum.

The prototype for the K5 giants is HD~173511.  As with K4,
these spectra show a variety of SiO band strengths, OH
band structure in LL2, and fringing in LL1.

\subsection{Measuring the SiO absorption band} \label{s.sio} % Sec. 6.2

\begin{figure} % Fig. 13
\includegraphics[width=3.4in]{\figpath 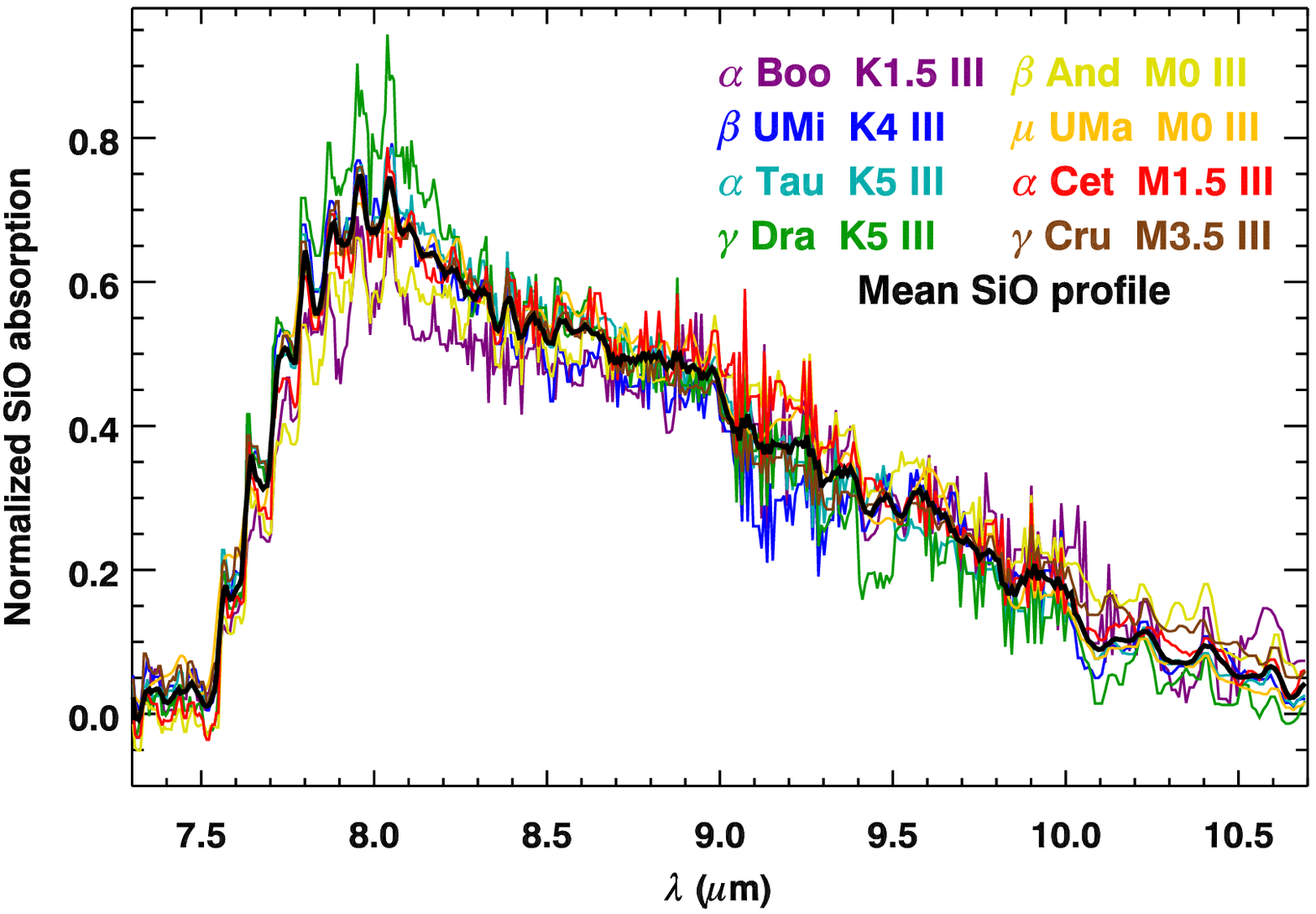}
\caption{Constructing a mean SiO absorption profile from the
SWS spectra of several standard stars.  We have 
median-smoothed the spectra before plotting to reduce the 
noise to reveal more of the spectral details.  While the 
shape of the SiO band varies from source to source, no
dependence with spectral class is apparent.\label{f.sioprof}}
\end{figure}

\begin{deluxetable}{lcccc} % Table 5
\tablenum{5}
\tablecolumns{5}
\tablewidth{0pt}
\tablecaption{Observed SiO measurements}
\label{t.sio}
\tablehead{
  \colhead{Source}      & \colhead{Fitted}                 & 
  \colhead{Fitted}      & \colhead{Eq.\ Width} & \colhead{($B-V$)$_0$} \\
  \colhead{ }           & \colhead{T (K)\tablenotemark{a}} & 
  \colhead{$A_v$ (mag)} & \colhead{SiO (\mum)} & \colhead{(mag)}
}
\startdata
HD 41371   & 4600 & 0.28 & 0.045 $\pm$ 0.003 & 0.925 $\pm$ 0.010 \\
HR 7042    & 3850 & 0.18 & 0.014 $\pm$ 0.005 & 0.935 $\pm$ 0.009 \\
HD 51211   & 3400 & 0.11 & 0.027 $\pm$ 0.000 & 0.969 $\pm$ 0.009 \\
HR 6606    & 4150 & 0.03 & 0.023 $\pm$ 0.003 & 1.006 $\pm$ 0.009 \\
HR 6348    & 4300 & 0.02 & 0.025 $\pm$ 0.003 & 1.015 $\pm$ 0.009 \\
HR 2712    & 3600 & 0.03 & 0.027 $\pm$ 0.010 & 1.030 $\pm$ 0.009 \\
\\
HR 1815    & 5350 & 0.17 & 0.045 $\pm$ 0.003 & 1.054 $\pm$ 0.009 \\
HD 59239   & 4250 & 0.10 & 0.052 $\pm$ 0.008 & 1.061 $\pm$ 0.010 \\
HD 156061  & 3100 & 0.36 & 0.053 $\pm$ 0.003 & 1.066 $\pm$ 0.010 \\
HR 6790    & 3550 & 0.01 & 0.036 $\pm$ 0.010 & 1.072 $\pm$ 0.009 \\
HD 39567   & 3350 & 0.12 & 0.070 $\pm$ 0.004 & 1.108 $\pm$ 0.011 \\
HR 7341    & 4100 & 0.00 & 0.053 $\pm$ 0.004 & 1.126 $\pm$ 0.010 \\
\\
HD 52418   & 4150 & 0.00 & 0.052 $\pm$ 0.004 & 1.141 $\pm$ 0.010 \\
HD 130499  & 3900 & 0.00 & 0.042 $\pm$ 0.005 & 1.152 $\pm$ 0.009 \\
HD 39577   & 3600 & 0.00 & 0.051 $\pm$ 0.007 & 1.152 $\pm$ 0.011 \\
HD 115136  & 3400 & 0.00 & 0.044 $\pm$ 0.004 & 1.155 $\pm$ 0.010 \\
HD 50160   & 4900 & 0.03 & 0.111 $\pm$ 0.015 & 1.234 $\pm$ 0.011 \\
\\
HD 56241   & 3350 & 0.21 & 0.155 $\pm$ 0.013 & 1.321 $\pm$ 0.012 \\
HD 42701   & 4000 & 0.00 & 0.131 $\pm$ 0.005 & 1.370 $\pm$ 0.009 \\
HD 44104   & 3300 & 0.00 & 0.185 $\pm$ 0.004 & 1.385 $\pm$ 0.013 \\
HD 23593   & 3750 & 0.05 & 0.156 $\pm$ 0.001 & 1.388 $\pm$ 0.010 \\
HD 214873  & 3550 & 0.02 & 0.133 $\pm$ 0.003 & 1.389 $\pm$ 0.012 \\
\\
HD 19241   & 4000 & 0.02 & 0.196 $\pm$ 0.013 & 1.411 $\pm$ 0.011 \\
HD 99754   & 3450 & 0.02 & 0.195 $\pm$ 0.006 & 1.419 $\pm$ 0.011 \\
HD 166780  & 3900 & 0.00 & 0.171 $\pm$ 0.005 & 1.421 $\pm$ 0.011 \\
HD 214046  & 3450 & 0.04 & 0.227 $\pm$ 0.017 & 1.446 $\pm$ 0.012 \\
HD 38214   & 4150 & 0.18 & 0.245 $\pm$ 0.008 & 1.451 $\pm$ 0.019 \\
BD+62 1644 & 4150 & 0.07 & 0.213 $\pm$ 0.006 & 1.451 $\pm$ 0.025 \\
\\
HD 53561   & 3450 & 0.00 & 0.169 $\pm$ 0.023 & 1.454 $\pm$ 0.014 \\
HD 15508   & 3900 & 0.00 & 0.197 $\pm$ 0.007 & 1.455 $\pm$ 0.013 \\
HD 173511  & 3900 & 0.00 & 0.218 $\pm$ 0.002 & 1.467 $\pm$ 0.011 \\
HD 34517   & 3250 & 0.00 & 0.222 $\pm$ 0.012 & 1.557 $\pm$ 0.012 \\
HD 39608   & 3300 & 0.00 & 0.208 $\pm$ 0.003 & 1.557 $\pm$ 0.012 \\
\enddata
\tablenotetext{a}{Temperature of fitted Engelke function, but 
this quantity has limited physical significance.  See 
Section~\ref{s.sio}.}
\end{deluxetable}

To measure the equivalent width of the SiO absorption band at 
8~\mum\ in the K giants, it is necessary to also determine 
the interstellar extinction and the temperature of the 
continuum simultaneously.  Using the optical photometry of 
the stars to determine $A_v$ proved unfeasible, because the 
reddening vector in $UBV$ color space follows too closely the 
effect of changing temperature for K giants.  Sources with 
high extinction have absorption from interstellar silicate 
dust at 10~\mum, which tends to extend the SiO absorption to 
longer wavelengths and adds to its apparent equivalent width.

We fitted the SiO band strength, $A_v$, and continuum 
temperature simultaneously, minimizing the $\chi^2$ residuals 
from 6.8 to 11.2~\mum.  Our $A_v$ estimates are based on the
local silicate extinction profiles by \cite{ct06}.  We
determined the continuum temperature by forcing an Engelke
function through the spectral continuum in the vicinity of
12~\mum\ and at $\sim$6.5--7.0~\mum.  

For the purposes of determining $\chi^2$, we used a mean SiO
profile constructed from the SWS spectra of eight K and M 
giants \citep[recalibrated by][]{eng06}.  Table~\ref{t.cohen}
lists ten K and M giants for which recalibrated SWS spectra
exist.  We used all except $\beta$~UMi, which has no
measurable SiO absorption, and $\beta$~Peg, which \cite{eng06}
reject as a trustworthy standard due to its photometric
variability.  Figure~\ref{f.sioprof} presents the eight
resulting SiO profiles, fitted with Engelke functions using
the temperatures determined by \cite{eng92}.  The spectra
do show some variation in the shape of the SiO profile, but
no systematic dependence with spectral class, leading us to 
take the mean of all eight to use for our IRS spectra.

While iterating, our minimum step size was 0.01 mag for $A_v$ 
and 50 K for $T$.  Table~\ref{t.sio} presents the results.
The fitted continuum temperatures in Table~\ref{t.sio} have 
limited physical significance because they are sensitive to 
errors in the slope of the spectrum in the SL module commonly 
produced by the typical pointing errors (SPITE, 
Section~\ref{s.obs}).

\subsection{SiO, color and spectral class} \label{s.ksp} % Sec. 6.3

\begin{figure} % Fig. 14
\includegraphics[width=3.4in]{\figpath 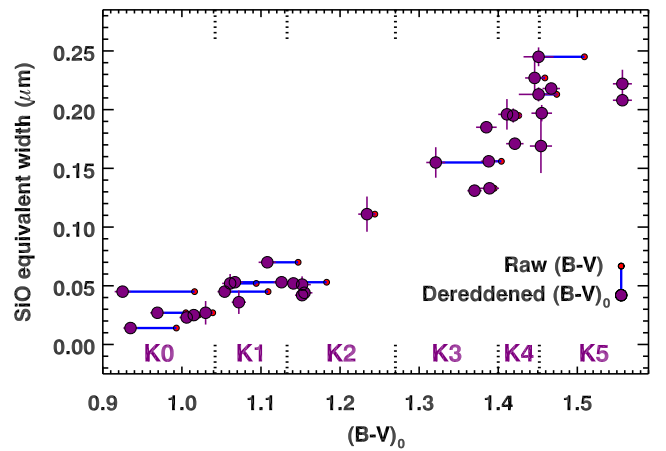}
\caption{The equivalent width of the SiO absorption band at
8~\mum\ as a function of $B-V$ color.  The larger circles 
plot $(B-V)_0$, while the horizontal lines and smaller 
circles depict the raw $B-V$ colors.\label{f.siobv}}
\end{figure}

\begin{figure} % Fig. 15
\includegraphics[width=3.4in]{\figpath 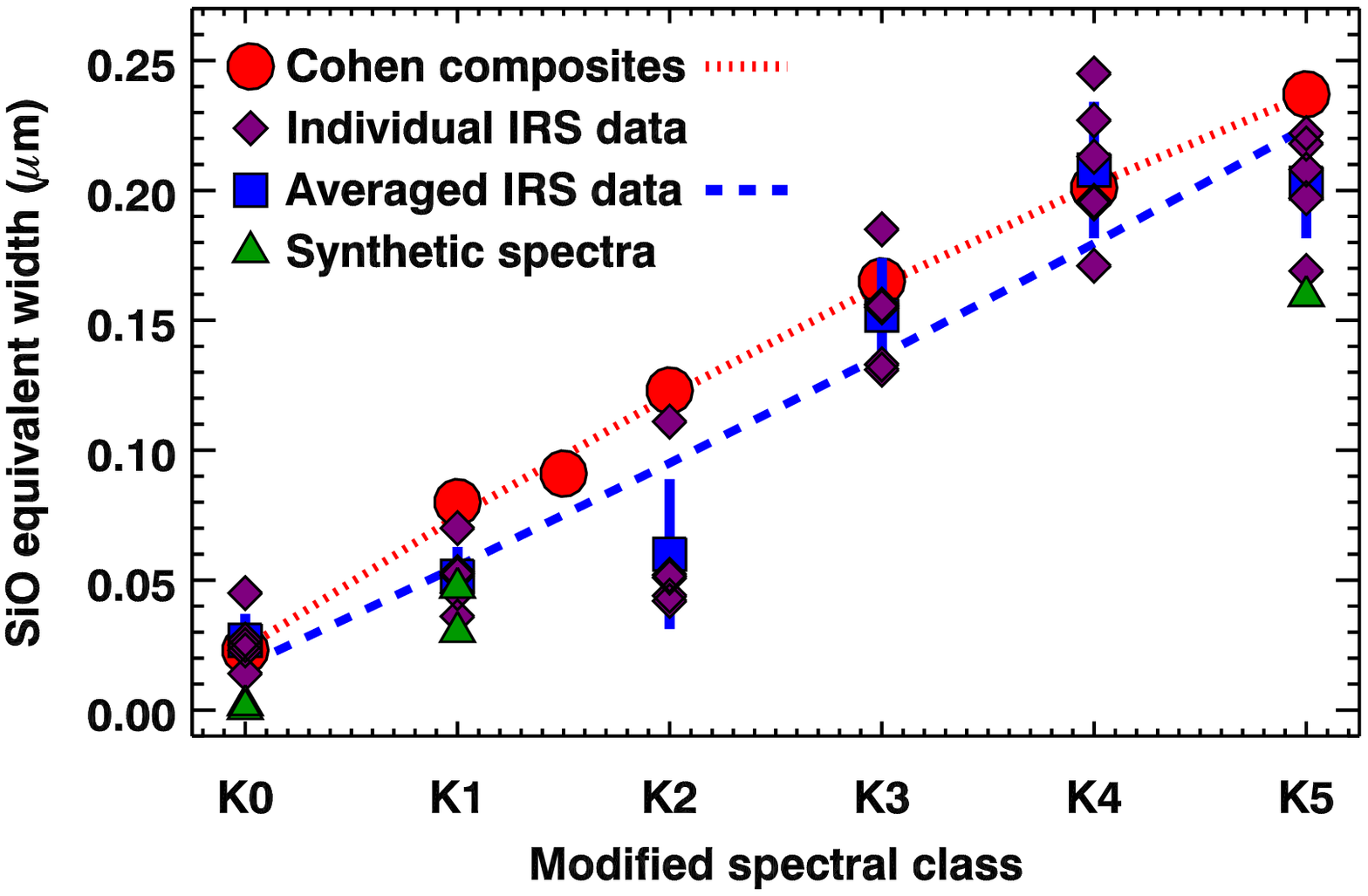}
\includegraphics[width=3.4in]{\figpath 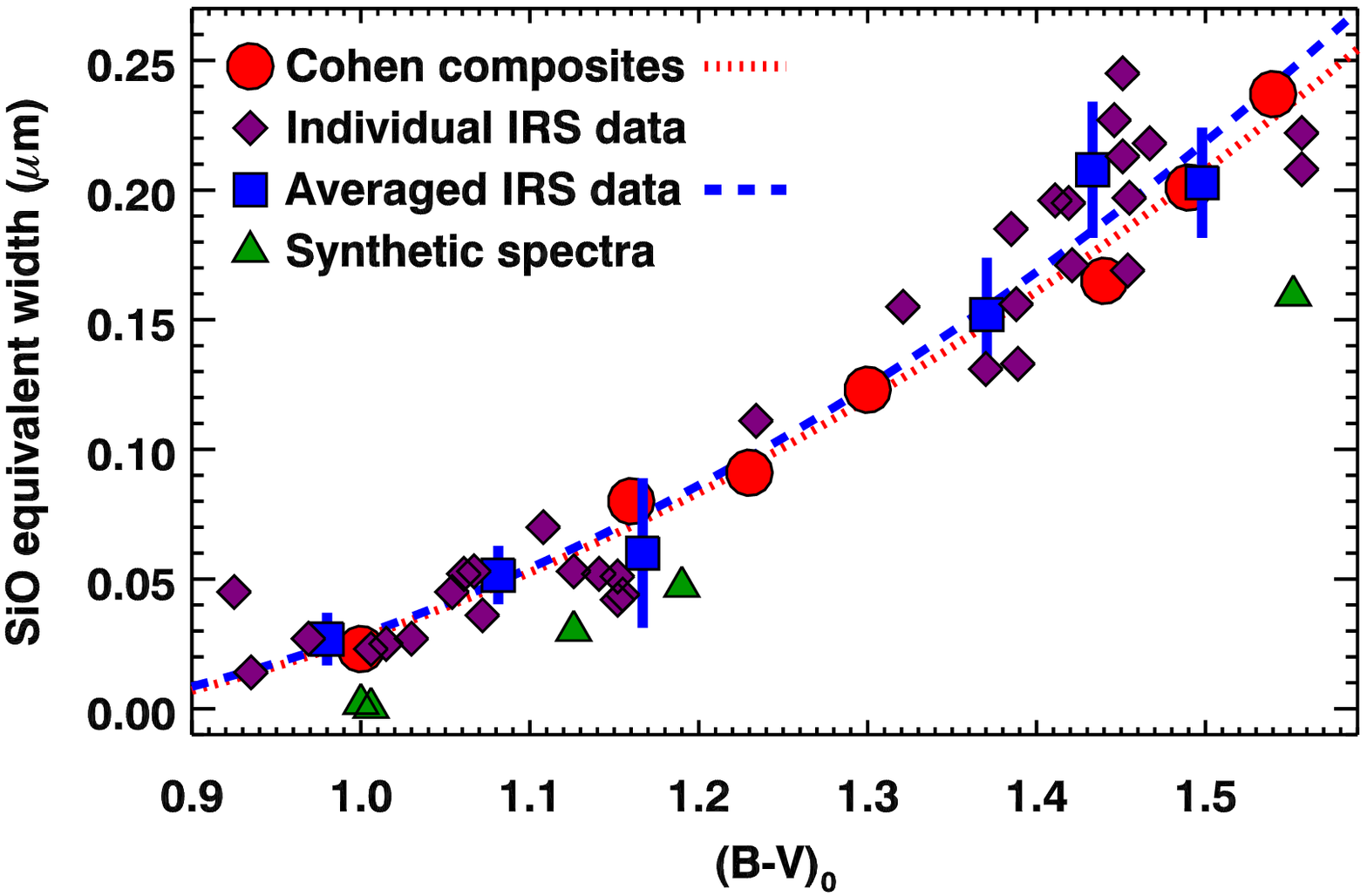}
\caption{The equivalent width of the SiO band as a function
of modified spectral class (top) and dereddened $B-V$ 
(bottom).  The Cohen composite spectra are also plotted as
red circles, while the individual IRS sources appear as
blue diamonds.  The average for each subclass are ploted as
purple squares, along with their standard deviations.  The
dotted and dashed lines are quadratics fitted to the Cohen
composites and averaged IRS data, respectively.\label{f.sio}}
\end{figure}

\begin{deluxetable}{lc} % Table 6
\tablenum{6}
\tablecolumns{2}
\tablewidth{0pt}
\tablecaption{Synthetic SiO measurements}
\label{t.synsio}
\tablehead{
  \colhead{Source}      & \colhead{Eq.\ Width of SiO (\mum)}
}
\startdata
$\delta$ Dra & 0.0019 \\
HR 6606      & 0.0004 \\
HR 7341      & 0.0301 \\
$\xi$ Dra   & 0.0471 \\
$\gamma$ Dra & 0.1594 
\enddata
\end{deluxetable}

One of our objectives was to probe how well the optical 
spectral class of a star predicted the SiO band strength in
the mid-infrared.  This question drove our desire to observe
several targets at each subclass, giving us a sample
large enough to compare variations in SiO band strength from
one subclass to the next, and also within a given subclass.
Discrepancies in classifications from different 
spectroscopists can lead to some amibiguity when deciding
which of our observed stars belong to which subclass.

Prior to launch, M.\ Cohen delivered spectral templates to
the Cornell team for most of the planned standards.  The
spectral subclass used to determine the template depended 
more on how well the photometry for the target in question 
fit the available templates, leading to some 
reclassifications and further ambiguity.

We have modified the optical spectral classes based on the 
dereddened $(B-V)_0$ colors in the right-most column in
Table~\ref{t.sio}.  These are based on the interstellar 
extinction $A_V$ determined in Section~\ref{s.sio} and a 
$B-V$ color generated from photometry in the Tycho~2 catalog 
as described in Section~\ref{s.phot}.  We require that the 
sequence from K0 to K5 increases monotonically with 
$(B-V)_0$, and we have defined the breaks between subclasses 
to minimize the changes required from the literature.  This 
approach generally works, but two issues stand out.  First, 
most of the K2 giants in our sample have colors close to K1, 
except for HD~50160, which is the single source in the 
sizable gap between HD~155136 (K2) and HD~56241 (K3).  
Second, the colors of K4 and K5 giants form a smooth 
continuum, forcing a fairly arbitrary separation.  Placing
the boundary between BD+62~1644 (K4) and HD~53561 (K5) meets 
minimizes changes to earlier spectral classifications.  In 
all, we have changed ten of 33 spectral classes from what 
appears in the literature.  These have modified spectral 
classes in {\bf bold} in Table~\ref{t.kstd}.

Figure~\ref{f.siobv} shows how the equivalent width of the 
SiO absorption band depends on $B-V$ color.  It also shows
the effect of dereddening the $B-V$ color, based on the $A_V$
values determined from fitting the SiO profile in the 
previous section.  The crowding at K1--K2 and K4--K5 is 
readily apparent.  The uneven spacing for the subclasses
illustrates quite nicely the difficulty in equating spectral
classes determined spectroscopically with the photometry.

Figure~\ref{f.sio} plots the equivalent width of the SiO band
as a function of both the modified spectral class and 
$(B-V)_0$ color.  The Cohen composites are $\beta$~UMi (K0), 
$\alpha$~Boo (K1.5), $\alpha$~Hya (K3), and $\alpha$~Tau 
(K5).  The points for K1, K2, and K4 are based on weighted 
means of the above four sources.  The averaged IRS data are 
for each modified spectral subclass.  We fitted quadratics to 
the Cohen composites and averaged IRS data and plotted these
as well.  The apparent difference between the fits in the top 
and bottom panel is due to the differences in colors between 
the sources used for the Cohen composites and in our sample.  
The K1.5 giant $\alpha$~Boo is redder than all but one of the 
K2 giants in our sample, and the $B-V$ color for the K2 
composite is closer to our K3 sources.  The bottom panel shows
that plotting versus color instead of spectral class 
eliminates this problem, and the resulting distributions are 
virtually identical.

We can also measure the SiO band strength in the synthetic
spectra of the five giants considered by 
\cite{dec04}.\footnote{This group includes two stars 
classified in the literature as G9~III, but we will consider 
them as part of the K-giant sample.}  These sources include 
two of the low-resolution IRS standards examined in 
Section~\ref{s.stdk2} (HR~6606 and HR~7341), two bright K 
giants observed in LL with the IRS ($\xi$~Dra and 
$\delta$~Dra), and $\gamma$ Dra, which served as the primary 
standard for the \iso/SWS, but was much too bright for the 
low-resolution IRS modules (it also appears in 
Table~\ref{t.bright}).

We have generated an optical $B-V$ color for the bright 
giants in Table~\ref{t.bright} as for the fainter giants in 
Table~\ref{t.sio}.  The bright giants are close enough that 
we can assume $A_v$=0.  Modifying the spectral classes based 
on $B-V$ would shift four of the six bright giants by one
subclass, which points to a subtle disconnect between the
temperatures of the different layers producing the continuum, 
atomic absorption lines, and molecular absorption bands.  

Table~\ref{t.synsio} gives the SiO equivalent widths from the
synthetic spectra, and Figure~\ref{f.sio} compares them to 
the observed band strengths.  In the top panel, we have 
used the modified spectral classes for the sources (in Tables
\ref{t.kstd} and \ref{t.bright}).  The synthetic data are 
always below the fitted polynomials and generally below the 
observed data.  The $B-V$ colors in the bottom panel are from 
Tables~\ref{t.sio} and \ref{t.bright}, and here all of the 
synthetic data are below the lower envelope defined by the 
observed data.  These results build on the more qualitative 
comparisons made in Section~\ref{s.stdk2}.  Previously, the 
best example of this disagreement was $\gamma$~Dra 
\citep{pri02}.  The synthetic band is weaker than observed in 
spectra from the \iso/SWS \citep[as can be seen in Fig.~3 
of][]{dec04}.  While the differences in band strength are not 
dramatic, they are large enough to propagate an SiO emission 
artifact into the larger database if the synthetic spectra 
are used as truth spectra.

Figure~\ref{f.sio} shows that $(B-V)_0$ is a better predictor 
of SiO band strength than spectral class.  The reason may lie 
in what the spectral class and $B-V$ color are measuring.  
For K giants, the determination of the spectral subclass is 
based on the relative strengths of atomic lines from metals.  
The $B-V$ color measures the shape of the continuum, and it 
is strongly influenced by the H$^-$ ion \citep{wil39}.  Our 
results suggest that the H$^-$ ion better traces the region 
in which the molecular bands arise than the metallic lines 
do.

\subsection{OH absorption} \label{s.oh} % Sec. 6.4

\begin{deluxetable*}{lrllrclcc} % Table 7
\tablenum{7}
\tablecolumns{9}
\tablewidth{0pt}
\tablecaption{Bright K giants}
\label{t.bright}
\tablehead{
  \colhead{Source}      & \colhead{HR} & \colhead{Sp. Type} &
  \colhead{Ref.\tablenotemark{a}}      & \colhead{$(B-V)_0$\tablenotemark{b}} & 
  \colhead{Sp. Class}   & \colhead{F$_{\nu}$ at 12} &
  \colhead{IRS}         & \colhead{Synthetic} \\
  \colhead{ }           & \colhead{ }  & \colhead{ } &
  \colhead{ }           & \colhead{ }  & 
  \colhead{from color\tablenotemark{c}} & 
  \colhead{\mum\ (Jy)\tablenotemark{d}} & \colhead{Obs.} & \colhead{Spectrum}
}
\startdata
$\delta$ Dra & 7310 & G9 III & many   & 1.000 & {\bf K0} &  14.6 &  4      & Y \\
42 Dra       & 6945 & K2 III & many   & 1.182 & {\bf K1} &   5.1 &  2      &   \\
$\xi$ Dra    & 6688 & K2 III & many   & 1.190 & {\bf K1} &  11.7 & 44\tablenotemark{e} & Y \\
HR 5755      & 5755 & K5 III & SIMBAD & 1.447 & {\bf K4} &   3.6 &  2      &   \\
$\gamma$ Dra & 6705 & K5 III & many   & 1.552 &      K5  & 106.9 & \nodata & Y \\
HR 420       &  420 & K5 III & MSS-75 & 1.548 &      K5  &   5.0 &  2      &   \\
\enddata
\tablenotetext{a}{MSS-75 = Michigan catalogue, vol.\ 1 \citep{mss75}, SIMBAD =
  in SIMBAD, but with no further references, many = many references given in
  SIMBAD, all of which agree.}
\tablenotetext{b}{A$_v$ assumed to be 0.}
\tablenotetext{c}{Changes are in bold; see Section~\ref{s.ksp}.}
\tablenotetext{d}{Photometry from the IRAS PSC \citep{psc} and color 
  corrected by dividing by 1.45.}
\tablenotetext{e}{Only 21 spectra coadded.}
\end{deluxetable*}

\begin{deluxetable*}{llrrrr} % Table 8
\tablenum{8}
\tablecolumns{6}
\tablewidth{0pt}
\tablecaption{OH band strengths}
\label{t.oh}
\tablehead{
  \colhead{Source} & \colhead{EW of combined} & 
  \multicolumn{4}{c}{Relative contribution from bands at} \\
  \colhead{ }      & \colhead{Bands (nm)} & \colhead{14.7~\mum} & 
  \colhead{15.3~\mum} & \colhead{16.0~\mum} & \colhead{16.7~\mum}
}
\startdata
\\
\multicolumn{6}{c}{Synthetic spectra}
\\
$\delta$ Dra &  3.12 $\pm$ \nodata & 0.17 $\pm$ \nodata & 0.25 $\pm$ \nodata & 0.28 $\pm$ \nodata & 0.31 $\pm$ \nodata \\
HR 6606      &  2.95 $\pm$ \nodata & 0.16 $\pm$ \nodata & 0.25 $\pm$ \nodata & 0.28 $\pm$ \nodata & 0.31 $\pm$ \nodata \\
HR 7341      &  5.60 $\pm$ \nodata & 0.16 $\pm$ \nodata & 0.24 $\pm$ \nodata & 0.29 $\pm$ \nodata & 0.32 $\pm$ \nodata \\
$\xi$ Dra    &  6.43 $\pm$ \nodata & 0.15 $\pm$ \nodata & 0.24 $\pm$ \nodata & 0.29 $\pm$ \nodata & 0.32 $\pm$ \nodata \\
$\gamma$ Dra &  7.25 $\pm$ \nodata & 0.14 $\pm$ \nodata & 0.23 $\pm$ \nodata & 0.30 $\pm$ \nodata & 0.34 $\pm$ \nodata \\
\\
\multicolumn{6}{c}{IRS low-resolution standards}
\\
HR 6606      &  6.75 $\pm$ 0.43 & 0.20 $\pm$ 0.03 & 0.25 $\pm$ 0.04 & 0.13 $\pm$ 0.02 & 0.42 $\pm$ 0.04 \\
HR 6348      &  6.27 $\pm$ 0.41 & 0.17 $\pm$ 0.03 & 0.33 $\pm$ 0.03 & 0.23 $\pm$ 0.03 & 0.27 $\pm$ 0.04 \\
HR 7341      &  9.05 $\pm$ 0.40 & 0.20 $\pm$ 0.02 & 0.24 $\pm$ 0.02 & 0.18 $\pm$ 0.02 & 0.38 $\pm$ 0.03 \\
HD 166780    & 13.92 $\pm$ 0.67 & 0.16 $\pm$ 0.02 & 0.31 $\pm$ 0.02 & 0.26 $\pm$ 0.02 & 0.27 $\pm$ 0.03 \\
HD 173511    & 13.70 $\pm$ 0.31 & 0.16 $\pm$ 0.01 & 0.30 $\pm$ 0.01 & 0.26 $\pm$ 0.01 & 0.28 $\pm$ 0.01 \\
\\
\multicolumn{6}{c}{IRS observations of bright K giants}
\\
$\delta$ Dra &  8.05 $\pm$ 0.33 & 0.30 $\pm$ 0.02 & 0.26 $\pm$ 0.03 & 0.19 $\pm$ 0.01 & 0.25 $\pm$ 0.01 \\
42 Dra       & 10.06 $\pm$ 0.47 & 0.21 $\pm$ 0.03 & 0.37 $\pm$ 0.02 & 0.24 $\pm$ 0.02 & 0.18 $\pm$ 0.03 \\
$\xi$ Dra    & 10.17 $\pm$ 0.45 & 0.20 $\pm$ 0.02 & 0.25 $\pm$ 0.03 & 0.31 $\pm$ 0.02 & 0.23 $\pm$ 0.03 \\
HR 5755      & 14.95 $\pm$ 0.63 & 0.20 $\pm$ 0.02 & 0.23 $\pm$ 0.02 & 0.24 $\pm$ 0.02 & 0.33 $\pm$ 0.02 \\
HR 420       & 18.85 $\pm$ 0.61 & 0.14 $\pm$ 0.01 & 0.24 $\pm$ 0.02 & 0.26 $\pm$ 0.00 & 0.35 $\pm$ 0.03 \\
\enddata
\end{deluxetable*}

\begin{figure} % Fig. 16
\includegraphics[width=3.4in]{\figpath 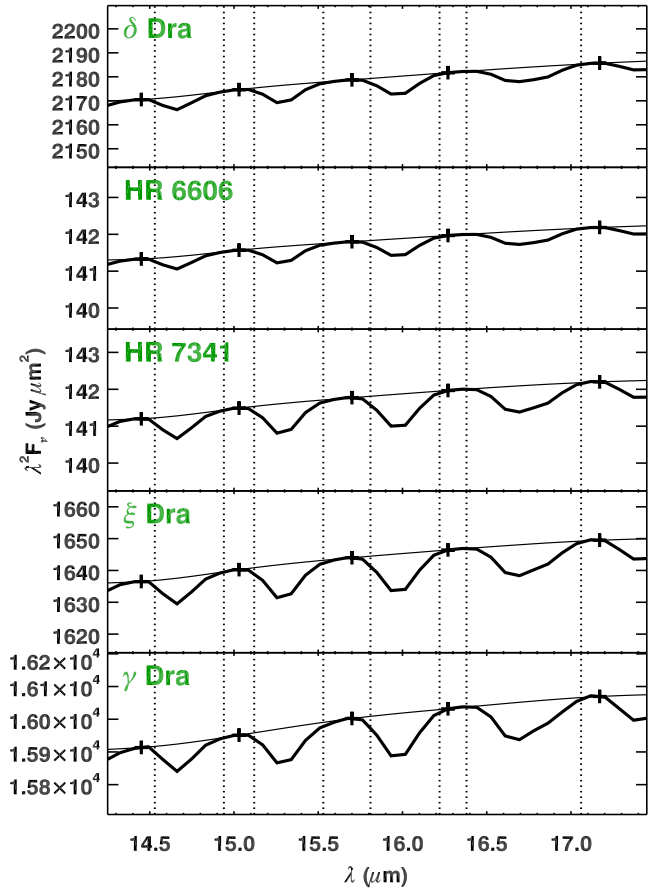}
\caption{Extracting the four OH bands in the 14.5--17~\mum\ 
region of the synthetic spectra of K giants by \citet[][rebinned
to the IRS resolution]{dec04}.
The thicker lines trace the actual spectra, while the thinner 
lines traces the fitted spline, which runs through the plus 
signs.  The vertical dashed lines mark the wavelength limits 
for integrating the equivalent width of each band.  All five 
panels have a vertical extent of 3.1\% of the 
mean.\label{f.ohm}}
\end{figure}

\begin{figure} % Fig. 17
\includegraphics[width=3.4in]{\figpath 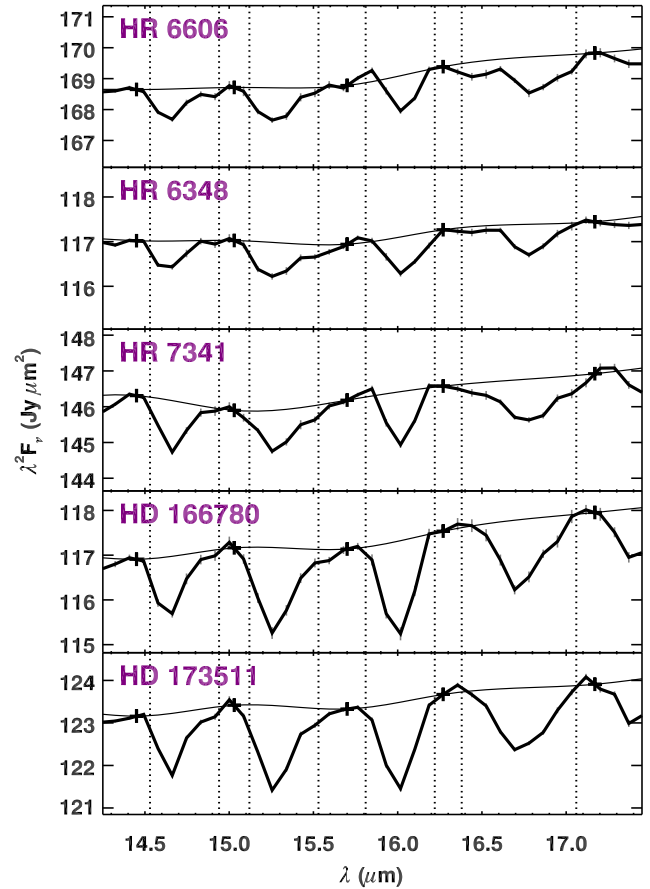}
\caption{The OH spectrum of the IRS low-resolution standards.
Lines and symbols are defined as in Fig.~\ref{f.ohm}, and as 
before, the vertical range of each panel is 3.1\% of the 
mean.\label{f.oh1}}
\end{figure}

\begin{figure} % Fig. 18
\includegraphics[width=3.4in]{\figpath 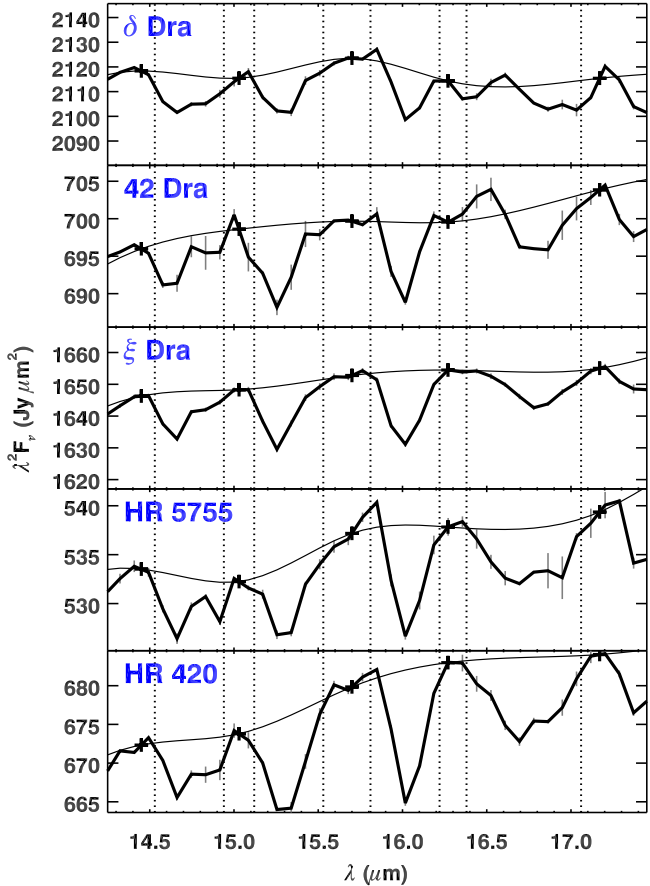}
\caption{The OH spectrum of the bright K giants observed with
the IRS LL module.  Lines, symbols, and vertical extent of 
the panels are as for Fig.\ref{f.ohm} and 
\ref{f.oh1}.\label{f.oh2}}
\end{figure}

\begin{figure} % Fig. 19
\includegraphics[width=3.4in]{\figpath 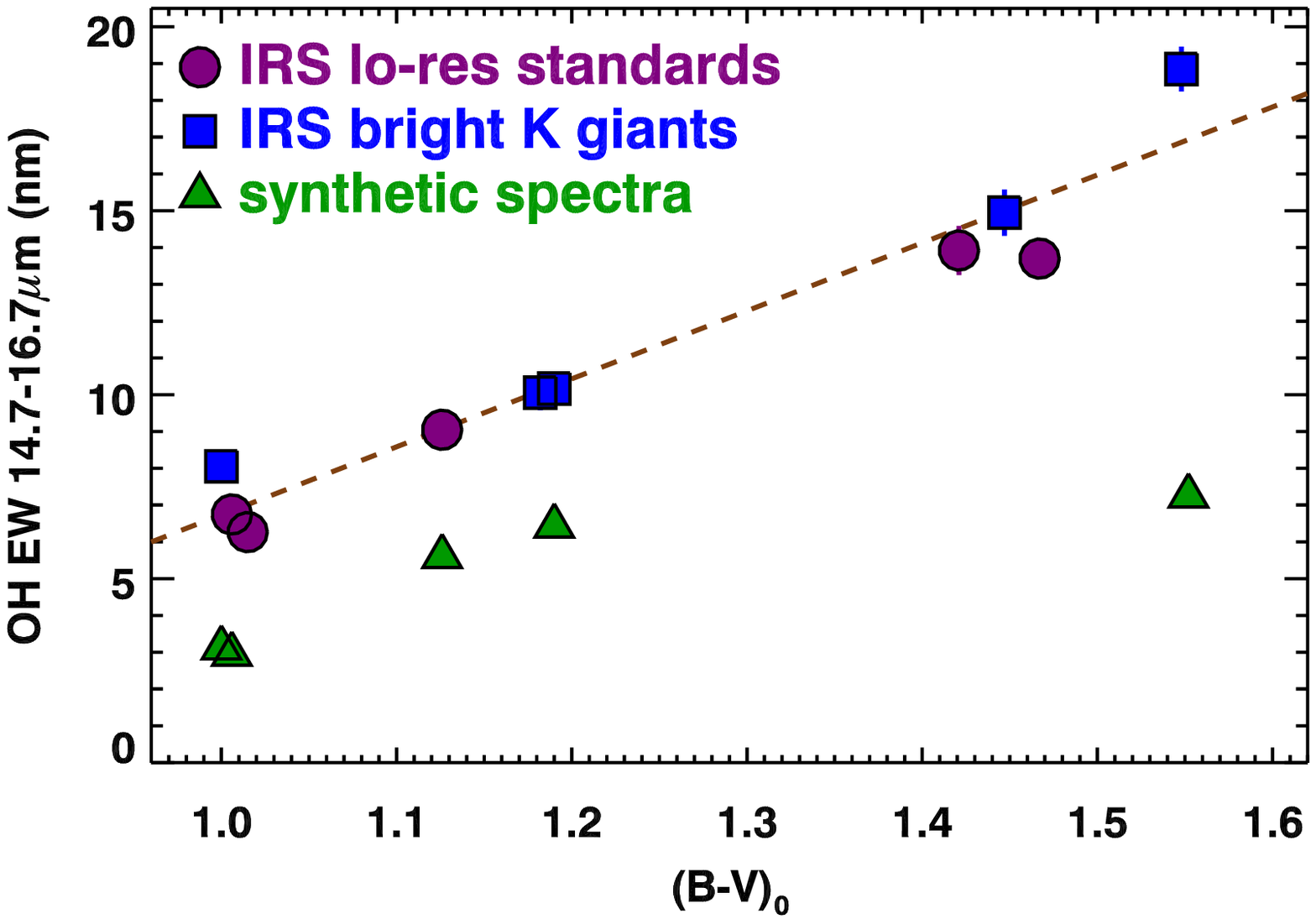}
\caption{The sum of the equivalent width of the four OH bands
between 14.5 and 17~\mum, plotted as a function of $(B-V)_0$.
The uncertainties are generally smaller than the symbols.  The 
dashed line is a least-squares fit to the IRS data.  The OH 
bands in the observed IRS spectra are consistently stronger 
than in the synthetic spectra, and the difference increases as 
the stars grow cooler.\label{f.ohbv}}
\end{figure}

The spectral templates and synthetic spectra of K giants 
behave differently past $\sim$12~\mum.  The spectral 
templates assumed that the long-wavelength behavior of a
giant could be modeled with an Engelke function, which falls
off smoothly with increasingly wavelength, while the 
synthetic spectra show corrugation from OH absorption bands,
which can grow as strong as $\sim$1\% in the 14--18~\mum\
region.  

The observed IRS spectra confirm the presence of OH band
structure.  To examine how the OH bands vary with
effective temperature and compares to its predicted 
structure in the synthetic spectra, we focus on the spectral 
region where they are strongest, which is covered by LL2.  
In particular, we concentrate on the strongest four bands 
centered at 14.7, 15.3, 16.0, and 16.7~\mum.

We base our analysis on the five K giants observed repeatedly
by the IRS in both SL and LL (HR~6348, HD~166780, HD~173511, 
HR~6606, and HR~7341, see Section~\ref{s.truth}, \ref{s.stdk1}, 
and \ref{s.stdk2}) and five brighter K giants observed with
LL, but not SL.  These include $\xi$~Dra, which served as our 
high-resolution standard, and four other IRS targets listed 
in Table~\ref{t.bright}, which gives their basic optical and
infrared properties.  With the exception of $\xi$~Dra, these 
stars were observed only 2--4 times, but they are bright 
enough to produce spectra with sufficient SNR for analysis
of the OH bands.

Figure~\ref{f.ohm} presents the LL2 portion of the synthetic
spectra and illustrates how we fit a spline to the continuum
and measured the equivalent width of the OH bands.  The
available low-resolution synthetic spectra were released 
before multiple changes to the wavelength calibration.  To 
match the currently used wavelength grid and resolution, we 
regridded the spectra, using a gaussian convolution.  A 
comparison of Figure~\ref{f.ohm} to the actual IRS 
measurements of the low-resolution standards in 
Figure~\ref{f.oh1} and the brighter giants
in Figure~\ref{f.oh2} shows that we are comparing spectra
with similar resolutions.  The dotted vertical lines in
the figures mark the wavelength ranges over which we
integrated the equivalent widths; they are fixed for all
of the spectra, despite a slight shift between the synthetic
and actual spectra.  These shifts are too small to account
for the differences between the two sets of data.

Table~\ref{t.oh} presents the total equivalent widths of the 
four bands (in nm) and the individual contributions (as a 
fraction of the total).  In order to propagate errors to
estimate the uncertainty in the equivalent widths, we have
attempted to remove the photometric component from the
uncertainties in the spectra.  These arise because of
fluctuations in slit throughput from one spectrum to the
next and dominate the spectroscopic (or point-to-point)
uncertainties.

Figure~\ref{f.ohbv} plots the total measured OH equivalent
width as a function of $(B-V)_0$
for the synthetic spectra, the low-resolution standards, and
the brighter K giants.  Both sets of giants observed with the
IRS show the same relation of steadily increasing equivalent
width as a function of $B-V$ color.  The line fitted in the
figure gives:
\begin{equation} % Eq. 3
  EW {\rm (nm)} = -11.7 + 18.5 (B-V)
\end{equation}

The synthetic spectra show consistently weaker OH bands.  For 
the four spectra with $B-V < 1.2$ (or spectral classes 
earlier than K3), the relation could be described with a
similar slope, but a lower y-intercept, with a mean shift of
$-$3.7$\pm$0.2 nm.  For all four, we are comparing synthetic 
spectra and observed IRS spectra of the same sources.  If the 
one point for $\gamma$~Dra at $B-V$ = 1.55 is characteristic 
of the behavior at later spectral classes, then the slope of 
the OH dependence in the synthetic spectra decreases for 
redder sources, while the IRS data are consistent with an 
unchanging slope at all $B-V$ colors.

\cite{vm04} compared the strength of the OH bands in this
wavelength range as observed by the \iso/SWS for several of
the giants listed in Table~\ref{t.cohen} with synthetic 
spectra, and they also found that the observed bands were
stronger than predicted by the models.  High-resolution
spectroscopy of OH lines in $\alpha$ Boo by \cite{ryd02} 
uncovered a similar problem.  They found that a cooler
upper atmosphere in the model produced stronger OH lines
and better reproduced the observations.  \cite{vm04} noted 
that the rising opacity of the H$^-$ ion with wavelength 
would push the continuum and the region probed by the 
molecular bands to larger radii at longer wavelengths.  
This in turn would produce the need lower temperatures and
stronger bands.  It also explains how synthetic spectra which
fit the observed data so well at wavelengths below 6~\mum\
\citep[e.g.][]{dec03a,dec03b} can be less successful at 
longer wavelengths.  \cite{de07} noted that a change that
using different absorption coefficients for H$^-$ in their
models changed the SiO band depth by 3\%, reinforcing the
importance of this ion to the emergent spectrum.

The power of the synthetic spectra lies in their prediction
of the presence of the OH bands.  Early spectral templates 
used by us relied on smooth Engelke functions in this 
wavelength range, even though \cite{vm04} and \cite{dec04} 
had shown that these bands were present in the spectra, the 
latter using IRS data.  It is unfortunate that the Cornell 
calibration did not reflect this fact until later in the 
mission.

\subsection{Engelke functions} \label{s.ll} % Sec. 6.5

\begin{figure} % Fig. 20
\includegraphics[width=3.4in]{\figpath 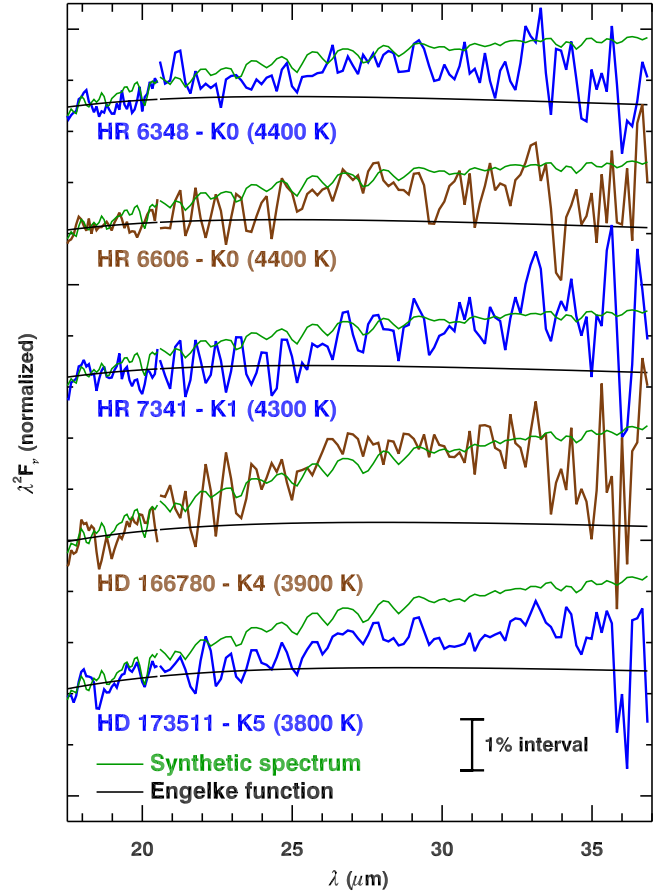}
\caption{A comparison of the shape of our low-resolution
spectra at longer wavelengths to Engelke functions and 
synthetic spectra.  The spectrum of HR~6348 is the truth 
spectrum (Section~\ref{s.truth}); the others are coadded 
spectra.  The synthetic spectra and Engelke functions are 
normalized to match the IRS data at 17.5--18.5~\mum.  The 
synthetic spectra are for HR~6606 (K0), HR~7341 (K1), and 
$\gamma$~Dra (K4--5).  The temperatures of the Engelke 
functions are given in parentheses next to the adopted 
spectral classes of each star.  A 1\% interval in the data is 
shown below the bottom spectrum.  The IRS data are more 
consistent with the long-wavelength shape of the synthetic 
spectra than the Engelke functions, but enough variation 
exists to prevent any firm conclusions.\label{f.tail}}
\end{figure}

Figure~\ref{f.tail} compares the IRS data at longer 
wavelengths to Engelke functions and synthetic spectra.  The
temperatures of the Engelke functions are based on the
temperatures of the adopted spectral classes.  We do not have
synthetic spectra for all five sources, so we used the 
synthetic spectrum of HR~6606 for both it and HR~6348, while
for HD~166780 and HD~173511, we used $\gamma$~Dra.  Both the
synthetic spectra and Engelke function are scaled to the mean
of the IRS data between 17.5 and 18.5~\mum.  

The corrugation in the LL1 spectra of HR~6606, HR~7341, and 
HD~166780 (to the red of 20.5~\mum) is most likely due to
fringing, but in the spectrum of HD~173511, the period and
phase track the OH band structure surprisingly well.  The
difference in band strength compared to the synthetic 
spectrum roughly matches that measured in LL2 
(Section~\ref{s.oh}).

The differences between the shape of the synthetic spectra 
and the Engelke functions are only $\sim$1\% from 17.5 to 
$\sim$35~\mum, and it is impressive that the IRS data are 
generally between them.  The IRS data for HD~166780 follow 
the shape of the synthetic spectrum for all wavelengths below 
$\sim$33~\mum.  Past $\sim$30~\mum, the data quickly become 
less trustworthy due to decreasing signal from the stars and 
responsivity in the Si:Sb detectors.  For HR~6606 and 
HR~7341, the data diverge from the Engelke functions at 
$\sim$25~\mum\ and follow the synthetic spectra to 
$\sim$33~\mum.  The IRS data for HR~6348 are more ambiguous, 
although they are generally closer to the synthetic spectrum.  
If we follow the continuum between OH bands in HD~173511, 
the spectrum generally stays right between the two.  Thus 
three of the spectra are more consistent with the shape of 
the synthetic spectrum, and none are more consistent with the 
shape of the Engelke function.  Perhaps the firmest 
conclusion to be drawn is that the differences between the 
shapes of the synthetic spectra and the corresponding 
Engelke functions at these wavelengths are within $\sim$1\%.

\section{Photometry} \label{s.phot} % Sec. 7.0

\begin{deluxetable*}{lcccrc} % Table 9
\tablenum{9}
\tablecolumns{6}
\tablewidth{0pt}
\tablecaption{K giant photometry}
\label{t.kphot}
\tablehead{
  \colhead{Target} & \colhead{$B_T$} & \colhead{$V_T$} & \colhead{$V_0$} &
  \colhead{$F_{22}$ (mJy)\tablenotemark{a}} & \colhead{$V -$ [22]}
}
\startdata
% target             B_T                 V_T           V_o      F_22    V-[22]
HD 41371   &  8.381 $\pm$ 0.016 & 7.204 $\pm$ 0.010 & 6.811 &  116.5~ & 2.194 \\
HR 7042    &  7.349 $\pm$ 0.015 & 6.201 $\pm$ 0.010 & 5.911 & (233.0) & 2.046 \\
HD 51211   &  8.069 $\pm$ 0.015 & 6.906 $\pm$ 0.010 & 6.685 &  143.2~ & 2.291 \\
HR 6606    &  7.159 $\pm$ 0.015 & 5.982 $\pm$ 0.009 & 5.839 &  341.3~ & 2.389 \\
HR 6348    &  7.440 $\pm$ 0.015 & 6.256 $\pm$ 0.010 & 6.123 &  234.7~ & 2.266 \\
HR 2712    &  7.797 $\pm$ 0.015 & 6.590 $\pm$ 0.009 & 6.445 &  201.3~ & 2.421 \\
\\
HR 1815    &  7.929 $\pm$ 0.015 & 6.632 $\pm$ 0.009 & 6.339 &  220.9~ & 2.416 \\
HD 59239   &  8.135 $\pm$ 0.016 & 6.858 $\pm$ 0.010 & 6.637 &  175.8~ & 2.466 \\
HD 156061  &  8.680 $\pm$ 0.017 & 7.288 $\pm$ 0.011 & 6.796 &  202.8~ & 2.780 \\
HR 6790    &  7.667 $\pm$ 0.015 & 6.414 $\pm$ 0.010 & 6.285 &  236.7~ & 2.437 \\
HD 39567   &  9.049 $\pm$ 0.018 & 7.704 $\pm$ 0.011 & 7.456 &   90.4~ & 2.563 \\
HR 7341    &  7.760 $\pm$ 0.016 & 6.442 $\pm$ 0.010 & 6.317 &  293.1~ & 2.701 \\
\\
HD 52418   &  8.341 $\pm$ 0.016 & 7.003 $\pm$ 0.010 & 6.876 &  172.8~ & 2.686 \\
HD 130499  &  8.074 $\pm$ 0.015 & 6.722 $\pm$ 0.010 & 6.594 &  218.3~ & 2.658 \\
HD 39577   &  8.584 $\pm$ 0.017 & 7.232 $\pm$ 0.010 & 7.104 &  118.1~ & 2.501 \\
HD 115136  &  7.994 $\pm$ 0.016 & 6.639 $\pm$ 0.010 & 6.510 &  229.6~ & 2.629 \\
HD 50160   &  9.190 $\pm$ 0.018 & 7.720 $\pm$ 0.011 & 7.550 &  109.3~ & 2.863 \\
\\
HD 56241   &  9.472 $\pm$ 0.019 & 7.816 $\pm$ 0.011 & 7.446 &  172.1~ & 3.252 \\
HD 42701   &  8.467 $\pm$ 0.015 & 6.836 $\pm$ 0.010 & 6.679 & (338.4) & 3.219 \\
HD 44104   &  9.470 $\pm$ 0.020 & 7.819 $\pm$ 0.011 & 7.659 &  155.3~ & 3.354 \\
HD 23593   &  9.200 $\pm$ 0.017 & 7.525 $\pm$ 0.011 & 7.313 &  216.4~ & 3.367 \\
HD 214873  &  8.579 $\pm$ 0.018 & 6.915 $\pm$ 0.010 & 6.734 & (343.4) & 3.290 \\
\\
HD 19241   &  8.948 $\pm$ 0.017 & 7.256 $\pm$ 0.010 & 7.072 &  256.0~ & 3.309 \\
HD 99754   &  9.111 $\pm$ 0.017 & 7.408 $\pm$ 0.010 & 7.222 &  230.5~ & 3.346 \\
HD 166780  &  9.198 $\pm$ 0.018 & 7.501 $\pm$ 0.011 & 7.336 &  237.0~ & 3.489 \\
HD 214046  &  9.456 $\pm$ 0.019 & 7.710 $\pm$ 0.011 & 7.499 &  202.8~ & 3.483 \\
HD 38214   & 10.083 $\pm$ 0.027 & 8.273 $\pm$ 0.012 & 7.914 &  149.2~ & 3.565 \\
BD+62 1644 & 10.549 $\pm$ 0.035 & 8.784 $\pm$ 0.014 & 8.541 &   77.4~ & 3.479 \\
\\
HD 53561   &  9.335 $\pm$ 0.022 & 7.596 $\pm$ 0.012 & 7.426 &  211.2~ & 3.454 \\
HD 15508   &  9.552 $\pm$ 0.020 & 7.811 $\pm$ 0.011 & 7.641 &  172.1~ & 3.447 \\
HD 173511  &  9.156 $\pm$ 0.018 & 7.400 $\pm$ 0.011 & 7.228 &  247.1~ & 3.427 \\
HD 34517   &  9.248 $\pm$ 0.018 & 7.376 $\pm$ 0.010 & 7.188 & (387.5) & 3.876 \\
HD 39608   &  9.413 $\pm$ 0.019 & 7.542 $\pm$ 0.011 & 7.355 & (296.6) & 3.752 \\
\enddata
\tablenotetext{a}{Values in parentheses are determined from the spectra.}
\end{deluxetable*}

\begin{deluxetable*}{lcccrrr} % Table 10
\tablenum{10}
\tablecolumns{7}
\tablewidth{0pt}
\tablecaption{A dwarf photometry}
\label{t.aphot}
\tablehead{
  \colhead{Target} & \colhead{$B_T$} & \colhead{$V_T$} & 
  \colhead{$V_0$\tablenotemark{a}} & \colhead{$(B-V)_0$\tablenotemark{a}} & 
  \colhead{$F_{22}$ (mJy)\tablenotemark{b}} & \colhead{$V -$ [22]}
}
\startdata
%  target               B_T                 V_T            V        B-V       F_22     V-[22]
HR 1014        & 6.203 $\pm$ 0.014 & 6.041 $\pm$ 0.009 & 6.022 &    0.142 &  (41.7) &    0.290 \\
$\nu$ Tau      & 3.939 $\pm$ 0.014 & 3.888 $\pm$ 0.009 & 3.882 &    0.040 &  234.5~ &    0.024 \\
$\eta^1$ Dor   & 5.681 $\pm$ 0.014 & 5.688 $\pm$ 0.009 & 5.690 & $-$0.012 &  (38.9) & $-$0.120 \\
HD 46190       & 6.709 $\pm$ 0.014 & 6.618 $\pm$ 0.010 & 6.607 &    0.076 &  (26.2) &    0.368 \\
21 Lyn         & 4.594 $\pm$ 0.014 & 4.607 $\pm$ 0.009 & 4.610 & $-$0.018 &  114.5~ & $-$0.027 \\
\\
26 UMa         & 4.515 $\pm$ 0.014 & 4.468 $\pm$ 0.009 & 4.463 &    0.036 &  153.4~ &    0.144 \\
HR 4138        & 4.766 $\pm$ 0.014 & 4.715 $\pm$ 0.009 & 4.709 &    0.040 &  121.4~ &    0.137 \\
$\tau$ Cen     & 3.913 $\pm$ 0.014 & 3.848 $\pm$ 0.009 & 3.841 &    0.053 &  273.4~ &    0.149 \\
$\xi^1$ Cen    & 4.854 $\pm$ 0.014 & 4.828 $\pm$ 0.009 & 4.826 &    0.017 &  108.4~ &    0.130 \\
HR 5467        & 5.818 $\pm$ 0.014 & 5.816 $\pm$ 0.009 & 5.817 & $-$0.004 &   37.4~ & $-$0.034 \\
\\
HR 5949        & 6.259 $\pm$ 0.014 & 6.299 $\pm$ 0.010 & 6.305 & $-$0.041 &  (28.4) &    0.154 \\
$\delta$ UMi   & 4.380 $\pm$ 0.014 & 4.340 $\pm$ 0.009 & 4.336 &    0.030 &  156.8~ &    0.041 \\
HD 163466      & 7.085 $\pm$ 0.015 & 6.875 $\pm$ 0.010 & 6.850 &    0.188 &   23.5~ &    0.495 \\
HR 7018        & 5.671 $\pm$ 0.014 & 5.718 $\pm$ 0.009 & 5.725 & $-$0.047 &   37.1~ & $-$0.135 \\
$\lambda$ Tel  & 4.775 $\pm$ 0.014 & 4.829 $\pm$ 0.009 & 4.837 & $-$0.054 &   86.6~ & $-$0.102 \\
\\
HD 165459      & 7.020 $\pm$ 0.015 & 6.875 $\pm$ 0.010 & 6.858 &    0.126 &  (27.0) &    0.652 \\
29 Vul         & 4.774 $\pm$ 0.014 & 4.798 $\pm$ 0.009 & 4.802 & $-$0.027 &   91.1~ & $-$0.082 \\
$\epsilon$ Aqr & 3.769 $\pm$ 0.014 & 3.758 $\pm$ 0.009 & 3.758 &    0.004 &  265.8~ &    0.036 \\
$\mu$ PsA      & 4.572 $\pm$ 0.014 & 4.502 $\pm$ 0.009 & 4.494 &    0.057 &  151.3~ &    0.160 \\
$\alpha$ Lac   & 3.784 $\pm$ 0.014 & 3.758 $\pm$ 0.009 & 3.756 &    0.017 &  252.0~ & $-$0.024 \\
\enddata
\tablenotetext{a}{Assumed to be unreddened.}
\tablenotetext{b}{Values in parentheses are determined from the spectra, after 
  correcting for the 24~\mum\ dip.}
\end{deluxetable*}

\begin{deluxetable*}{ccrrrr} % Table 11
\tablenum{11}
\tablecolumns{6}
\tablewidth{0pt}
\tablecaption{Photometric transformations and color relations}
\label{t.trans}
\tablehead{
  \colhead{Relation} & \colhead{Range} & \multicolumn{4}{c}{Polynomial coefficients}
}
\startdata
$B-V$ vs.\ $B_T-V_T$   & ~~~1.10 -- 1.90 &    0.099 &    0.779 &       & \\
$B-V$ vs.\ $B_T-V_T$   & $-$0.10 -- 0.25 & $-$0.006 &    0.890 & 0.155 & \\
$V-V_T$ vs.\ $B_T-V_T$ &        all      &    0.001 & $-$0.013 & 0.055 & $-$0.020 \\
$V-$ [22] vs.\ $B-V$   &        all      & $-$0.008 &    2.380 &       & \\
$V-$ [22] vs.\ $B-V$   & ~~~0.90 -- 1.60 & $-$0.240 &    2.561 &       & \\
$V-$ [22] vs.\ $B-V$   & $-$0.06 -- 0.20 & $-$0.004 &    2.737 &       & \\
\enddata
\end{deluxetable*}

\begin{figure} % Fig. 21
\includegraphics[width=3.4in]{\figpath 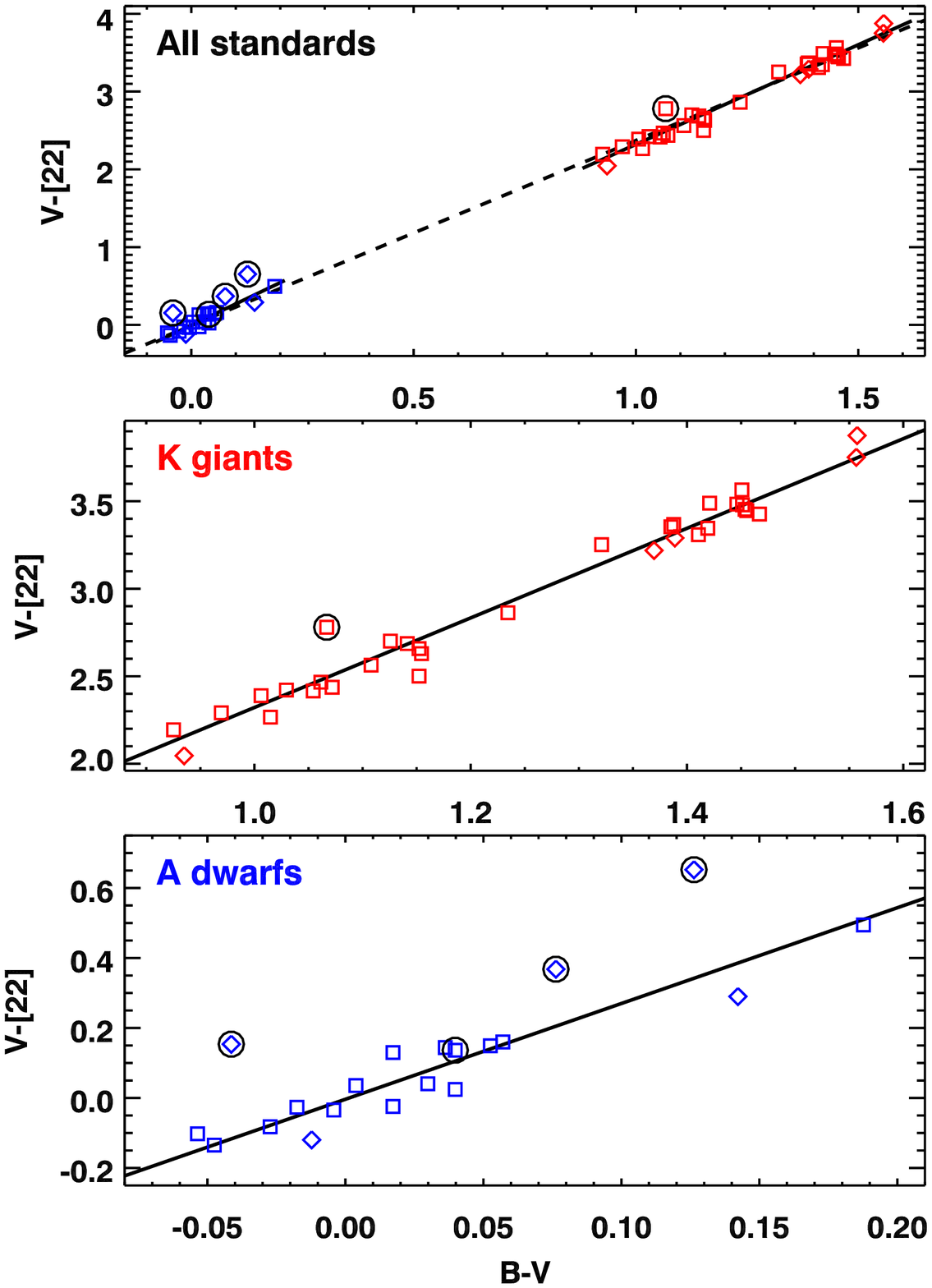}
\caption{$V-[22]$ as a function of $B-V$.  The top panel
presents the data for all standards and the fitted lines.  
The dashed line is fitted to all of the data (after 
excluding outliers), while the solid line segments are 
fitted to the K giants and A dwarfs separately.  Squares
depict sources with Red PU photometry, while diamonds 
depict sources where [22] is based on the spectra.  Data 
with possible debris disks are circled.  The bottom two 
panels zoom in on the K giants and A dwarfs.  The vertical 
scale of the A dwarfs in the bottom panel is half that of 
the K giants.\label{f.bv22}} 
\end{figure}

Tables~\ref{t.kphot} and \ref{t.aphot} present the optical
and 22-\mum\ PU photometry for our sample.  The Tycho data
are from the Tycho~2 catalog \citep{hog00}.  
Table~\ref{t.trans} gives the transformations to convert to 
Johnson $B$ and $V$, which we derived using data presented 
by \citet[][Table 2]{bes00}.  The K giants lie in the range 
1.10 $\le B_T-V_T \le$ 1.90, and they follow a linear
relation with $B-V$.  The A dwarfs lie in the range $-$0.10 
$\le B_T-V_T \le$ 0.25, and for these a quadratic was 
necessary to transform to $B-V$.  \cite{bes00} found that a
cubic could fit $V-V_T$ as a function of $B_T-V_T$ for all
colors.

We found no evidence of interstellar extinction in the IRS
spectra of the A dwarfs, which is consistent with their close 
distances compared to the K giants.  Consequently, we have 
used their $(B-V)$ colors and $V$ magnitudes without 
correction.  The $V_0$ magnitudes for the K giants in 
Table~\ref{t.kphot} are based on the extinctions in 
Table~\ref{t.sio}.

The Red PU photometry is labelled as $F_{22}$ in 
Tables~\ref{t.kphot} and \ref{t.aphot}.  The photometry for 
HR~7018 differs slightly from the value given by 
\citep{sl11b} because it adds 7 observations late in the
cryogenic mission not included earlier.  To convert to a 
magnitude, we have assumed a zero-magnitude flux density
of 8.19 Jy at 22.35~\mum, scaled from 7.30 Jy at 23.675~\mum\
for MIPS-24.

Values for $F_{22}$ in parentheses are for sources with no 
available photometry; instead, we estimated $F_{22}$ from the 
spectra in a 1-\mum\ window centered on 22.35~\mum.  For the 
A dwarfs, when the photometry is estimated from the spectra, 
it is {\it after} correcting for the 24~\mum\ dip seen in 
some faint spectra, as described below 
(Section~\ref{s.adwarf}).

Table~\ref{t.trans} includes linear relations of $V-[22]$ on 
$B-V$ fitted to all of our standards, along with separate 
relations for the K giants and A dwarfs.  All of the lines 
are based on least-square fitting in two passes, with the 
second pass excluding data more than 0.10 mag from the line  
fitted in the first pass.

Figure~\ref{f.bv22} illustrates the dependence of $V-[22]$ on 
$B-V$ for the different samples.  It is somewhat surprising 
that a single line fits all of the data as well as it does.  
The differences in the two lines fitted separately to the A 
dwarfs and K giants suggests that a cubic might fit all 
colors even better, but with no data for spectral classes 
F or G, we will just consider the separately fitted lines.

We have already noted the red excess in HD~156051 suggestive
of a debris disk.  Four of the A dwarfs also show possible
debris disks.  Figure~\ref{f.bv22} circles those data; they
account for the worst outliers from the fitted lines.  
Excluding these five sources, we find a standard deviation of
0.073 mag about the fitted line for K giants and 0.053 mag
for the A dwarfs.  Both groups behave well.

The y-intercept of the relation between $B-V$ and $V-[22]$ 
for the A dwarfs is $-$0.0036, close to the expected value of 
zero and smaller than the standard deviation (0.053) and
uncertainty in the mean (0.013).  This result verifies that 
the MIPS calibration at 24~\mum\ \citep{rie08} is consistent 
with the expected behavior of an A0 dwarf.

\section{The A dwarfs} \label{s.adwarf} % Sec. 8.0

\begin{figure} % Fig. 22
\includegraphics[width=3.4in]{\figpath 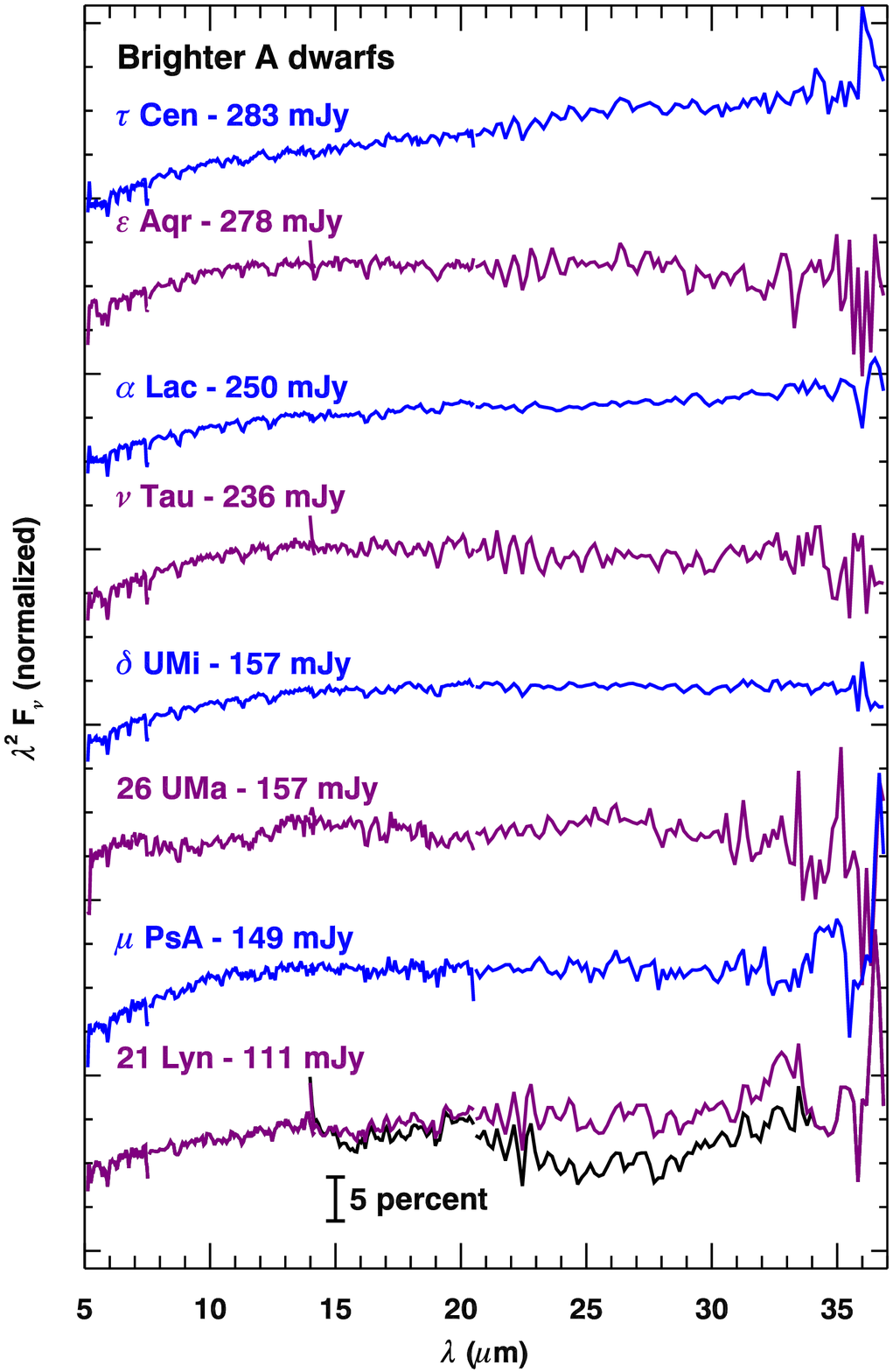}
\caption{The IRS spectra of the brighter A dwarfs observed
as standards, plotted in Rayleigh-Jeans units.  The spectrum
of 21 Lyn has been corrected for the 24~\mum\ dip, with the
uncorrected spectrum shown in black.  The flux densities 
given after the names of the sources are the values at
22.35~\mum, as measured from the (corrected) 
spectra.\label{f.a1}}
\end{figure}

\begin{figure} % Fig. 23
\includegraphics[width=3.4in]{\figpath 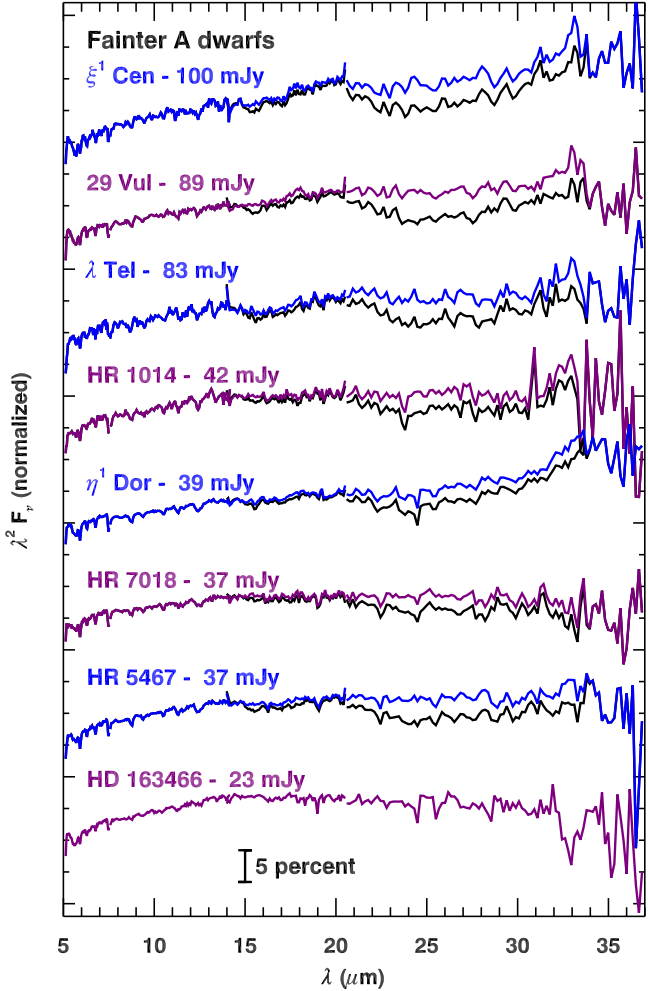}
\caption{The IRS spectra of the fainter A dwarfs observed
as standards, plotted in Rayleigh-Jeans units.  Traces and
flux densities are as described in 
Fig.~\ref{f.a1}.\label{f.a2}}
\end{figure}

\begin{figure} % Fig. 24
\includegraphics[width=3.4in]{\figpath 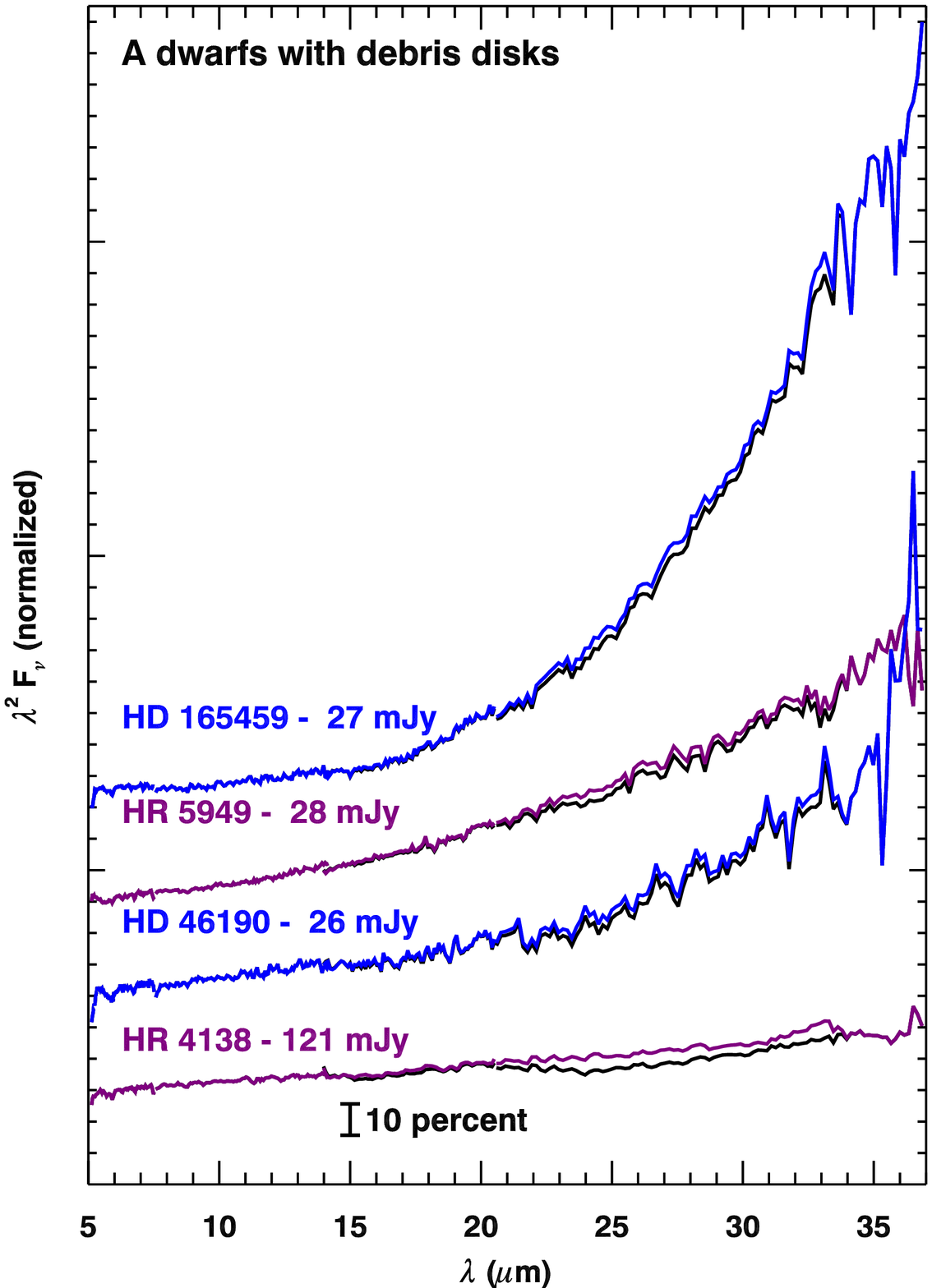}
\caption{The IRS spectra of the A dwarfs with excesses from
possible debris disks, plotted in Rayleigh-Jeans units.  
Traces and flux densities are as described in 
Fig.~\ref{f.a1}.\label{f.a3}}
\end{figure}

Our sample of calibrators includes 20 A dwarfs.  Their hotter
effective temperatures make them easier to model and thus
potentially better standards than K giants.  However, they 
are far less common in the bright infrared sky.  And their
spectral properties have proven to be full of surprises.
\cite{egg84} quotes W.~P.\ Bidelman:  ``I do not know, nor 
have I ever seen, a normal A0 star.''  

Vega is a case in point.  IRAS discovered a previously 
unsuspected debris disk \citep{aum84}.  Its narrow absorption 
lines arise from a pole-on configuration, as first speculated 
by \cite{gra85}.  Later work substantiates this scenario.  We 
are seeing a fast rotator pole-on, leading to a range of 
effective temperatures across the disk of the star 
\citep{hil10}.  It is rather ironic that Vega originally 
defined zero magnitude in most photometric systems.

While we used the available infrared photometry to select
against obvious debris disks, we were still concerned that
contamination from debris disks, binarity, or rotation and
inclination effects might limit the utility of any star as a
standard.  Thus we observed many candidates for standards.

Figures~\ref{f.a1} and \ref{f.a2} present the spectra of the 
A dwarfs observed as potential IRS standards, ordered by 
their brightness at 22~\mum, while Figure~\ref{f.a3} 
segregrates the four spectra showing red excesses from
possible debris disks.  Several others show fainter red 
excesses.  We have already described $\alpha$~Lac as one
example.  These spectra are considered in 
Section~\ref{s.debris} below.

Most of the fainter spectra suffer from an artifact known as
the 24~\mum\ dip, as can be seen in Figure~\ref{f.a2}.  Our 
spectra show an impact in both LL2 and LL1, with the apparent 
detector output pushed down slightly at $\sim$15--18~\mum\ 
and more significantly at $\sim$22-30~\mum.  An apparent 
upward cusp results at $\sim$19-20~\mum.

To correct this artifact, we follow the method of 
\cite{slo04b}, who first identified the problem when 
investigating the cool red excess around HD~46190.  We 
have assumed that HR~5467 and HR~7018 should have identical 
shapes in LL2 and LL1 to $\delta$~UMi.  We averaged their 
spectra, divided by $\delta$~UMi, normalized the result, and 
smoothed it.  Due to the degrading SNR at longer wavelengths, 
we set the correction = 1 past 34~\mum.  To apply the 
correction to a given spectrum, we use the ratio of spectral 
emission at 24--25~\mum\ to 19--20~\mum\ (in Rayleigh-Jeans 
units) to estimate the strength of the 24~\mum\ dip and scale
the correction accordingly.  For the four A dwarfs with 
possible debris disks, we use spectra with similar 
brightnesses to estimate the scaling, but as 
Figure~\ref{f.a3} shows, the effect is only noticeable for 
HR~4138, which has the weakest excess of the four.  

We assumed that the strength of the artifact depends on the 
brightness of the spectrum for the debris-disk candidates, 
which is disputable.  For example, the faintest of our A 
dwarfs, HD 163466, shows no sign of the 24~\mum\ dip.  Except 
for this one source, all of the A dwarfs with $F_{22} <$ 
115~mJy require a correction, but the two K giants fainter 
than this limit (HD 39567 and BD+62 1644) show no convincing 
evidence of the artifact (see Figures~\ref{f.k1} and 
\ref{f.k4}).\footnote{BD+62 1644 does show a small deficit in 
LL1, but no cusp at $\sim$20~\mum.}  Nor do the two K giants 
in the 115--120 mJy range (HD 41371 and HD 39577; 
Figures~\ref{f.k0} and \ref{f.k2}).

Early in the mission, we found that some of the A dwarfs were
brighter at longer wavelengths than expected based on the 
spectral templates supplied by M.\ Cohen.  We reported these
results informally \citep[e.g.][although the published 
abstract is non-committal]{slo04a}.  Based on the photometric
analysis in Section~\ref{s.phot}, we can now conclude that 
the A dwarfs are at least as well behaved as the K giants.  
What has changed is our current focus on using the Tycho 
photometry to estimate expected flux densities at 22~\mum.
Previously, we were also considering near-infrared 
photometry, primarily from 2MASS \citep{skr06}, as well as
longer-wavelength data from the IRAS surveys, but our sources 
are bright enough to saturate 2MASS, and the precision of the
other infrared data does not appear to be sufficient for our
purposes.  We can now state that the A dwarfs are just as 
well behaved photometrically as the K giants, provided one
has filtered out the possible debris disks.

The 24~\mum\ dip remains an unsolved calibration problem with
IRS data.  While we did not this problem in the four K giants
in our sample with $F_{22}$ $\la$ 120~mJy, we are hesitant to
draw any conclusions, because our of small sample.  
\cite{ard10} applied a correction to all sources in their 
sample fainter than 100~mJy (at 24~\mum).  Better understanding
of this artifact awaits a more rigorous analysis of the fainter
stellar spectra observed by the IRS.

\section{Debris disks} % Sec. 9.0

\subsection{Sources with strong red excesses} \label{s.debris} % Sec. 9.1

\begin{figure} % Fig. 25
\includegraphics[width=3.4in]{\figpath 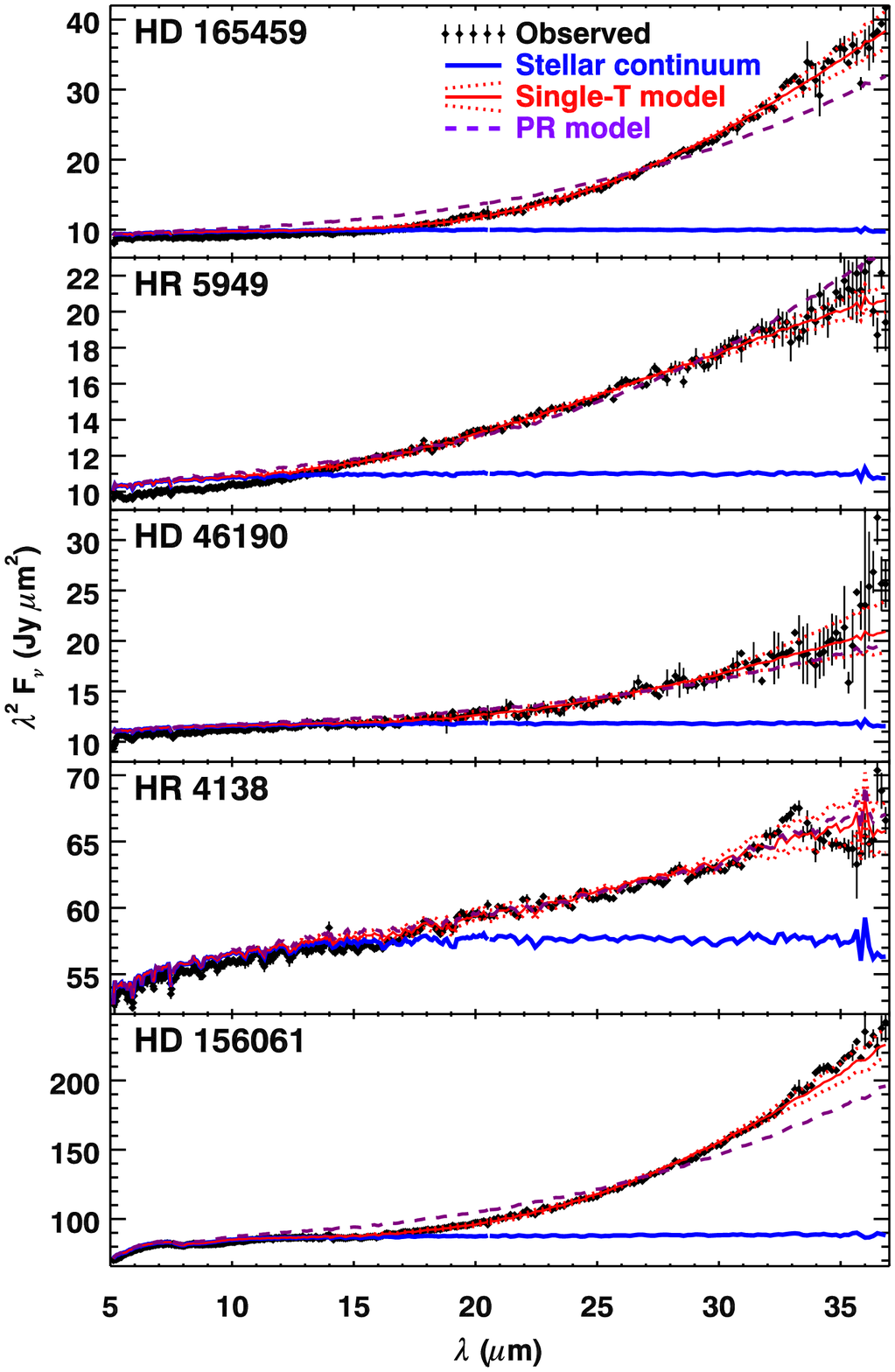}
\caption{Possible debris disks among the IRS standards.  Each
panel depicts the IRS spectrum with black diamonds, the 
assumed continuum as a solid blue line, the 
single-temperature model with a solid red line, the
$\pm$1-$\sigma$ temperature range with dotted red lines,
and the Poynting-Robertson model with a dashed purple line.
This last model is only viable for HR~4138, which has the
excess with the lowest contrast in the 
sample.\label{f.debris}}
\end{figure}

\begin{deluxetable}{lcrrrrl} % Table 12
\tablenum{12}
\tablecolumns{7}
\tablewidth{0pt}
\tablecaption{Stellar properties of candidate debris disks}
\label{t.stellar}
\tablehead{
  \colhead{Source} & \colhead{Parallax} & \colhead{$T_{*}$} & 
  \colhead{BC} & \colhead{$R_*$} & \colhead{$M_*$} & \colhead{$L_*$} \\
  \colhead{ }      & \colhead{(mas)}    & \colhead{(K)}   & 
  \colhead{(mag)} & \colhead{(R$_\odot$)} & \colhead{(M$_\odot$)} & 
  \colhead{(L$_\odot$)} 
}
\startdata
% target         parallax        T_*        BC    R_*    M_*       L_*
HD 165459 & 11.26 $\pm$ 0.28 &  8420 & $-$0.158 & 1.8  & 2.1 & 13 $\pm$ 1 \\
HR 5949   &  6.89 $\pm$ 0.40 & 10010 & $-$0.408 & 2.5  & 3.1 & 72$^{+9}_{-8}$ \\
HD 46190  & 11.92 $\pm$ 0.27 &  8840 & $-$0.183 & 2.0  & 2.3 & 15 $\pm$ 1 \\
HR 4138   & 12.52 $\pm$ 0.17 &  9140 & $-$0.217 & 2.1  & 2.5 & 79 $\pm$ 2 \\
HD 156061 &  5.95 $\pm$ 0.64 &  4540 & $-$0.544 & 17~~ & (2) & 69$^{+18}_{-13}$ \\
\enddata
\end{deluxetable}

\begin{deluxetable*}{lccrrccccc} % Table 13
\tablenum{13}
\tablecolumns{10}
\tablewidth{0pt}
\tablecaption{Dust properties of candidate debris disks}
\label{t.dust}
\tablehead{
  \colhead{Source} & \colhead{Continuum} & \colhead{Contrast at} &
  \multicolumn{2}{c}{$T_{dust}$ (K)} & \multicolumn{2}{c}{$D_{dust}$ (AU)} & 
  \multicolumn{2}{c}{$L_{dust}$ (10$^{-4}$ L$_{\odot}$)} & 
  \colhead{Min. $M_{dust}$} \\
  \colhead{ }      & \colhead{range (\mum)} & \colhead{30~\mum\ (\%)} &
  \colhead{Best} & \colhead{1-$\sigma$ range} & 
  \colhead{Best} & \colhead{1-$\sigma$ range} & 
  \colhead{Best} & \colhead{1-$\sigma$ range} & 
  \colhead{(10$^{-4}$ M$_{\Moon}$)}
}
\startdata
% target    cont. contrast    Tdust          Ddust          Ldust          mass
HD 165459 & 14--18 & 134 &  97 &  89--105 & 32 & 27--38 & 5.7 & 5.2--6.4 & 5.4 \\
HR 5949   &  9--16 &  61 & 161 & 148--176 & 23 & 19--27 & 7.5 & 7.3--8.0 & 2.7 \\
HD 46190  & 13--17 &  40 & 106 &  84--136 & 32 & 20--52 & 1.7 & 1.5--2.4 & 0.8 \\
HR 4138   & 14--18 &  10 & 140 & 114--175 & 21 & 13--31 & 1.8 & 1.8--2.0 & 0.4 \\
HD 156061 & 12--16 &  74 &  98 &  92--104 & 85 & 75--96 & 98  &  92--108 & 760 \\
\enddata
\end{deluxetable*}

When selecting potential standards, we avoided sources with 
excesses as measured by \iras\ at 12 and 25~\mum.   
Nonetheless, five of the spectra in our sample show strong 
red excesses which could arise from debris disks.  These 
spectra represent 20\% of our A dwarfs and 3\% of our K 
giants.  By comparison, a MIPS survey at 24 and 70~\mum\ 
found debris disks around 32\% of 160 stars with spectral 
classes between B6 and A7.  Given the small size of 
our sample, the rate of incidence is similar.  While we
tried to avoid such stars, the \iras\ 25~\mum\ filter was too 
far to the blue and not sensitive enough for our purposes.

Table~\ref{t.stellar} gives the stellar properties of the five
sources with possible debris disks.  The parallaxes are from 
\cite{vl07}.  The effective temperatures and the bolometric 
corrections are interpolated using the $B-V$ colors 
(Tables~\ref{t.kphot} and \ref{t.aphot}) and data from 
\citet[][Table 5]{kh95}.  Radii and masses for the A dwarfs
and all data for HD~156061 are interpolated similarly with 
data from \citet[][ApQ4]{cox00}.  Stellar luminosities are 
based on the apparent $V$ magnitudes (Tables~\ref{t.kphot} 
and \ref{t.aphot}), bolometric corrections, and parallaxes.  
The uncertainties reflect the uncertainties in parallax.

MIPS 24~\mum\ photometry confirms the red excesses in three of
our sample stars, HD~46190, HR~4138 and HD~165459 
\citep{eng07}.  Photometry at 70~\mum\ exists only for 
HD~165459 \citep{su06}.  For temperatures above $\sim$40~K, 
the 70~\mum\ point falls on the Rayleigh-Jeans side of the 
dust emission, and without it, our analysis is limited to the 
Wien side.  Consequently, we have only fitted two simple 
models to the data:  a single-temperature Planck function, 
simulating a population of large grains all at the same 
distance from the central star, and a model where 
Poynting-Robertson (PR) drag is feeding particles into the 
disk, which leads to an excess following the relation 
$F_{\nu} \propto \lambda$ \citep{jur04}.

For each source, we fit the spectrum of a naked star with
a similar spectral class, using $\delta$~UMi for the four
A dwarfs and HR~7341 for HD~156061.  We have minimized the
$\chi^2$ difference between model and spectrum from 18 to 
34~\mum, allowing both the temperature of the excess and the
wavelength range over which we fitted the stellar continuum 
to vary.  We also found the temperature that produced a 
$\chi^2$ value twice the minimum to estimate the 1-$\sigma$ 
spread in the likely dust temperature.  Figure~\ref{f.debris} 
illustrates the process, and Table~\ref{t.dust} reports the 
wavelength ranges used to fit the continuum, the contrast
between excess and continuum at 30~\mum, and the resulting 
dust temperatures.

Figure~\ref{f.debris} shows a fitted model based on PR
drag using the same stellar continuum as for the 
single-temperature model.  Only for HR~4138 is the PR-drag 
model as good as the single-temperature model.  This star
has an excess with a lower contrast than the other four 
debris disks, consistent with arguments by \cite{wya05}
that PR-drag should only be important in the least massive 
debris disks, which would be those with the lowest contrast.

If we assume that the dust grains are blackbodies in
radiative equilibrium, then their distance from the central
star follows the relation:
\begin{equation}
  D_{dust} = \frac{R_*}{2} \left( \frac{T_{dust}}{T_*} \right)^2.
\end{equation}
Table~\ref{t.dust} gives the distances to the dust grains 
($D_{dust}$ for the best-fitted dust temperature and the 
$\pm$1-$\sigma$ temperatures to either side.  It also gives 
the corresponding dust luminosities, found by summing the 
single-temperature blackbodies fitted through the excesses at 
the wavelengths indicated.  We can estimate a minimum mass 
using the relation given by \cite{jur99}:
\begin{equation}
  M_{dust} > \frac{L_{dust} D_{dust}^2}{G M_* c}
\end{equation}
Table~\ref{t.dust} reports the lowest minimum mass from the
range of likely temperatures.

Some caveats are in order.  Several nearby debris disks show 
clear evidence for a hot inner disk \citep[see][and 
references therein]{su13}, and our spectra show some evidence
of a warm dust component.  Our estimates of the location of 
the dust assume that the grains radiate as blackbodies, which 
is a reasonable approximation for large grains.  Grains 
smaller than several microns will be warmer at a given 
distance and for a fixed temperature would have to be further 
away than our estimated distances.  

HD~46190 illustrates the limits of our analysis for faint 
sources.  \cite{slo04b} examined a version of its spectrum 
based on a much earlier calibration, and they found a dust 
temperature of 81 K, compared to the current estimate of 
106 K.  The temperature change has moved the dust inward, 
from 82 AU to 32 AU.  While we are confident that a debris 
disk is present, our quantitative results depend strongly on
the calibration and our ability to correct for the 24~\mum\
dip.

HD~156061 is not the first red giant associated with a 
possible debris disk.  \cite{jur99} examined eight red giants 
with red excesses, based on their IRAS colors, and concluded
that debris disks were better explanations of the excesses 
than dust outflows or the Pleaides effect (which he described
as ``cirrus hot spots").  Our dust-mass estimate is based on
an assumed stellar mass of 2~M$_{\odot}$.

\cite{slo04b} noted that HD~46190 lies within the Local 
Bubble, making the Pleaides effect an unlikely explanation 
for the red excess in its spectrum.  The same is true for 
HR~4138 and HD~165459.  HR~5949 and HD~156061 lie beyond the 
boundaries of the Local Bubble.  HR~5949, with a distance of 
145~pc and galactic coordinates (91$\degr$, +45$\degr$) lies 
right in a bank of material identified in the 
three-dimensional maps by \citet[][Figure 6]{lal03}.  
HD~156061 is 180~pc away with galactic coordinates (0$\degr$, 
+7$\degr$), putting it in a crowded field within several 
degrees of the Pipe Nebula.  Furthermore, red images from the 
Palomar Sky Survey show dust lanes within 10$\arcmin$ of 
HD~156061, but the images show no evidence for extended 
reflection from dust in the immediate vicinity of either
HD~156061 or HR~5949.  Inspection of the IRS spectral images
reveals no evidence for extended emission in LL.  We conclude
that the Pleaides effect is not likely for any of the five
sources, although direct high-resolution imaging would be a
more definitive test.

\subsection{Sources with weak red excesses} \label{s.weak} % Sec. 9.2

\begin{figure} % Fig. 26
\includegraphics[width=3.4in]{\figpath 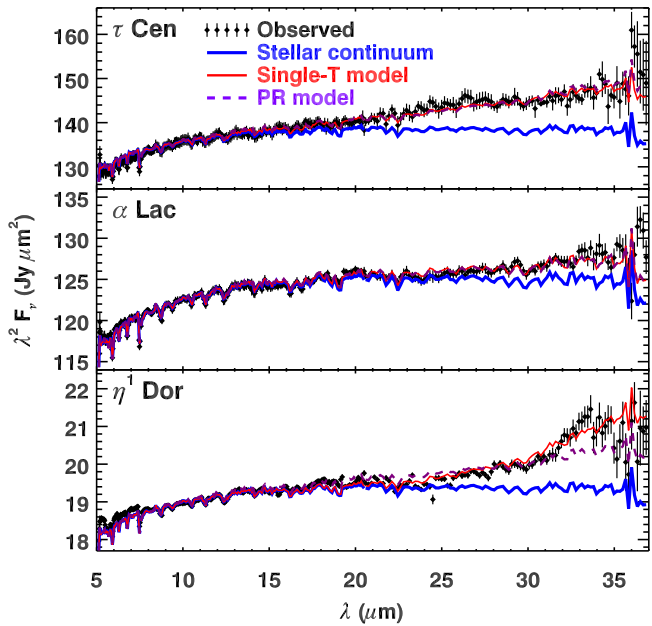}
\caption{IRS spectra of three A dwarfs with weak excesses,
showing fitted Planck functions and PR models.  The spectra
are noisy enough that the temperatures of the Planck 
functions cannot be relied on.  For $\tau$~Cen, the PR model
fits the excess slightly worse than the Planck function, 
while for $\alpha$~Lac, it is a slight improvement.  Nothing
works well for $\eta^1$~Dor, but the Planck function is the
better of the two.\label{f.excess}}
\end{figure}

Several other spectra in Figures~\ref{f.a1} and \ref{f.a2} 
show evidence of weak red excesses.  The spectra of 
$\tau$~Cen, $\alpha$~Lac, and $\eta^1$~Dor have sufficient 
SNRs for further analysis.  Other spectra suggestive of weak 
excesses include 21~Lyn, $\xi^1$~Cen, 29~Vul, $\lambda$~Tel, 
and HR~1014.  Most of this latter group show artifacts 
remaining after our attempt to correct for the 24~\mum\ dip,
making further analysis inadvisable.

Figure~\ref{f.excess} plots the spectra of $\tau$~Cen, 
$\alpha$~Lac, and $\eta^1$~Dor, along with fitted Planck
functions and PR models as done for the five sources with
stronger red excesses.  For all three sources, the 
temperature of the fitted Planck function depends strongly
on the wavelength range chosen for fitting the continuum,
and we limit our comments to the strength of the excess.

The spectrum of $\tau$~Cen shows an excess of 4.6\% at 
30~\mum.  This spectrum is based on seven separate
observations, is bright, and consistently has a red excess.
The spectrum of $\alpha$~Lac shows a 1.3\% excess at 
30~\mum, and it is based on 19 pointings.  As already
discussed, this low-contrast excess is repeatable and has 
forced us to modify our method of using $\alpha$~Lac and 
$\delta$~UMi to calibrate HR~6348.

The faint spectrum of $\eta^1$ Dor makes a good test case.  
It is based on 44 separate spectra, and the weak red excess 
has appeared consistently throughout the observations.  Some 
of the spectral structure may be due to an incomplete 
correction for the 24~\mum\ dip, and at the longest 
wavelengths, a poor SNR.  Despite these problems, the 
red excess has a shape consisent with a debris disk, and
with a 3.5\% excess at 30~\mum, is stronger than in 
$\alpha$~Lac.

\section{Summary and conclusions} % Sec. 10.0

For the spectral calibration of the IRS, we observed a large
sample of potential standards.  This approach proved wise, as 
our pre-launch plans required considerable adjustment during
the cryogenic mission.

Our sample of K giants confirms what \cite{her02} found with
the \iso/SWS data:  the strength of the SiO bands increases 
as the giants grow cooler, but with considerable scatter. 
This scatter persisted in the IRS sample despite our care in 
limiting the luminosity class to giants (``III'').  
Metallicity might explain the scatter, but that is an open
question.  The scatter in band strength limits the 
accuracy of the spectral templating method for K giants, 
which assumes that the infrared spectral properties can be 
predicted from the optical spectral class.  

The $B-V$ color predicts the SiO band strength better than 
the optical spectral class.  The optical spectral classes
are based on atomic lines, which arise from layers of the
envelope beneath those responsible for the absorption bands.

\cite{pri02} found that the synthetic spectra do not predict 
the strength of the SiO absorption band at 8~\mum\ with
sufficient accuracy to be used as truth spectra for 
calibration.  We confirm that the synthetic spectra 
consistently underpredict the actual SiO band strength.

Thus the use of either synthetic spectra or spectral 
templates as truth spectra would introduce artifacts at the 
level of a few percent into the IRS database.  With no 
reliable truth spectra for the K giants and no suitable A 
dwarfs available in the continuous viewing zones, we repeated 
the steps taken by \cite{coh92a} and used A dwarfs, which are 
more easily modeled, to calibrate the K giants, which are 
more easily observed, mitigating in the process for the 
hydrogen recombination lines and a slight red excess in one
of our A dwarfs.  The resulting truth spectrum for HR~6348 
has a spectroscopic fidelity better than 0.5\%, except in the 
immediate vicinity of the Pfund-$\alpha$ line at 7.5~\mum\
and where the SNR becomes an issue at the longest 
wavelengths.  This fidelity does not include photometric 
uncertainties or uncertaintites in overall spectral shape.

The K giants show OH band structure in the 14--18~\mum\
region with band depths of up to $\sim$3\% of the continuum
(at the resolution of LL).  We failed to include these bands
in earlier versions of the calibration, despite their clear
presence in synthetic spectra and their confirmation with the
IRS \citep{dec04}.  This mistake led to artifacts visible in
spectra with high SNRs in earlier versions of the Cornell
calibration.

The disagreement between the molecular bands in the synthetic
and observed spectra increases from the SiO band at 8~\mum\ 
to the OH bands at 14--18~\mum.  This result is consistent 
with arguments by \cite{vm04} that the H$^-$ ion could be 
driving the effective absorbing layer higher above the 
photosphere with increasing wavelength, due to its increasing 
opacity.

The Engelke function only models the continuum of K giants,
and so must be used with caution when constructing truth
spectra.  The long-wavelength behavior of the
synthetic spectra and Engelke functions differ by only 
$\sim$1\%, and the IRS data are unable to clearly distinguish
between the two.

Our concerns about the use of synthetic spectra and spectral
templates as truth spectra are limited to spectroscopic
calibration, where small deviations between assumptions and
reality can have a noticeable impact on data.  For
photometry, the resulting errors are closer to tolerances 
because of the lower spectral resolution, as long as one 
avoids filters which coincide with strong molecular bands.  
Thus the use of spectral templates to calibrate IRAC or 
synthetic spectra to calibrate photometry from the {\it 
Herschel Space Observatory} should not face the problems
described in this paper.

When we decided to concentrate on A dwarfs and K giants to
calibrate the IRS, we chose not to consider solar analogues,
because their spectra included both atomic lines and
molecular bands.  This logic was flawed, because
the atomic lines weaken from A to G, and the absorption 
bands weaken from K to G.  The relative weakness of both
in solar analogues will make uncertainties in their strength
and shape far less significant than for the classes of 
standards we chose.  For the IRS, the real problem with solar 
analogues was their rarity.  None with appropriate brightness
were available in the parts of the sky continuously visible 
from {\it Spitzer}.

The primary danger presented by A dwarfs is debris disks.
Screening by the available photometry proved insufficient,
and even then, the case of $\alpha$~Lac demonstrates the
need for caution.  While the apparent red excess is weak, 
it is still strong enough to affect the calibration at the 
longest wavelengths.

New missions bring new challenges.  The sensitivity of the
{\it James Webb Space Telescope} will force a new set of
standards.  \cite{gb11} described a plan to use a combination 
of white dwarfs, A dwarfs, and solar analogs, all of which 
are more reliably modeled than K giants.  Observations of 
K giants should still be included for cross-calibration 
with previous missions.  

The {\it Stratospheric Observatory for Infrared Astronomy
(SOFIA)} has the opposite difficulty.  Its lower sensitivity
has forced it to rely on much brighter standards like those
listed in Table~\ref{t.cohen}.  The recalibration of the
\iso/SWS spectra of several of these sources by \cite{eng06} 
should provide truth spectra which will avoid both the 
limitations of synthetic spectra and the lower SNR in the 
original data which served as the prototypes for the spectral 
templates.

\acknowledgements

The observations presented were made with the {\it Spitzer
Space Telescope}, which is operated by the Jet Propulsion
Laboratory, California Institute of Technology, under NASA
contract 1407.  NASA provided support of this work through
contract 1257184 issued by JPL/Caltech.  We are deeply
grateful to Lee Armus, the leader of the IRS Instrument 
Support Team at the SSC, for his support of the calibration 
effort and his determination to observe a sufficient sample 
of standards to ensure the quality of the IRS data.  The 
entire astronomy community will continue to benefit from this 
legacy.  The late Stephan Price deserves special mention for 
his tireless support of infrared calibration throughout his 
career.  His spirit underlies the philosophy of this paper.  
Numerous members of the IRS Team have helped with the 
calibration of the IRS, including D.~Devost, K.~I.~Uchida, 
J.~Bernard-Salas, H.~W.~W.~Spoon, and V.~Lebouteiller, then 
at Cornell, J.~Van~Cleve, then at Ball Aerospace Corp., and 
P.~W.~Morris, S.~Fajardo-Acosta, J.~G.~Ingalls, 
C.~J.~Grillmair, P.~N.~Appleton, and P.~Ogle, then at the 
SSC.  L.~Decin and M.~Cohen provided synthetic spectra and 
spectral templates prior to launch, which were essential to 
the progress we have made in calibration.
P.~S.\ Nerenberg, M.~S.\ Keremedjiev, and D.~A.\ Ludovici
contributed to this project over the years as participants in 
the Research Experience for Undergraduates at Cornell funded 
by the NSF.  In addition, we thank the anonymous referee,
whose comments led to a considerably improved manuscript.
This research has made use of NASA's Astrophysics Data System, 
the SIMBAD database, operated at Centre de Donn\'{e}es 
astronomiques in Strasbourg, France, and the Palomar Sky 
Survey.


\begin{thebibliography}{}
\bibitem[Ardila et al.(2010)]{ard10} Ardila, D.~R., Van Dyk, S.~D., 
  Makowiecki, W., et al.\ 2010, \apjs, 191, 301
\bibitem[Aumann et al.(1984)]{aum84} Aumann, H.~H., Beichman, C.~A., 
  Gillett, F.~C., et al.\ 1984, \apjl, 278, L23
% de Jong, T., Houck, J.~R., Low, F.~J., Neugebauer, G., 
% Walker, R.~G., Wesselius, P.~R.\ 
\bibitem[Beichman et al.(1988)]{psc} Beichman, C.~A., Neugebauer, G., Habing, 
  H.~J., Clegg, P.~E., \& Chester, T.~J.\ 1988, Infrared Astronomical 
  Satellite (IRAS) Catalogs and Atlases Explanatory Supplement (Pasadena:  JPL)
\bibitem[Bessell(2000)]{bes00}  Bessell, M.\ 2000, \pasp, 112, 773
\bibitem[Buscombe \& Morris(1958)]{bm58} Buscombe, W., \& Morris, P.~M.\ 
  1958, \mnras, 118, 609
\bibitem[Buscombe \& Morris(1960)]{bm60} Buscombe, W., \& Morris, P.~M.\ 
  1960, \mnras, 121, 263
\bibitem[Chiar \& Tielens(2006)]{ct06} Chiar, J.~E., \& Tielens, A.~G.~G.~M.\
  2006, \apj, 637, 774
\bibitem[Cohen et al.(1992a)]{coh92a} Cohen, M., Walker, R.~G., Barlow, M.~J., 
  \& Deacon, J.~R.\ 1992, \aj, 104, 1650
\bibitem[Cohen et al.(1992b)]{coh92b} Cohen, M., Walker, R.~G., \&
  Witteborn, F.~C.\ 1992, \aj, 104, 2030
\bibitem[Cohen et al.(1992c)]{coh92c} Cohen, M, Witteborn, F.~C., 
  Carbon, D.~F., et al.\ 1992, \aj, 104, 2045
% Augason, G., Wooden, D., Bregman, J., \& Goorvitch, D.\
\bibitem[Cohen et al.(1995)]{coh95} Cohen, M., Witteborn, F.~C., Walker, 
  R.~G., Bregman, J.~D.,\& Wooden, D.~H. 1995, \aj, 110, 275
\bibitem[Cohen et al.(1996a)]{coh96a} Cohen, M., Witteborn, F.~C., Bregman, 
  J.~D., et al.\ 1996, \aj, 112, 241
% Wooden, D.~H., Salama, A. \& Metcalfe, L.\
\bibitem[Cohen et al.(1996b)]{coh96b} Cohen, M., Witteborn, F.~C., Carbon, 
  D.~F., et al.\ 1996, \aj, 112, 2274
% Davies, J.~K., Wooden, D.~H. \& Bregman, J.~D.\
\bibitem[Cohen et al.(1999)]{coh99} Cohen, M., Walker, R.~G., Carter, B., 
  et al.\ 1999, \aj, 117, 1864 % Hammersley, P., Kidger, M., \& Noguchi, K.\
\bibitem[Cohen et al.(2003)]{coh03} Cohen, M., Megeath, S.~T., Hammersley,
  P.~L., Mart\'{i}n-Luis, F., \& Stauffer, J.\ 2003, \aj, 125, 2645
\bibitem[Cowley et al.(1969)]{cow69} Cowley, A., Cowley, C., Jaschek, M., 
  \& Jaschek, C.\ 1969, \aj, 74, 375
\bibitem[Cox (2000)]{cox00} Cox, A.~N., ed., 2000, Allen's Astrophysics
  Quantities, 4th Ed., (New York:  Springer)
\bibitem[Decin et al.(2003a)]{dec03a} Decin, L., Vandenbussche, B., Waelkens,
  C., et al.\ 2003a, \aap, 400, 679
\bibitem[Decin et al.(2003b)]{dec03b} Decin, L., Vandenbussche, B., Waelkens,
  C., et al.\ 2003b, \aap, 400, 709
\bibitem[Decin et al.(2004)]{dec04} Decin, L., Morris, P.~W., Appleton, P.~N.,
  Charmandaris, V., Armus, L., \& Houck, J.~R.\ 2004, \apjs, 154, 408
\bibitem[Decin \& Eriksson(2007)]{de07} Decin, L., \& Eriksson, K.\ 2007, \aap,
  472, 1041
\bibitem[de Veaucouleurs(1957)]{dev57} de Veaucouleurs, A.\ 1957, \mnras, 117, 
  449
\bibitem[Eggen(1950)]{egg50} Eggen, O.~J.\ 1950, \apj, 111, 414
\bibitem[Eggen(1957)]{egg57} Eggen, O.\ 1957, \aj, 62, 45
\bibitem[Eggen(1960)]{egg60} Eggen, O.~J.\ 1960, \mnras, 120, 448
\bibitem[Eggen(1962)]{egg62} Eggen, O.~J.\ 1962, Roy.\ Obs.\ Bull.\ 51
\bibitem[Eggen(1984)]{egg84} Eggen, O.~J.\ 1984, \apjs, 55, 597
\bibitem[Engelbracht et al.(2007)]{eng07} Engelbracht, C.~W., Blaylock, M., 
  Su, K.~Y.~L., et al.\ 2007, \pasp, 119, 994
\bibitem[Engelke(1992)]{eng92} Engelke, C.~W.\ 1992, \aj, 104, 1248
\bibitem[Engelke et al.(2006)]{eng06} Engelke, C.~W., Price, S.~D., \&
  Kraemer, K.~E.\ 2006, \aj, 132, 1445 % Paper XVI
\bibitem[Evans et al.(1964)]{eva64} Evans, D.~S., Laing, J.~D., Menzies, A., \& 
  Stoy, R.~H., 1964, Roy.\ Obs.\ Bull.\ 85, 207
\bibitem[Gordon \& Bohlin(2011)]{gb11} Gordon, K.~D., \& Bohlin, R., Am.\ 
  Astron.\ Soc.\ Meeting 217, 254.23
\bibitem[Gray(1985)]{gra85} Gray, R.~O.\ 1985, \jrasc, 79, 237
\bibitem[Gray et al.(2006)]{gra06} Gray, R.~O., Corbally, C.~J., Garrison, 
  R.~F., et al.\ 2006, \aj, 132, 161
%\bibitem[Gulliver et al.(1994)]{gul94} Gulliver, A.~F., Hill, G., \& Adelman,
%  S.~J.\ 1994, \apjl, 429, L81
\bibitem[Halliday(1955)]{hal55} Halliday, I.\ 1955, \apj, 122, 222
\bibitem[Heras et al.(2002)]{her02} Heras, A.~M., Shipman, R.~F., Price, S.~D.,
  et al.\ 2002, \aap, 394, 539
%  de Graauw, Th., Walker, H.~J., Jourdain de Muizon, M., Kessler, M.~F.,
%  Prusti, T., Decin, L., Vandenbussche, B., \& Waters, L.~B.~F.~M.\
\bibitem[Higdon et al.(2004)]{hig04} Higdon, S.~J.~U., Devost, D., Higdon,
  J.~L., et al.\ 2004, \pasp, 116, 975
\bibitem[Hill et al.(2010)]{hil10} Hill, G., Gulliver, A.~F., \& Adelman,
  S.~J.\ 2010, \apj, 712, 250
\bibitem[Hog et al.(2000)]{hog00} Hog, E., Fabricius, C., Marakov, V.~V., et
  al.\ 2000, \aap, 500, 583
\bibitem[Houck et al.(2004)]{hou04} Houck, J.~R., Roellig, T.~L., van Cleve,
  J., et al.\ 2004, \apjs, 154, 18
\bibitem[Houk(1978)]{mss78} Houk, N.\ 1978, Michigan Catalogue of 
  Two-Dimensional Spectral Types for the HD Stars.  Vol.\ 2 (Ann Arbor, MI:  
  Univ.\ of Michigan)
\bibitem[Houk \& Cowley(1975)]{mss75} Houk, N., \& Cowley, A.~P.\ 1975, Univ.\ 
  of Michigan Catalogue of Two-Dimensional Spectral Types for the HD Stars.  
  Vol.\ I (Ann Arbor, MI:  Univ.\ of Michigan)
\bibitem[Houk \& Smith-Moore(1988)]{mss88} Houk, N., \& Smith-Moore, M.\ 1988, 
  Michigan Catalogue of Two-dimensional Spectral Types for the HD Stars.  
  Vol.\ 4 (Ann Arbor, MI:  Univ.\ of Michigan)
\bibitem[Jaschek \& Jaschek(1990)]{jj90} Jaschek, C., \& Jaschek, M.\ 1990,
  The Classification of Stars (Cambridge, UK:  Cambridge Univ.\ Press)
\bibitem[Johnson \& Morgan(1953)]{jm53} Johnson, H.~L., \& Morgan, W.~W.\ 
  1953, \apj, 117, 313
\bibitem[Jura(1999)]{jur99} Jura, M.\ 1999, \apj, 515, 706
\bibitem[Jura et al.(2004)]{jur04} Jura, M., et al.\ 2004, \apjs, 154, 453
\bibitem[Kenyon \& Hartmann(1995)]{kh95} Kenyon, S.~J., \& Hartmann, L.\
  1995, \apjs, 101, 117
\bibitem[Kurucz(1979)]{kur79} Kurucz, R.~L.\ 1979, \apjs, 40, 1
\bibitem[Lallement et al.(2003)]{lal03} Lallement, R., Welsh, B.~Y., Vergely,
  J.~L., Crifo, F., \& Sfeir, D.\ 2003, \aap, 411, 447
\bibitem[Lebouteiller et al.(2010)]{leb10} Lebouteiller, V., Bernard-Salas, J.,
  Sloan, G.~C., \& Barry, D.~J.\ 2010, \pasp, 122, 231
\bibitem[Levato(1972)]{lev72} Levato, H.\ 1972, \pasp, 84, 584
\bibitem[Moore \& Paddock(1950)]{mp50} Moore, J.~H., \& Paddock, G.~F.\ 1950, 
  \apj, 112, 48
\bibitem[Morgan et al.(1953)]{mor53} Morgan, W.~W., Harris, D.~L., \& 
  Johnson, H.~L.\ 1953, \apj, 118, 92
\bibitem[Moshir et al.(1992)]{fsc} Moshir, M., et al.\ 1992, Explanatory 
  Supplement to the IRAS Faint Source Survey, ver.\ 2, JPL D-10015 8/92 
  (Pasadena: JPL).
\bibitem[Price et al.(2002)]{pri02} Price, S.~D., Sloan, G.~C., \& Kraemer,
  K.~E.\ 2002, \apjl, 565, L55
\bibitem[Rieke et al.(2008)]{rie08} Rieke, G.~H., Blaylock, M., Decin, L., et 
  al.\ 2008, \aj, 135, 2245
%\bibitem[Roman(1952)]{rom52} Roman, N.~G.\ 1952,, \apj, 116, 122
\bibitem[Ryde et al.(2002)]{ryd02} Ryde, N., Lambert, D.~L., Richter, M.~J.,
  \& Lacy, J.~H.\ 2002, \apj, 580, 447
\bibitem[Schaeidt et al.(1996)]{sch96} Schaeidt, S.~G., Morris, P.~W., Salama,
  A., et al.\ 1996, \aap, 315, L55
\bibitem[Shipman et al.(2003)]{shi03} Shipman, R.~F., et al.\ 2003, in The 
  Calibration Legacy of the ISO Mission, ed.\ L.\ Metcalfe, A.\ Salama, 
  S.~B.\ Peschke, \& M.~F.\ Kessler, ESA SP-481, 107 (Noordwijk, The 
  Netherlands: ESA)
\bibitem[Skrutskie et al.(2006)]{skr06} Skrutskie, M.~F., et al.\ 2006, \aj,
  131, 1163
\bibitem[Slettebak(1954)]{sle54} Slettebak, A.\ 1954, \apj, 119, 146
\bibitem[Sloan et al.(2004b)]{slo04b} Sloan, G.~C., Charmandaris, V.,
  Fajardo-Acosta, S.~B., Shupe, D.~L., Morris, P.~W., Su, K.~Y.~L., Hines,
  D.~C., Rho, J., \& Engelbracht, C.~W.\ 2004, \apjl, 614, L77
\bibitem[Sloan et al.(2003)]{slo03} Sloan, G.~C., Nerenberg, P.~S., \&
  Russell, M.~R.\ 2003, IRS-TR 03001:  The Effect of Spectral Pointing-Induced
  Throughput Error on Data from the IRS (Ithaca, NY:  Cornell)
\bibitem[Sloan \& Ludovici(2011a)]{sl11a} Sloan, G.~C., \& Ludovici, D.\ 
  2011a, IRS-TR 11001: Temporal Responsivity Variations on the Red Peak-Up 
  Sub-Array (Ithaca, NY:  Cornell)
\bibitem[Sloan \& Ludovici(2011b)]{sl11b} Sloan, G.~C., \& Ludovici, D.\ 
  2011b, IRS-TR 11002: Calibration of the Acquisition Images from the Red 
  Peak-Up Sub-Array (Ithaca, NY:  Cornell)
\bibitem[Sloan \& Ludovici(2012a)]{sl12a} Sloan, G.~C., \& Ludovici, D.\ 
  2012a, IRS-TR 12001:  Spectral Pointing-Induced Througput Error and Spectral 
  Shape in Short-Low Order 1 (Ithaca, NY:  Cornell)
\bibitem[Sloan \& Ludovici(2012b)]{sl12b} Sloan, G.~C., \& Ludovici, D.\ 
  2012b, IRS-TR 12002: Constructing a Short-Low Truth Spectrum of the Standard 
  Star HR 6348 (Ithaca, NY:  Cornell)
\bibitem[Sloan \& Ludovici(2012c)]{sl12c} Sloan, G.~C., \& Ludovici, D.\ 
  2012c, IRS-TR 12003: Constructing Low-Resolution Truth Spectra of the 
  Standard Stars HR 6348 and HD 173511 (Ithaca, NY:  Cornell)
\bibitem[Sloan et al.(2004a)]{slo04a} Sloan, G.~C., Morris, P.~W., 
  Fajardo-Acosta, S.~B., Charmandaris, V., Shupe, D.~L., Engelbracht, C.~W., 
  Hines, D.~C., Rho, J., \& Su, K.~Y.~L.\ 2004, \baas, 36, 722.
\bibitem[Stephenson(1960)]{ste60} Stephenson, C.~B.\ 1960, \aj, 65, 60
\bibitem[Su et al.(2006)]{su06} Su, K.~Y.~L., et al.\ 2006, \apj, 653, 675
\bibitem[Su et al.(2013)]{su13} Su, K.~Y.~L., et al.\ 2013, \apj, 763, 118
  % Rieke, G.~H., Malhotra, R., Stapelfeldt, K.~R., Meredith Hughes, A., ...
\bibitem[van Leeuwen(2007)]{vl07} van Leeuwen, F.\ 2007, Hipparcos:  The New
  Reduction of the Raw Data, Astrophysics and Space Science Library, 350,
  (Berlin:  Springer)
\bibitem[Van Malderen et al.(2004)]{vm04} Van Malderen, R., Decin, L., Kester,
  D., et al.\ 2004, \aap, 414, 677
\bibitem[Werner et al.(2004)]{wer04} Werner, M.~W., Roellig, T.~L., Low, F.~J.,
  et al.\ 2004, \apjs, 154, 1
\bibitem[Wildt(1939)]{wil39} Wildt, R.\ 1939, \apj, 90, 611
\bibitem[Wyatt(2005)]{wya05} Wyatt, M.~C.\ 2005, \mnras, 433, 1007
\bibitem[Yoss(1961)]{yos61} Yoss, K.~M.\ 1961, \apj, 134, 809
\end{thebibliography}
\end{document}